\documentclass{aa}  
\usepackage{graphicx}
\usepackage{txfonts}
\usepackage[normalem]{ulem}
\usepackage[colorlinks=true, linkcolor=blue, citecolor=blue, filecolor=blue, urlcolor=blue]{hyperref}
\usepackage{adjustbox}

\begin{document}

   \title{\textsc{GalSBI-SPS}: a stellar population synthesis-based  galaxy population model for cosmology and galaxy evolution applications}

   \author{Luca Tortorelli
          \inst{1}
          \and
          Silvan Fischbacher \inst{2}
          \and 
          Daniel Gr\"un \inst{1,3}
          \and
          Alexandre Refregier \inst{2}
          \and
          Sabine Bellstedt \inst{4}
          \and \\
          Aaron S. G. Robotham \inst{4}
          \and
          Tomasz Kacprzak \inst{5}
          }

   \institute{Universit\"ats-Sternwarte, Fakult\"at f\"ur Physik, Ludwig-Maximilians-Universit\"at M\"unchen, Scheinerstr. 1, 81679 M\"unchen, Germany\\
              \email{luca.tortorelli@physik.lmu.de}
         \and
            Institute for Particle Physics and Astrophysics, ETH Zurich, Wolfgang-Pauli-Strasse 27, CH-8093 Zurich, Switzerland
         \and
             Excellence Cluster ORIGINS, Boltzmannstr. 2, 85748 Garching, Germany
         \and
            ICRAR, The University of Western Australia, 7 Fairway, Crawley WA 6009, Australia
         \and
         Swiss Data Science Center, Paul Scherrer Institute, Forschungsstrasse 111, 5232 Villigen, Switzerland
             }

    \date{Received 31 May 2025; accepted 21 September 2025}
 
  \abstract
   {Next-generation photometric and spectroscopic surveys will enable unprecedented tests of the concordance cosmological model and of galaxy formation and evolution. Fully exploiting their potential requires a precise understanding of the selection effects on galaxies and biases on measurements of their properties, required, above all, for accurate estimates of redshift distributions. The forward-modelling of galaxy surveys offers a powerful framework to simultaneously recover galaxy redshift distributions and characterise the observed galaxy population.}
   {We present \textsc{GalSBI-SPS}, a new stellar population synthesis (SPS)-based galaxy population model developed for cosmological and galaxy evolution studies. The model generates realistic galaxy catalogues, which we use to forward-model Hyper-Suprime Cam (HSC) observations in the COSMOS field.}
   {\textsc{GalSBI-SPS} samples galaxy physical properties from analytical parametrisations informed by GAMA, DEVILS, and literature data, computes galaxy magnitudes with the generative SED package \textsc{ProSpect}, and simulates HSC images in the COSMOS field with \textsc{UFig}. We measure photometric properties consistently in real data and simulations. We compare redshift distributions, photometric and physical properties to observations and to those from the phenomenological \textsc{GalSBI} model.}
   {\textsc{GalSBI-SPS} reproduces the observed $g,r,i,z,y$ magnitude, colour, and size distributions down to $i \le 23$ with good accuracy. Median differences in magnitudes and colours remain below $0.14 \ \mathrm{mag}$, with the model covering the full colour space spanned by HSC data. Galaxy sizes are overestimated by $\sim 0.2\arcsec$ on average and some tension exists in the $g-r$ colour distribution, but the latter is comparable to that seen in the phenomenological \textsc{GalSBI} model. Redshift distributions show a mild positive offset ($0.01 \lesssim \Delta \bar{z} \lesssim 0.08$) in the mean. \textsc{GalSBI-SPS} qualitatively reproduces the stellar mass–star formation rate and size–stellar mass relations seen in COSMOS2020 data.}
  {\textsc{GalSBI-SPS} provides a realistic, survey-independent description of the galaxy population at a Stage-III-like depth using only literature-based parameters. Its predictive power is expected to improve significantly when constrained against deep observed data using simulation-based inference, thereby providing accurate redshift distributions satisfying the stringent requirements set by Stage IV surveys.}


   \titlerunning{\textsc{\textsc{GalSBI-SPS}}}

   \maketitle

\section{Introduction}
\label{sect:introduction}

We are entering a transformative era for cosmology and galaxy evolution, driven by current and upcoming wide-field photometric and spectroscopic surveys. The recent baryon acoustic oscillation (BAO, \citealt{Blake2003,Seo2003}) measurements from three years of operation of the Dark Energy Spectroscopic instrument (DESI) challenged the standard $\Lambda$CDM model by pointing towards a Universe with a time-evolving dark energy equation of state \citep{DESIY3}. Meanwhile, cosmic shear measurements from the complete Kilo-Degree Survey (KiDS-Legacy, \citealt{Wright2025b}) have found the $S_8 \equiv \sigma_8 \sqrt{\Omega_{\mathrm{m}}/0.3}$ parameter to be consistent with Planck Legacy constraints \citep{Wright2025a,Stolzner2025}, potentially resolving the longstanding `$S_8$ tension' \citep{Abdalla2022,Amon2022,Dalal2023,DES-KiDS,DiValentino2025} between early and late-time cosmological probes. Euclid has recently released the Quick data release 1 \citep{EUCLIDQ1}, based on $63.1 \ \mathrm{deg}^2$ of the Euclid Deep Fields \citep{Mellier2024} to nominal wide-survey depth. These data span a wide range of science cases, including studies of the star-forming \citep{Enia2025} and passive \citep{Cleland2025} galaxy populations, as well as the identification of strong gravitational lensing systems from individual galaxies \citep{Rojas2025} to massive clusters \citep{Bergamini2025}. Looking ahead, upcoming surveys such as the Vera C. Rubin Observatory’s Legacy Survey of Space and Time (hereafter, Rubin-LSST, \citealt{Ivezic2019}) and spectroscopic programs like 4MOST \citep{Dejong2019} are expected to deliver the most precise constraints to date on the large-scale structure (LSS) of the Universe. These will significantly improve measurements of key cosmological parameters while also enabling detailed studies of structure growth and galaxy evolution at low \citep{Driver2019} and intermediate redshifts \citep{Iovino2023}.

Next-generation surveys will probe a regime where the statistical uncertainties of measurements are subdominant compared to the systematic ones. Maximising the scientific return therefore requires a precise characterisation of measurement biases and incompleteness. An exemplary case is the use of weak gravitational lensing (see \citealt{Mandelbaum2018} for a comprehensive review) as a powerful probe of the LSS. Its ability to constrain key cosmological parameters, such as the clustering amplitude of the matter power spectrum, $\sigma_8$, and the matter density parameter, $\Omega_{\mathrm{m}}$, is highly sensitive to uncertainties in the redshift distribution of source galaxies, especially the mean redshift \citep{Amon2022,vandenbusch2022,Li2023,Dalal2023}. Three notable recent studies highlight this challenge.  The first is an improved redshift calibration method and sample that significantly enhanced the cosmic shear analysis of the KiDS-Legacy dataset \citep{Wright2025c}. The second involves the Hyper-Suprime Cam (HSC) Year 3 cosmic shear analysis \citep{Dalal2023} that struggled to tighten constraints over the Year 1 analysis despite a three-fold increase in sky area, due to the wide prior placed on their higher redshift tomographic bins. The third example is that of the Dark Energy Survey (DES) that discarded the two highest redshift bins in the MagLim sample, with the redshift distribution estimates being the dominant systematic error in DES weak lensing measurements \citep{DESY3}. Given these limitations, Stage IV experiments \citep{Albrecht2006}, such as Euclid and Rubin-LSST, have set stringent requirements on the systematic uncertainty in the mean redshift of each source tomographic bin. For the Rubin-LSST weak lensing analysis, the systematic uncertainty should not exceed mean redshift errors of $|\delta z|<0.002 \times (1 + z)$ in the year 1 analysis, and $0.001 \times (1 + z)$ in the year 10 analysis \citep{LSSTDESC2018}. However, current photometric redshift (photo-z) calibration methods from Stage III dark energy experiments \citep{Myles2021,Wright2020,vandenbusch2022} still fall short of these goals. They yield uncertainties on the mean redshift that are degraded by spectroscopic incompleteness and outliers in the calibration redshifts, sample variance, blending and astrophysical systematics that cause systematic errors in the mean and standard deviation of redshift to still be one order of magnitude greater than what is required for Stage IV experiments (see \citealt{NewmanGruen2022} for a comprehensive review). Efforts are on-going in calibrating the colour-redshift relation to explicit target the colour manifold spanned by Stage IV surveys \citep{Masters2015,Masters2017,Masters2019,Guglielmo2020,Stanford2021,Saglia2022,Gruen2023,McCullough2024}. However, these observational efforts need to be complemented with an accurate assessment of their incompleteness in the targeted galaxy population, especially at faint magnitudes, where a significant gap remains in the spectroscopic coverage. Similar issues arise when using survey data to study the evolution of the galaxy population. In photometric surveys, redshift uncertainties hinder the accurate reconstruction of galaxy evolution across cosmic time, particularly at high redshift. In spectroscopic surveys, the complicated spectroscopic selections often lead to a difficult assessment of the stellar mass completeness or the contamination of a different population of objects with respect to that selected via photometry, potentially biasing the conclusions one draws from the analysis of these samples.

The forward-modelling of photometric and spectroscopic galaxy surveys is a powerful approach with which one can tackle both the problem of accurate galaxy redshift distribution estimates and the characterisation of the galaxy population underlying the observations of a survey. This methodology involves constructing a detailed, end-to-end model of a galaxy survey. Key components are a realistic galaxy population model (i.e. the intrinsic distribution of redshift-evolving physical properties of galaxies), a model for the galaxy spectral energy distributions (SEDs) that connects the galaxy physical properties to their emitted light, a mapping of the intrinsic properties and fluxes to the observed ones by means of image and spectra simulators, and informative observed data against which compare and constrain the model. By forward-modelling all relevant observational and instrumental effects and calibrating model parameters using simulation-based inference (SBI, \citealt{Cranmer2020}) methods such as Approximate Bayesian Computation (ABC, \citealt{Akeret2015}),  the galaxy redshift distributions constructed from observed data can be replaced by the ones that have been accurately determined via simulations in a cosmological analysis. Moreover, forward-modelling enables the simultaneous inference of the underlying relations between galaxy physical properties, offering a consistent framework for studying galaxy evolution across cosmic time. Compared to other modelling approaches, such as semi-analytic models or hydrodynamical simulations, forward-modelling offers two main advantages: it allows for fast realisations of mock galaxy catalogues, enabling efficient exploration of high-dimensional parameter spaces and of the detailed simulation of the observational and measurement process, and it has, by construction, matched properties distributions with the observations it aims to reproduce at all redshifts. 

The concept of forward-modelling galaxy surveys for cosmological applications was first introduced in \cite{Refregier2014} and further developed in a series of subsequent studies \citep{Herbel2017,Tortorelli2018,Tortorelli2020,Fagioli2018,Fagioli2020,Tortorelli2021, Bruderer2016,Moser2024,Fischbacher2025a}. These works adopted a common phenomenological galaxy population model, which has been progressively refined by incorporating deeper photometric data across a broader range of wavebands, including narrow-band filters \citep{Tortorelli2018,Tortorelli2021}, and tested against spectroscopic data from SDSS \citep{Fagioli2018,Fagioli2020}. A major strength of this approach is its ability to accurately simulate both photometric and spectroscopic datasets by means of fast and realistic image (\textsc{UFig}, \citealt{Berge2013,Bruderer2016,Fischbacher2024ufig}) and spectra simulators (\textsc{USpec}, \citealt{Fagioli2018,Fagioli2020}; Tortorelli et al. ,  in prep.). These tools allow for the possibility of consistently apply the exact same measurement pipelines and selection cuts on data and simulations. This ensures a robust and consistent comparison of observables and properly accounts for systematics introduced by the measurement process itself, including, for example, the impact of rejecting blended sources, a source of bias expected to significantly affect cosmological constraints from Stage IV surveys (see e.g. \citealt{Ramel2024} and references therein). Using this methodology, \cite{Kacprzak2020} performed a cosmic shear analysis with DES Y1 data, while \cite{Tortorelli2020,Tortorelli2021} measured the B-band galaxy luminosity function from the Canada-France-Hawaii Telescope Legacy Survey \citep{Cuillandre2012}, and inferred galaxy physical properties from the narrow-band SEDs in the Physics of the Accelerating Universe  survey \citep{Navarro-Girones2024}. 

The latest iteration of this phenomenological model,  named \textsc{GalSBI}, has been made publicly available in \citet{Fischbacher2025a,Fischbacher2024b}.  This model builds upon the foundational works of \citet{Herbel2017,Tortorelli2018,Tortorelli2020,Kacprzak2020,Tortorelli2021,Moser2024}, incorporating analytical parametrisations of galaxy luminosity functions, morphologies, and SEDs.  A key innovation introduced in \citet{Fischbacher2025a} is the development of an emulator that directly generates catalogues with observed-like galaxy properties from the true input ones. This emulator, composed of a neural network classifier and a normalising flow, bypasses the computationally expensive need to simulate survey images using \textsc{UFig}, significantly accelerating the forward-modelling process. This approach enabled the authors to explore 40 different configurations of the model and analysis choices. Combined with a highly informative dataset, such as the HSC deep fields \citep{Aihara2018}, the authors placed very tight constraints on \textsc{GalSBI} parameters, with this phenomenological model being a realistic representation of the galaxy population visible in a Stage IV precursor dataset like HSC. This is testified by the strong agreement between in the magnitude, colour and size distributions down to $i \le 25$ and by the agreement of the simulated tomographic and non-tomographic redshift distributions with those derived from COSMOS photo-$z$ catalogues \citep{Weaver2022}, computed using both the \textsc{EAZY} \citep{Brammer2008} and \textsc{LePhare} \citep{Arnouts1999,Ilbert2006} codes, in a Stage-III setup. Some tension is still present in a Stage-IV setup, a gap that we aim at filling with a well constrained physical model of the galaxy population.

Alternative forward-modelling approaches to the one adopted by \textsc{GalSBI} have recently emerged in the literature, offering complementary strategies that differ in several key aspects. Some of these aspects are the treatment of the noise model for the data, the modelling of galaxy SEDs, and the way galaxy properties are sampled from the underlying population model. For example, \cite{Sudek2022} used the \textsc{SkyPy} framework \citep{Amara2021} to forward-model photometric data and investigate the sensitivity of redshift distributions to assumptions about galaxy demographics. Their approach is based on the works of \cite{Herbel2017,Fagioli2018,Fagioli2020,Tortorelli2018,Tortorelli2020,Tortorelli2021}, and relies on empirically derived SED templates, which are typically governed by a small set of parameters. A more physically motivated alternative is the use of evolutionary stellar population synthesis (hereafter, SPS,  \citealt{Tinsley1980,Pickles1985,Bruzual2003,Maraston2005,Conroy2009,Vazdekis2016,Conroy2013,Iyer2025}), a technique which exploits the knowledge of stellar evolution to model the starlight component of galaxy SEDs by combining the emission due to stars using a stellar initial mass function (IMF,  \citealt{Salpeter1955,Chabrier2003,Kroupa2002}) and a star-formation history (SFH) according to given isochrones and atmosphere libraries.  SPS, when combined with models for the chemical evolution, dust attenuation and emission, gas and active galactic nuclei (AGN) emission,  allows for directly connecting the physical properties of galaxies to their emitted light by accounting for all these major SED modelling components. Though computationally more expensive,  SPS offer a richer and more flexible description of galaxy spectra.  Well-established SPS-based codes, such as \textsc{FSPS} \citep{Conroy2009,Conroy2010}, have been integrated into galaxy population models like \textsc{PROSPECTOR-$\alpha$} \citep{Leja2017} and \textsc{PROSPECTOR-$\beta$} \citep{Wang2023} to generate galaxy SEDs from the physical properties drawn from their model priors. \textsc{PROSPECTOR-$\beta$}, in particular, was used in \cite{Tortorelli2024} to forward model the 9 band $u,g,r,i,Z,Y,J,H,Ks$ photometry from the KiDS-VIKING \citep{Wright2019} survey and evaluate the impact of SED modelling choices on forward modelling-based redshift distribution estimates. Another notable example is the forward-modelling framework developed by \citet{Alsing2023} and \citet{Leistedt2023}, which also uses SPS, but accelerates magnitude predictions via an emulator of \textsc{FSPS}, named \textsc{SPECULATOR} \citep{Alsing2020}. This framework replaces full image or spectral simulations with a data-driven noise model, enabling efficient estimation of redshift distributions for photometric surveys. The initial implementation of their model was based on analytical parametrisations of the galaxy population \citep{Alsing2023}. In follow-up works \citep{Alsing2024,Thorp2024,Thorp2025}, the authors replaced this analytical model with a diffusion model (\textsc{pop-cosmos}) that was introduced to learn the joint distribution of galaxy physical properties by matching observed magnitudes in 26 bands from the COSMOS 2020 catalogue \citep{Weaver2022}. The pre-trained \textsc{pop-cosmos} model can also be used as an efficient Bayesian method for estimating individual photo-zs and galaxy properties, leveraging GPU-accelerated affine-invariant ensemble sampler to achieve fast posterior sampling \citep{Thorp2024}. \citet{Hahn2023,Hahn2024} introduced the Bayesian SED modelling framework \textsc{PROVABGS} to forward-model the DESI Bright Galaxy Sample (BGS; \citealt{Hahn2023BGS}), enabling inference of both individual and population-level SPS parameters using the \textsc{PopSED} framework \citep{Li2024}. In a parallel development, \citet{Alarcon2023} proposed a parametric model for galaxy assembly histories, which, in conjunction with the differentiable SPS code \textsc{DSPS} \citep{Hearin2023}, was used to construct the extra-galactic forward model \textsc{Diffsky} and to populate the OpenUniverse2024 simulation suite \citep{OpenUniverse2025} with synthetic galaxies.

In this paper, we introduce a new SPS-based galaxy population model called \textsc{GalSBI-SPS}. Developing a realistic galaxy population model based on physical properties and SPS is crucial for the success of forward-modelling Stage IV cosmological experiments. Such a model allows for leveraging information from more constraining dataset (low-redshift samples, deep fields, or spectroscopic datasets pre-selected through photometry) and transfer that information to the fainter, magnitude-limited samples targeted by upcoming surveys. In a purely data-driven framework, such extrapolation beyond the training set is inherently uncertain. In a phenomenological model, this extrapolation is somewhat more viable, provided that the SED templates calibrated at low redshift remain valid at higher redshifts. In contrast, a physically motivated model like \textsc{GalSBI-SPS} enables a more principled and flexible extrapolation. By sampling galaxy physical properties from redshift-evolving galaxy stellar mass functions (GSMFs) and galaxy SFH priors, one can construct higher-redshift analogs of galaxies observed at low redshift, i.e. fainter and younger versions that retain physical consistency.  This capability is a direct consequence of using SPS coupled with models for the gas, dust and AGN emission and absorption, which allows for the generation of realistic galaxy SEDs across cosmic time based on galaxy physical properties.  As a result, SPS enables the detailed simulation of the faint and high redshift galaxy population that Stage IV surveys are going to target. Moreover, the integration of SPS-based forward-modelling with image and spectra simulators such as \textsc{UFig} and \textsc{USpec} places us in a unique position of being able to characterise the full selection function of a survey as a function of galaxy physical properties, instrumental configuration, target selection criteria, and observing conditions.

The prescriptions implemented in \textsc{GalSBI-SPS} are motivated by the findings in \cite{Tortorelli2024}, where we conducted a sensitivity analysis on how different SED modelling choices affect the mean and scatter of tomographic galaxy redshift distributions in SPS-based forward-modelling. In \cite{Tortorelli2024}, the joint distribution of physical galaxy properties were sampled from the \textsc{PROSPECTOR-$\beta$} \citep{Wang2023} model, which uses the \textsc{FSPS} code to compute intrinsic galaxy spectra and magnitudes. In contrast, the present work introduces a completely new set of analytical parametrisations from which we sample galaxy redshifts and physical properties, such as galaxy stellar masses, SFHs, gas-phase metallicities,  sizes, internal dust extinction, and emission from the central AGN. This is motivated by the fact that, in \textsc{PROSPECTOR-$\beta$}, galaxy SFHs, gas and dust properties are sampled from either uniform or data-driven priors, that are independent of other galaxy physical properties. The physical properties we sample from \textsc{GalSBI-SPS} are then passed to generative codes that makes use of SPS to generate galaxy SEDs and magnitudes. The generative SED code we use is the second noteworthy change with respect to \cite{Tortorelli2024}. We replace \textsc{FSPS} with the generative SED package \textsc{ProSpect}  \citep{Robotham2020} to generate galaxy SEDs. This is motivated by \textsc{ProSpect} execution speed, high-resolution capabilities and ability to input user-defined simple stellar populations (SSPs). The resulting intrinsic properties and magnitudes are then provided as input to the image simulator \textsc{UFig}, which we employ to simulate the entire COSMOS field footprint covered by HSC Deep data. We then process the observed and the simulated images using \textsc{Source Extractor} \citep{Bertin1996}, applying the same selection criteria as in \cite{Fischbacher2025a}. This pipeline allows us not only to test the \textsc{GalSBI-SPS} model against observed galaxies, detected by running \textsc{Source Extractor} on real HSC images, but also to carry out a robust cross-validation with \textsc{GalSBI}. Both the SPS and the phenomenological version of the model are included in the same publicly released \textsc{Python} package \textsc{galsbi}\footnote{https://cosmo-docs.phys.ethz.ch/galsbi/} \citep{Fischbacher2024b}.

In this work, we adopt parameter values for our analytical prescriptions that are either modelled on data from the Galaxy and Mass Assembly survey (GAMA,\citealt{Driver2011,Liske2015}) and Deep Extragalactic VIsible Legacy survey (DEVILS, \citealt{Davies2018}) or drawn from existing literature values. We do not perform SBI to constrain the model parameters against observational data, as the primary goal of this study is to introduce and motivate the set of prescriptions used in \textsc{GalSBI-SPS} to model the galaxy population.  SBI is able to leverage the enormous constraining power that results by jointly analysing photometric  and spectroscopic data.  Our forward-modelling framework uniquely enables such a joint analysis, thanks to the physically motivated galaxy population model and the ability to simulate both images and spectra from a single parent galaxy catalogue. Each dataset is in principle able to provide constraining power on specific aspects of the galaxy population model, e.g. photometric data are highly sensitive to the GSMF, while spectroscopic data constrains aspects of the gas and stellar population physics. Moreover, the use of emulators, such as the one introduced in \cite{Tortorelli2025} and \cite{Fischbacher2025a}, allows for rapidly testing various analytical prescriptions for galaxy properties and different analysis choices, circumventing the computational cost of simulating images and spectra across large survey footprints.

The paper is structured as follows. In Sect. \ref{sect:data}, we describe the datasets used to model the analytical prescriptions and whose catalogues are later used as a comparison for our simulated properties. Section \ref{sect:gal_pop_model} presents the galaxy population model and the SPS-based SED modelling code employed to generate galaxy SEDs. In Sect.  \ref{sect:hsc_image_sims}, we detail the forward-modelling of the HSC survey and the generation of realistic simulated images covering the COSMOS field.  Section \ref{sect:results} shows the comparison between observed and simulated photometric and physical galaxy properties. In Sect. \ref{sect:discussion}, we discuss the results of the comparison and the strategies with which one can constrain the physical model. Section \ref{sect:conclusions} summarises the main findings of this study. Throughout this work, we use $H_0 = 67.8 \ \mathrm{km\ s^{-1} \ Mpc^{-1}}$ in a flat $\Lambda$CDM cosmology with $\Omega_\mathrm{m} = 0.308$, which corresponds to the \textit{Planck} 15 cosmology \citep{Planck2015} used in \cite{Bellstedt2020,Bellstedt2021,Bellstedt2024}.

\section{Data}
\label{sect:data}

In this section, we present the observational datasets used in this work. The data are used in two steps. We first model the analytical prescriptions between galaxy physical properties using catalogues with spectroscopic redshifts. Then, we test whether the modelled relations allow us to reproduce magnitudes, colours, and sizes from deeper photometric observed data. The data with spectroscopic redshifts are used to model SFHs, dust, AGN and morphological parameters, while the other physical properties are drawn from analytical prescriptions taken from the literature.

\subsection{Galaxy And Mass Assembly Survey}

GAMA \citep{Driver2011,Liske2015}) is a spectroscopic survey conducted at the 3.9m Anglo-Australian Telescope using the AAOmega fibre-fed spectrograph. Completed in 2015, GAMA represents one of the largest spectroscopic surveys of the low redshift Universe ($z < 0.25$), comprising over $\sim 230000$ spectra with reliable redshifts. GAMA covers a total area of $\sim 286 \ \mathrm{deg}^2$ down to an $r$-band Petrosian magnitude of $r < 19.8 \ \mathrm{mag}$, distributed across three equatorial fields (G09, G12, G15) and two southern fields (G02, G23). GAMA is complemented by multi-wavelength imaging from the UV to the far-IR, making it a legacy dataset that has enabled a wide range of galaxy evolution studies. These include measurements of merger rates \citep{Robotham2014,Davies2015}, group catalogues \citep{Robotham2011,Taylor2020}, low-z galaxy population properties \citep{Driver2012,Loveday2012,Bellstedt2020,Bellstedt2021,Thorne2022b,Bellstedt2024}, and constraints on the low-z GSMF down to $10^{6.75} M_{\odot} \mathrm{h}_{70}^{-2}$ \citep{Baldry2012,Wright2018,Thorne2021,Driver2022,Sbaffoni2025}.

In this work, we use the \textsc{ProSpect} value added catalogue\footnote{https://www.gama-survey.org/dr4/schema/table.php?id=721} used in \cite{Bellstedt2020,Bellstedt2021,Bellstedt2024}, which is part of the GAMA fourth data release (DR4, \citealt{Driver2022}). This catalogue provides stellar masses, star-formation rates (SFRs), gas-phase metallicities, and dust parameters derived with \textsc{ProSpect} in SED-fitting mode for $233,833$ galaxies in the GAMA equatorial and G23 fields. The sample spans the redshift range $0.002 < z < 4.7$, with median redshift $Me(z) = 0.225$ and $95\%$ of galaxies below $z<0.5$. The stellar population properties were obtained with \textsc{ProSpect} using SSPs produced with the \cite{Bruzual2003} evolutionary SPS code, the \cite{Chabrier2003} IMF, the skewed Normal SFH parametrisation \citep{Robotham2020}, a linearly evolving gas-phase metallicity \citep{Robotham2020}, the \cite{Charlot2000} dust attenuation model and dust emission modelled following \cite{Dale2014}. To ensure our calibration sample is representative of the underlying population at fixed stellar mass, we apply the spectroscopic completeness limit from \cite{Robotham2014}, which avoids biases toward bright, star-forming systems. Details of this selection are provided in Appendix \ref{appendix:gama_completeness}. The resulting stellar-mass-complete sample used in this work contains $101,784$ galaxies.

In Sect.  \ref{sect:morpho_params}, we additionally use the morphological measurements from the \cite{Kelvin2012} value added catalogue\footnote{https://www.gama-survey.org/dr4/schema/dmu.php?id=101}, which provides single-component S\'ersic profile fits for $221,373$ galaxies in the GAMA II equatorial fields. The fit was performed with \textsc{SIGMA}, an \textsc{R} wrapper that interfaces \textsc{Source Extractor}. \textsc{PSF Extractor} \citep{Bertin2011} and \textsc{GALFIT} \citep{PengGALFIT}. In this work, we use the the S\'ersic fits from SDSS $u,g,r,i,z$ imaging, although the GAMA database also includes fits in the near-IR from UKIDSS-LAS $YJHK$ and VIKING $ZYJHK$ imaging.

\subsection{Deep Extragalactic VIsible Legacy Survey}

DEVILS \citep{Davies2018} is a spectroscopic survey conducted with the same instrument as GAMA, the AAOmega fibre-fed spectrograph on the 3.9m Anglo-Australian Telescope, but designed to target intermediate redshift galaxies ($0.3 < z < 1.0$) with high spectroscopic completeness ($> 85 \%$) down to $Y < 21.2 \ \mathrm{mag}$. The DEVILS spectroscopic sample comprises $\sim 60,000$ objects over a combined area of $\sim 6 \ \mathrm{deg}^2$ across three deep extra-galactic fields: COSMOS, ECDFS, and XMM-LSS. DEVILS can be considered the higher redshift counterpart to GAMA, sharing key science goals such as characterising group and pair environments and investigating how these environments influence galaxy evolution over the last $8 \ \mathrm{Gyr}$. 

This work makes use of the galaxy physical properties measured with \textsc{ProSpect} using the spectroscopic and photometric data from the D10-COSMOS field of DEVILS \citep{Thorne2021,Thorne2022}. This catalogue has been used to measure the evolution of the GSMF and the stellar mass-SFR relation out to $z \sim 4.25$ \citep{Thorne2021}, and to constrain the evolution of the bolometric AGN luminosity function \citep{Thorne2022}. The catalogue includes stellar masses, SFRs, gas-phase metallicities, dust parameters and AGN luminosities for $494,084$ sources over the redshift range $0.0002 < z < 9$, with a median redshift of $Me(z) = 1.24$ and  $95\%$ of the sample below $z<4$. Roughly $\sim 24$k out of $494,084$ sources have spectroscopic redshifts from DEVILS, while for the remaining ones the redshifts have been estimated via photometry. We apply the stellar mass completeness criterion defined in \cite{Thorne2022}, yielding a final sample of $38,066$ galaxies. The selection procedure is detailed in Appendix \ref{appendix:devils_completeness}.

\begin{figure*}
   \centering
   \includegraphics[width=\hsize]{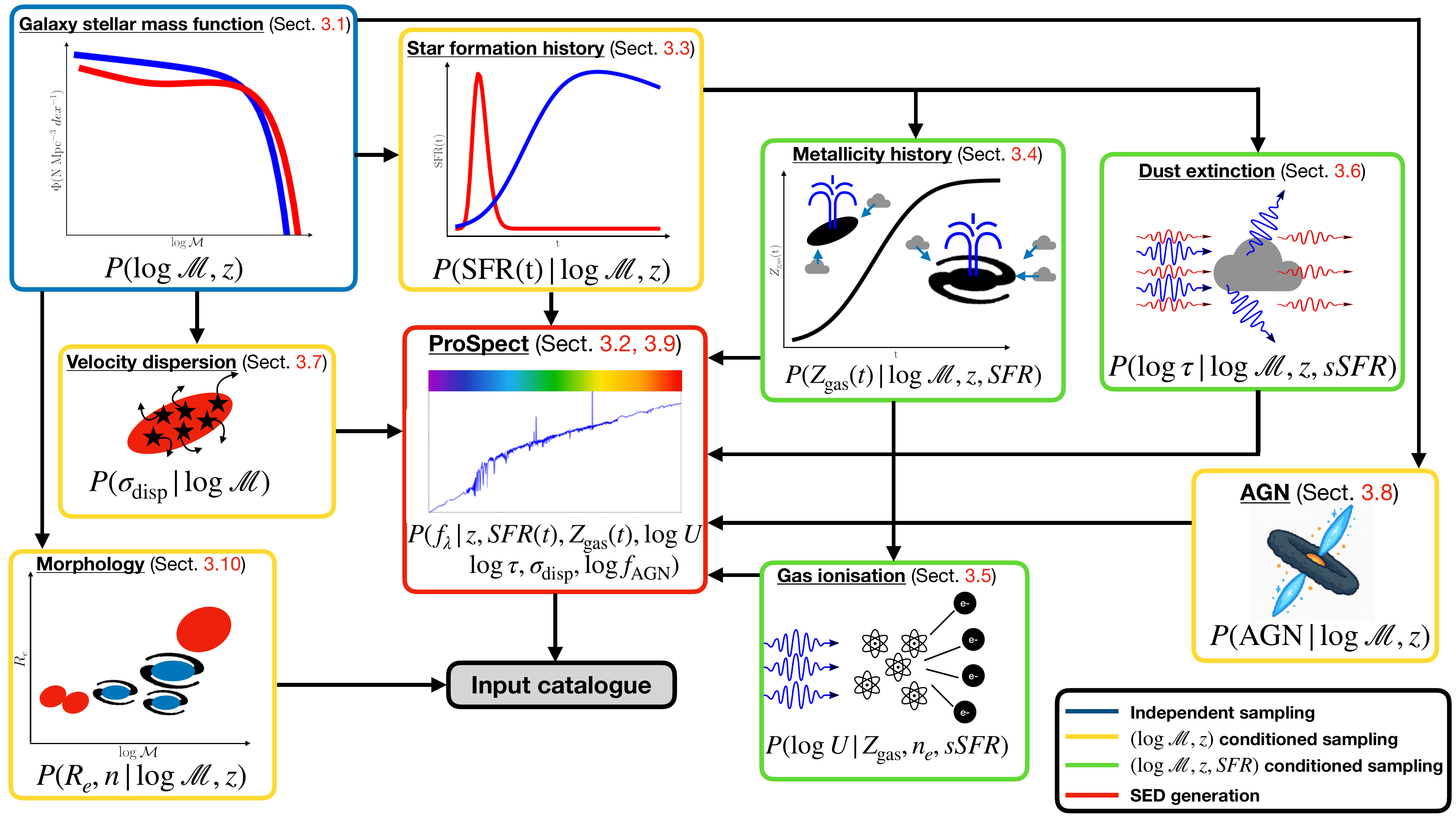}
      \caption{Flowchart describing the main components of the \textsc{GalSBI-SPS} galaxy population model. Galaxy physical properties are sampled hierarchically from analytical distributions and conditional relations, as detailed in Sect. \ref{sect:gal_pop_model}. The box colours reflect the hierarchy of dependencies. Blue boxes represent properties that are independently sampled, with the GSMF being the starting point of the sampling. Yellow boxes represent properties conditioned only on galaxy stellar mass and redshift (namely, the galaxy SFH, velocity dispersion,  morphology, and AGN). Green boxes represent properties conditioned simultaneously on galaxy stellar mass, redshift, and SFH (galaxy metallicity history, dust attenuation, and gas ionisation).  All sampled properties are then passed to \textsc{ProSpect} (red box) to generate galaxy SEDs and magnitudes. Together with the morphological properties they form the input catalogue for the forward-modelling. Each model component is described in its corresponding subsection of Sect. \ref{sect:gal_pop_model}.
      }
    \label{fig:flowchart_model}
\end{figure*}

\subsection{Hyper-Suprime Cam Subaru Strategic Program}

The HSC Subaru Strategic Program \citep{Aihara2018} is a deep multi-band imaging survey conducted with the 8.2-$\mathrm{m}$ Subaru telescope using the HSC camera, covering the $g,r,i,z,y$ filters with exquisite image quality and depth. The survey is structured into three sub-surveys of increasing photometric depth. The Wide survey, the shallowest one ($r \sim 26 \ \mathrm{mag}$), covers a large equatorial area of $1470 \ \mathrm{deg}^2$, while the Deep ($r \sim 27 \ \mathrm{mag}$) and the Ultra Deep ($r \sim 28 \ \mathrm{mag}$) surveys cover a smaller area of just $\sim 36 \ \mathrm{deg}^2$ in four different equatorial and northern pointings. HSC was designed to address a wide range of science goals, including weak-lensing tests of cosmology, galaxy formation and evolution over cosmic time, and the detection of high redshift galaxies.  Its depth, photometric precision, and high galaxy surface density make it an ideal precursor dataset for upcoming Stage IV surveys such as Rubin-LSST \citep{Ivezic2019}. 

In this work, we use the publicly available data from the HSC third data release \citep{Aihara2022} in the COSMOS field. This data are the same ones used in \cite{Fischbacher2025a} to constrain the phenomenological version of \textsc{GalSBI}, leveraging HSC constraining power for the higher redshift population. The HSC Deep and Ultra-deep (DUD) data overlaps with the COSMOS field and therefore with the COSMOS2020 \citep{Weaver2022} photo-z catalogue in what is labelled as the \textsc{COMBINED} footprint. The latter refers to the part of the COSMOS field where there is full coverage of optical/near-infrared data. HSC DUD in the COSMOS area contains 63 patches, out of which 56 are almost fully covered. We use these 56 patches as they are the same ones used in \cite{Moser2024,Fischbacher2025a}. Each patch corresponds to a $12 \times 12 \ \mathrm{arcmin}^2$ region, equivalent to a $4200 \times 4200$ pixel image given the HSC pixel scale of $0.168\arcsec$ per pixel.

\subsection{COSMOS2020}
\label{sect:cosmos}

COSMOS2020 \citep{Weaver2022} is the latest release of the COSMOS catalogue, comprising over $1.7$ million sources with multi-band photometry from far-UV to mid-infrared and stellar mass completeness down to $10^9$ $M_{\odot} \ \mathrm{h}_{70}^{-2}$ at $z \sim 3$. The COSMOS2020 catalogue represents the current state-of-the-art in photo-z catalogues thanks to its depth, wavelength coverage, and photometric accuracy. There are a total of four publicly available COSMOS2020 catalogues that differ in the code used to extract the photometry. Following \cite{Moser2024,Fischbacher2025a}, we use the \textsc{CLASSIC} catalogue, where the optical/near-infrared photometry is measured with \textsc{Source Extractor}, while the photo-zs are estimated via the template fitting codes \textsc{LePhare} \citep{Arnouts1999,Ilbert2006} and \textsc{EAZY} \citep{Brammer2008}. We apply several quality cuts to ensure reliable photometric and redshift estimates. We select sources lying in areas where the photometry is reliable and there is full coverage of optical/near-infrared data by means of the \textsc{FLAG\_COMBINED} parameter. Furthermore, we select objects for which \textsc{MAG\_AUTO}$<99$ in all band (ignoring in this study possible UV dropouts), \textsc{LePhare} or \textsc{EAZY} photo-zs are between $0$ and $8$, \textsc{LePhare} object type flag set to galaxy (\textsc{lp\_type}$=0$) and \textsc{Source Extractor} \textsc{FLAGS}$<4$.

\section{Galaxy population model}
\label{sect:gal_pop_model}

In this section, we present the SPS-based galaxy population model \textsc{GalSBI-SPS}. The prescriptions implemented in this new model are motivated by the insights from \cite{Tortorelli2024}. In that work we identified the SED modelling components that most significantly impact galaxy magnitudes and, consequently, redshift distribution estimates of colour-magnitude selected samples for cosmological applications. Unlike that previous work, we replace \textsc{FSPS} \citep{Conroy2009,Conroy2010} with the generative galaxy SED package \textsc{ProSpect}\footnote{https://github.com/asgr/ProSpect} \citep{Robotham2020}. 
\textsc{ProSpect} simulates far-ultraviolet-to-radio galaxy SEDs by modelling the emission from stars, gas, dust, and AGN. The SED modelling is self-consistent: the radiation produced by episodes of star formation heats up the dust grains that then cause re-emission of the absorbed energy at longer wavelengths.  The attenuation of starlight by the dust is simulated as a two component process.  Old stellar populations ($>10$ Myr) experience attenuation from a screen-like interstellar medium (ISM), while younger stellar populations ($<10$ Myr) are embedded in birth clouds,  experiencing additional attenuation.  A simple energy balance scheme is then used to produce nebular emission lines,  where the starlight emitted by young stars ($<10 \ \mathrm{Myr}$) at wavelengths shorter of the Lyman limit ($< 911.8 \ \AA$) determines the ionising flux for nebular emission. \textsc{ProSpect} is highly flexible, allowing users to input both parametric and non-parametric galaxy SFHs and to select among different stellar population libraries (SPL),  like those produced with the \citet{Bruzual2003} and EMILES \citep{Vazdekis2016} evolutionary SPS codes,  or custom generated with the newly released \textsc{ProGeny} SPL software package \citep{Robotham2025,Bellstedt2025}\footnote{The output of \textsc{ProGeny} is an SSP, while we refer to the family of related (by software) SSPs as an SPL.}. A notable feature of \textsc{ProSpect} is the ability, when used in conjunction with \textsc{ProGeny},  to generate single and composite stellar populations (CSPs) that make use of stellar age- or metallicity-dependent IMFs.  \textsc{ProGeny} also enables the generation of high-resolution SSPs for \textsc{ProSpect}, a crucial feature that will enable the use of \textsc{ProSpect} in spectrophotometric mode to jointly analyse photometric and spectroscopic data from the WAVES \citep{Driver2019} survey conducted with the 4MOST spectrograph \citep{Dejong2019}. The code's flexibility, ongoing development, high-resolution capabilities, and fast execution speed ($\sim20$–$30 \ \mathrm{ms}$ per galaxy SED on a Mac M1 CPU) strongly motivate its adoption as the generative SED engine for \textsc{GalSBI-SPS}. Importantly, the consistent use of \textsc{ProSpect} to analyse observed GAMA and DEVILS data, and to synthesise new galaxy SEDs in generative mode, ensures internal consistency in the physical assumptions underlying our galaxy population model and in the relation of physical to observable properties.

As in the phenomenological version of the model \citep{Fischbacher2025a},  galaxies in \textsc{GalSBI-SPS} are drawn from two overlapping populations of blue and red galaxies that share similar parametrisations, but differ in parameter values. Adding scatter to the physical relations allows the two populations to overlap, reflecting the observed continuum of galaxy properties, both in terms of SEDs and global physical characteristics. Figure \ref{fig:flowchart_model} presents a flowchart illustrating the main components of the model.  Briefly, \textsc{GalSBI-SPS} samples galaxy formed stellar masses and redshifts from two redshift-evolving GSMFs for blue and red galaxies.  It construct galaxy SFHs using parametrisations based on GAMA and DEVILS data, and samples the remaining physical properties,  necessary for realistic SED generation with \textsc{ProSpect}, conditioned on their respective relations to galaxy  stellar mass,  redshift,  and SFR.  Apparent magnitudes are obtained by redshifting,  dimming,  and integrating the generated SEDs over the relevant filter bands.  Finally,  each galaxy is assigned morphological properties,  including single S\'ersic indices,  sizes,  ellipticities,  and random spatial positions within the simulated field\footnote{The \textsc{Python} package \textsc{galsbi} allows for creating a simulated field through either an healpix map of user-defined NSIDE or by defining the central Right Ascension and Declination of the image coupled with the pixel scale and the number of pixels per image side. },  providing all the information required for realistic image simulations of un-clustered galaxies.

\subsection{Galaxy stellar mass function}
\label{sect:smf}

\textsc{GalSBI-SPS} samples formed stellar masses of galaxies $\log{\mathcal{M}}$ (in solar mass units $M_{\odot}$) and their redshifts $z$ from redshift-evolving GSMFs, modelled as the sum of two Schechter functions \citep{Schechter1976}. These two components, a low mass (`$\mathit{l}$' subscript) and a high mass (`$\mathit{h}$' subscript) ones, are assigned separate parameter values for blue (`$\mathit{b}$' subscript) and red (`$\mathit{r}$' subscript) galaxy populations. This double-Schechter parametrisation provides an accurate description of the statistical distribution of galaxy stellar masses in the local Universe (e.g. \citealt{Peng2010}) and up to $z\lesssim 3$ \citep{Leja2020}. At higher redshifts, the advantage of using a double function over a single Schechter function becomes less apparent \citep{Weaver2023}. However, we use a well motivated redshift evolution of the GSMF parameters that naturally accommodate the transition from a double to a single-Schechter form at high redshifts.  The GSMF defines the number density of galaxies per comoving volume $V_{\mathrm{C}}$ per logarithmic stellar mass bin at redshift $z$,  and is expressed in terms of logarithmic stellar mass $\log{\mathcal{M}}$ following the formalism of \cite{Weigel2016, Weaver2023}:
\begin{equation}
    \begin{split}
        \Phi(\log{\mathcal{M}}, z) =& \frac{\mathrm{d}N}{\mathrm{d}( \log{\mathcal{M}}) \mathrm{d}V_{\mathrm{C}}} = \\ &\mathrm{ln}(10) e^{-10^{\log{\mathcal{M}} - \log{\mathcal{M}^*(z)}}} \\
        & \times \left [ \phi_{\mathit{l}}^*(z) \left( 10^{\log{\mathcal{M}} - \log{\mathcal{M}^*(z)}} \right)^{\alpha_{\mathit{l}}+1} \right] \\
        & \times \left [ \phi_{\mathit{h}}^*(z) \left( 10^{\log{\mathcal{M}} - \log{\mathcal{M}^*(z)}} \right)^{\alpha_{\mathit{h}}+1} \right].
    \end{split}
\end{equation}
The characteristic stellar mass $\log{\mathcal{M^*}}$ is assumed to be the same for both Schechter components, as it is standard practice when fitting double Schechter functions \citep{Baldry2012, Muzzin2013,Tomczak2014}. $\phi^*_{\mathit{l}},\phi^*_{\mathit{h}}$ represents the low mass and high mass number density of galaxies at the characteristic stellar mass $\log{\mathcal{M^*}}$, while $\alpha_{\mathit{l}},\alpha_{\mathit{h}}$ represent the faint-end slopes of the two functions. We choose to allow for a redshift evolution of $\log{\mathcal{M^*}}, \phi^*_{\mathit{l}},\phi^*_{\mathit{h}}$, while we keep $\alpha_{\mathit{l}},\alpha_{\mathit{h}}$ fixed to the mean values reported for blue and red galaxy populations in Tables C.2 and C.3 of \citet{Weaver2023}.

\begin{figure}
   \centering
   \includegraphics[width=\hsize]{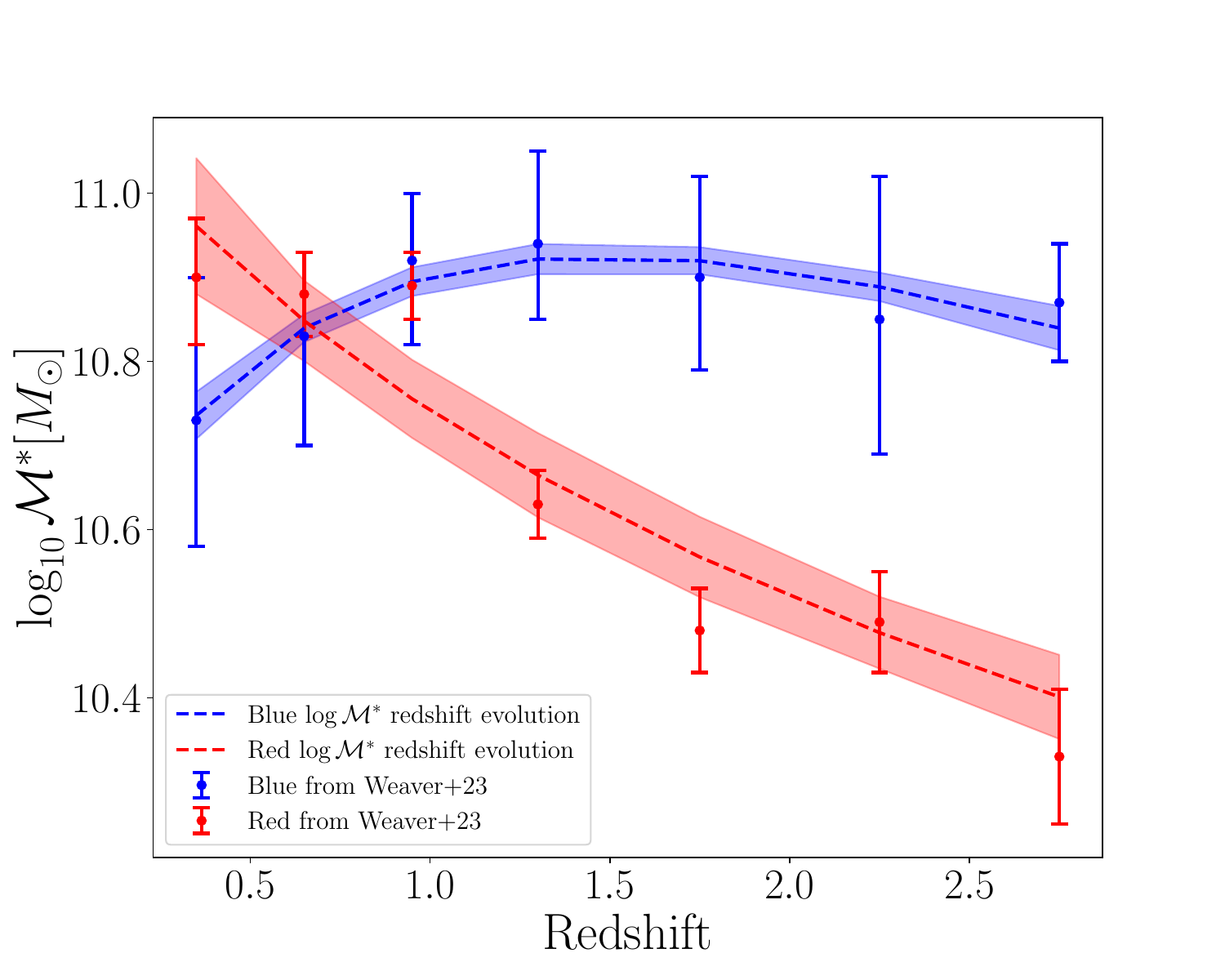}
      \caption{Redshift evolution of the characteristic stellar mass $\log{\mathcal{M^*}}$ for blue and red galaxies. Blue and red points represent the measurements from \cite{Weaver2023}. The blue and red dashed lines are built by fitting the measurements with equation \ref{eq:log_mstar_z_evo}. The blue and red bands represent the $1 \sigma$ uncertainty band on the fit.
              }
    \label{fig:logMstar_redshift_evolution}
\end{figure}

We parametrise the redshift evolution of the characteristic stellar mass as:
\begin{equation}
    \begin{split}
    \log{\mathcal{M^*}}(z) = &\log{\mathcal{M^*}}_0 + \log{\mathcal{M^*}}_1 \times \log{(1+z)} \ + \\ &\log{\mathcal{M^*}}_2 \times \log{(1+z)}^2 \ .
    \end{split}
    \label{eq:log_mstar_z_evo}
\end{equation}
The $\log{\mathcal{M^*}}_0,\log{\mathcal{M^*}}_1,\log{\mathcal{M^*}}_2$ values for both blue and red galaxy populations are obtained by fitting the redshift evolution of $\log{\mathcal{M^*}}$ reported in Tables C.2 (blue galaxies) and C.3 (red galaxies) of \cite{Weaver2023}. This functional form is motivated by the fact that it provides a good fit to the measurements in \citet{Weaver2023}. We limit our fit to $z \le 3$ galaxies, where a second Schechter component is necessary for describing the GSMF. Beyond this redshift, we extrapolate the best-fitting relations. Although extrapolation at $z > 3$ is sub-optimal, the vast majority of galaxies for a cosmological analysis in Stage IV surveys will lie below this redshift. Figure \ref{fig:logMstar_redshift_evolution} shows the evolution of $\log{\mathcal{M^*}}$ with redshift, based on the best-fitting parameters listed in the first three rows of Table \ref{table:sm_funct_parameters}, separately for blue and red galaxies.

\begin{figure}
   \centering
   \includegraphics[width=\hsize]{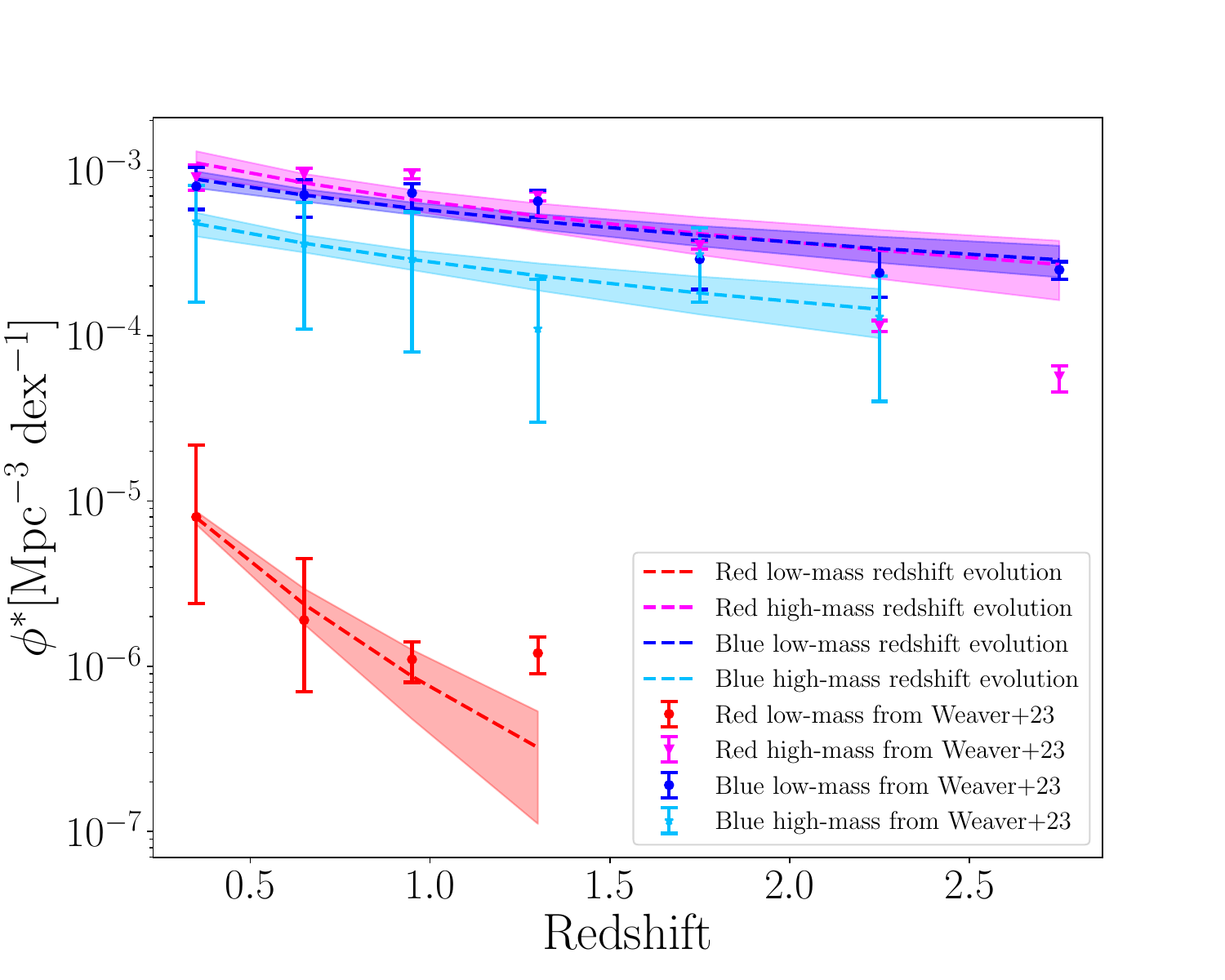}
      \caption{Redshift evolution of the characteristic density $\phi^*$ for blue and red galaxies. Blue, deep sky blue, magenta and red points represent the measurements from \cite{Weaver2023}. The Blue, deep sky blue, magenta and red dashed lines are built by fitting the measurements with equation \ref{eq:phi_star_z_evo}. The bands represent the $1 \sigma$ uncertainty band on the fit.
              }
    \label{fig:phi_star_redshift_evolution}
\end{figure}

The redshift evolution of the normalisation parameters $\phi^*_{\mathit{l}},\phi^*_{\mathit{h}}$ follows the same prescription adopted for luminosity functions in \cite{Moser2024},
\begin{equation}
 \phi^*_{\mathit{l,h}}(z) = \phi^*_{\mathrm{amp},\mathit{l,h}} \times (1 + z)^{\phi^*_{\mathrm{exp},\mathit{l,h}}} \ ,
 \label{eq:phi_star_z_evo}
\end{equation}
where the $\phi^*_{\mathrm{amp}},\phi^*_{\mathrm{exp}}$ differs for blue and red galaxies, and for the low-mass and high-mass components of the GSMF. The parameter values used in this work are obtained by fitting equation \ref{eq:phi_star_z_evo} to the values in Tables C.2 (blue galaxies) and C.3 (red galaxies) of \cite{Weaver2023}. Although not all fitted parameters are constrained over the full redshift range up to $z=3$ in \cite{Weaver2023}, we extrapolate the best-fitting relations when necessary. Similar considerations to the case of the high-redshift extension of $\log{\mathcal{M^*}}$ apply.  Figure \ref{fig:phi_star_redshift_evolution} shows the redshift evolution of $\phi^*_{\mathit{l}}$ and $\phi^*_{\mathit{h}}$ obtained with the best-fitting parameters listed in the fourth to seventh rows of Table \ref{table:sm_funct_parameters}, again separately for blue and red populations. The parametric form provides a good fit to the observed evolution for low- and high-mass blue galaxies and for low-mass red galaxies. A tension is observed for high-mass red galaxies at $z \gtrsim 2$, where the model under predicts the decline in number density. This could be due to either an intrinsic evolution that would require a more flexible parametrisation or the limited statistics for that population of galaxies in the COSMOS field.

\textsc{GalSBI-SPS} computes the redshift distribution of galaxies, $n(z)$, by integrating the redshift-stellar mass density $\varphi$ over a specified stellar mass range. The redshift-stellar mass density $\varphi$ is defined as:
\begin{equation}
    \varphi = \frac{\mathrm{d}N}{\mathrm{d}\log{\mathcal{M}} \ \mathrm{d}z \ \mathrm{d}\Omega} \ ,
\end{equation}
where $\mathrm{d} \Omega$ is the solid angle. Substituting the expression for the comoving volume element \citep{Hogg1999},
\begin{equation}
    \mathrm{d}V_{\mathrm{C}} = \frac{\mathrm{D_H}(1+z)^2\mathrm{D_A}^2}{\mathrm{E}(z)} \mathrm{d}z \ \mathrm{d}\Omega
\end{equation}
where $\mathrm{D_H}, \mathrm{D_A}$ are the Hubble distance and the angular diameter distance, respectively, and $\mathrm{E}(z) = \sqrt{\Omega_{\mathrm{M}} (1+z)^3 + \Omega_{\mathrm{k}}(1+z)^2 + \Omega_\Lambda}$, and using the definition of the GSMF $\Phi(\log{\mathcal{M}}, z)$, the redshift distribution for each Schechter component becomes:
\begin{equation}
    \begin{split}
        n(z) = & \int_{\log{\mathcal{M}_{\mathrm{min}}}}^{\log{\mathcal{M}_{\mathrm{max}}}} \varphi \ (\log{\mathcal{M}},z) \ \mathrm{d} \log{\mathcal{M}} = \\
        & \int_{\log{\mathcal{M}_{\mathrm{min}}}}^{\log{\mathcal{M}_{\mathrm{max}}}} \frac{\mathrm{D_H}(1+z)^2\mathrm{D_A}^2}{\mathrm{E}(z)}  \Phi(\log{\mathcal{M}}, z) \ \mathrm{d} \log{\mathcal{M}} = \\
        & \frac{\mathrm{D_H}(1+z)^2\mathrm{D_A}^2}{\mathrm{E}(z)} \phi^* \Gamma(\alpha + 1, 10^{\log{\mathcal{M_{\mathrm{min}}}} - \log{\mathcal{M}^*(z)}}) \  ,
    \end{split}
\end{equation}
where $\Gamma$ denotes the incomplete gamma function. Stellar mass sampling proceeds in two steps: first we draw redshifts from the $n(z)$, then we sample stellar masses from the conditional GSMF at the selected redshift. The total number of galaxies $N$ expected within a sky area $A$ (in $\mathrm{steradian}$) over a redshift range $[z_{\mathrm{min}}, z_{\mathrm{max}}]$ is given by:
\begin{equation}
    N = \int_{z_{\mathrm{min}}}^{z_{\mathrm{max}}} n(z) A \frac{\mathrm{D_H}(1+z)^2\mathrm{D_A}^2}{\mathrm{E}(z)} \mathrm{d}z \ .
\end{equation}
The total number of parameters for the low-mass and high-mass blue and red GSMFs is 18. The GSMF parameter values used in this work are listed in Table \ref{table:sm_funct_parameters}.  It is important to note that the logarithmic stellar mass $\log{\mathcal{M}}$ sampled from the GSMF in \textsc{GalSBI-SPS} represents the total stellar mass formed by the galaxy and all its progenitors throughout their lifetime.  This will be used to obtain the correct galaxy SFH to pass to \textsc{ProSpect}. Care should therefore be taken when directly comparing this GSMF with those quoted in the literature,  as the GSMFs derived from observations commonly refer to the galaxy surviving stellar masses. We do not sample the latter quantity as it depends on the specific SPL being used and is generally produced as an output quantity in an SPS-based code.  \textsc{ProSpect} allows for the computation of the surviving stellar mass,  however, since this will increase the runtime of the galaxy population sampling,  we develop a machine learning-based emulator that predicts the galaxy surviving stellar mass from its formed stellar mass,  SFH parameters and gas-phase metallicity with per-mille level accuracy.  We describe the emulator in Appendix \ref{appendix:surviving_mass_emulator}.

\begin{table}
\caption{GSMF parameter values for the red and blue galaxy populations.}
\centering
\begin{tabular}{lcc}
\hline\hline
Parameter  & Red & Blue\\
\hline
$\log{\mathcal{M}}_0$ & 11.132 & 10.428 \\
$\log{\mathcal{M}}_1$ & -1.322  & 2.757\\
$\log{\mathcal{M}}_2$ & 0.085  & -3.784\\
\hline
$\phi^*_{\mathrm{amp,}\mathit{l}} (\times 10^{-3})$ & 0.144 & 3.605\\ 
$\phi^*_{\mathrm{exp,}\mathit{l}}$ & -6.008 & -1.052\\
$\phi^*_{\mathrm{amp,}\mathit{h}} (\times 10^{-3})$ & 3.992 & 2.145 \\ 
$\phi^*_{\mathrm{exp,}\mathit{h}}$ & -0.847 & -1.358\\
\hline
$\alpha_{\mathit{l}}$ & -2.01 & -1.434\\
$\alpha_{\mathit{h}}$ & -0.52 & -0.123\\
\hline
\end{tabular}
\tablefoot{High-mass and low-mass GSMF parameter values for red and blue galaxies obtained by modelling the redshift evolution of $\log{\mathcal{M^*}}$ and $\phi^*$ from the values in tables C.2 and C.3 in \cite{Weaver2023}. The faint-end slopes $\alpha$ are computed as mean values across redshift from the same tables in \cite{Weaver2023}.}
\label{table:sm_funct_parameters}
\end{table}

\subsection{Stellar population model}
\label{sect:stellar_pop_model}

Evolutionary SPS \citep{Tinsley1980,Pickles1985,Bruzual2003,Maraston2005,Conroy2009,Vazdekis2016,Conroy2013} models the spectral emission from a group of unresolved stars, like those in a galaxy, by making use of the knowledge of stellar evolution.  The simplest flavour of an evolutionary SPS model is an SSP,  where it is assumed that all stars are coeval and share the same chemical composition.  Following the nomenclature of \cite{Conroy2013},  when combining SSPs wth a SFH, chemical evolution and a model for dust attenuation,  we can build CSPs that are complex representations of the integrated starlight emitted from galaxies. 

The basic ingredients to build an SSP are the stellar evolutionary tracks (isochrones),  stellar model atmospheres, and stellar IMF.  Isochrones trace the evolution of stars of given mass,  chemical composition and $\alpha/\mathrm{Fe}$ abundance,  through the various evolutionary phases that characterise the life of a star.  The key predictions that isochrones must convey are basic stellar parameters,  such as bolometric luminosities,  surface gravities,  effective temperatures (and potentially the effective physical size),  as functions of evolutionary timescales.  Stellar model atmospheres, which can be either empirical or theoretical, are instead the means by which we translate these basic stellar parameters into the emergent flux of stars.  Commonly the  predictions of stellar properties generated by isochrones have a much finer resolution than the coarser grids of available stellar spectra, therefore some degree of interpolation is required to generate flux predictions (see discussion in \citealt{Robotham2025}).  Predictions from isochrones and stellar atmospheres are then complemented by the choice of a stellar IMF, that is a weighting term that encodes the probability to form a star of given mass, thereby returning the distribution of stellar masses that were generated by a single burst of star formation.

Different SPS codes exist in the literature with which one can generate SSPs,  some notable ones being the evolutionary SPS codes of \cite{Bruzual2003,Maraston2005},  EMILES \citep{Vazdekis2016} and \textsc{FSPS} \citep{Conroy2009,Conroy2010}.  \textsc{ProSpect}, on the contrary, does not produce SSPs, but rather combines them with complex models of SFHs (and other SED emission components) to generate spectral outputs for simulations and fit observed galaxy SEDs.  Similar functionality is implemented by other SED fitting codes,  such as \textsc{MAGPHYS} \citep{daCunha2008} and \textsc{CIGALE} \citep{Boquien2019},  each with different emphasis on specific SED modelling choices.  The original implementation of \textsc{ProSpect} allows the user to choose between the SSPs generated with the evolutionary SPS codes of \cite{Bruzual2003} and \citep{Vazdekis2016} (EMILES).  However,  \cite{Robotham2025} recently released \textsc{ProGeny},  an SPL software package that allows for the creation of SSPs and their associated spectra.  \textsc{ProGeny} directly interfaces with \textsc{ProSpect} by creating custom, user-defined SSPs that are used to create CSP to model the starlight emission from galaxies. The notable addition, in comparison with other SPS-based codes,  is the possibility to produce SSPs with evolving IMFs, in terms of stellar age and/or metallicity,  and the ability to output spectra with higher wavelength resolution than traditional SSPs,  such as those of \cite{Bruzual2003}.

In \cite{Robotham2025,Bellstedt2025}, the authors compared the \textsc{ProGeny}-generated SSPs against those from other evolutionary SPS codes, as well as the impact of the isochrones, stellar atmospheres, and IMF choices on the recovery of galaxy properties using simulations and GAMA observations. They found that the predicted spectra are more strongly impacted by the choice of the isochrones rather than that of the stellar atmospheres or the IMF, which is also the major factor influencing the differences in the recovered SFHs for GAMA observed galaxies. They also tested the impact of \cite{Chabrier2003} and \cite{Kroupa2002} IMFs, either keeping them fixed or evolving them as function of metallicity \citep{Martin-Navarro2015}, finding a very small impact on recovered SFHs and a somewhat larger impact on galaxy stellar mass. This is similar to what has been found in \cite{Tortorelli2024}, where changing the IMF was leading to differences in the $i$-band fluxes that can be interpreted as differences in galaxy stellar mass when using the $i$-band flux as its proxy.

In our work, we use the \textsc{ProGeny} web tool to create our custom SSP by incorporating the fiducial choices of isochrones, stellar spectral atmospheres, and IMFs highlighted in \cite{Robotham2025,Bellstedt2025}.  We select the \textsc{MIST} isochrones \citep{Dotter2016} as they cover all phases of stellar evolution, including remnant stars, important for the UV upturn of older stellar populations.  They have high-resolution stellar evolution age cadences of $0.05 \ \mathrm{dex}$ between $10^5$ and $10^{10.3} \ \mathrm{yr}$,  with a $0.5 \  \mathrm{dex}$ metallicity gridding between $10^{-4}$ and $10^{0.5} \ Z_{\odot}$ and a mass range of $0.1$ to $3\times10^2 \ M_{\odot}$.  The stellar atmosphere consists of a combination of `base' and `extended' templates to ensure a good all-around coverage in temperature,  surface gravity and metallicity space.  The `base' templates come from the atmospheric spectra included with \textsc{FSPS} \citep{Conroy2018} (sometimes labelled as C3K).  They cover temperatures from $2\times10^3$ to $5\times10^4 \ \mathrm{K}$,  surface gravities from $10^{-1}$ to $10^{5.5} \ \mathrm{cm \ s^{-2} }$ on a logarithmic grid and logarithmic metallicities in solar units from $-2.1$ to $0.5$.  The `extended' template spectra come from the \cite{Allard2012} version of \textsc{PHOENIX} spectra \cite{Hauschildt1999} and are meant to extend the atmospheres coverage to very low mass stars and up to temperatures of $7 \times 10^4 \ \mathrm{K}$.  To extend the coverage of the atmospheres to the asymptotic giant branch (AGB),  white dwarf and Wolf Rayet phases of stellar evolution,  we further select the \textsc{PoWR} Wolf-Rayet library \citep{Todt2015},  the \cite{Lancon2002} AGB library (also used in \citealt{Maraston2005} and \citealt{Conroy2010}) and the \cite{Werner2003} white dwarf library.  We then choose the \cite{Chabrier2003} IMF as the last critical component for our custom SSP.  The latter covers a wavelength range from $5$ to $3.6\times10^8 \ \AA$,  re-binned to match the exact wavelength sampling of the 2019 version (CB19) of \cite{Bruzual2003,Plat2019}.  More details on the invididual components are provided in \cite{Robotham2025}.

\subsection{Galaxy star formation history}
\label{sect:sfh_history}

The galaxy SFH is a key ingredient for building a physically motivated model for the galaxy population that is based on SPS,   as it represents,  in the most general form,  the number of stars produced in a galaxy per unit time from the conversion of gas into stars.  Coupled with the IMF,  the galaxy SFH represents the amount of stellar mass produced per unit time.  In \textsc{ProSpect},  using the Chabrier IMF and the SSP produced with \textsc{ProGeny},  the SFH represents the weights of the weighted sum of SSPs with which CSPs are produced.  When designing a SFH parametrisation, we need to find the right trade-off between flexibility,  needed to capture the diversity of galaxy SFHs,  and simplicity,  to avoid difficulty in constraining with observations and introducing unnecessary parameter degeneracies.  The literature offers a wide range of SFH models.  The simplest is an instantaneous burst model,  where all the stellar mass is produced in a single burst of star-formation and the galaxy stellar population coincides with an SSP, where only the evolution of stars with a single metallicity is required.  The most commonly used time-evolving parametrisation is instead the $\tau$-model,  which assumes an exponentially declining SFR. However, numerous studies have shown that this form produces biased estimates of galaxy properties when applied to mock galaxies \citep{Simha2014,Pacifici2015,Ciesla2017,Carnall2018}. To overcome these limitations, researchers have developed more flexible parametrisations. These include the delayed-$\tau$ model \citep{Simha2014}, the double power-law SFH \citep{Alsing2023}, the step-wise SFHs \citep{Leja2019b,Wang2023}, and the skewed Normal SFH \citep{Robotham2020}. Non-parametric SFH models are also widely used, e.g. basis SFHs \citep{Hahn2023} constructed by applying non-negative matrix factorisation to Illustris SFHs \citep{Genel2014} or SFHs from semi-analytic models \citep{Lagos2019}. 

\begin{figure*}
   \centering
   \includegraphics[width=1\textwidth]{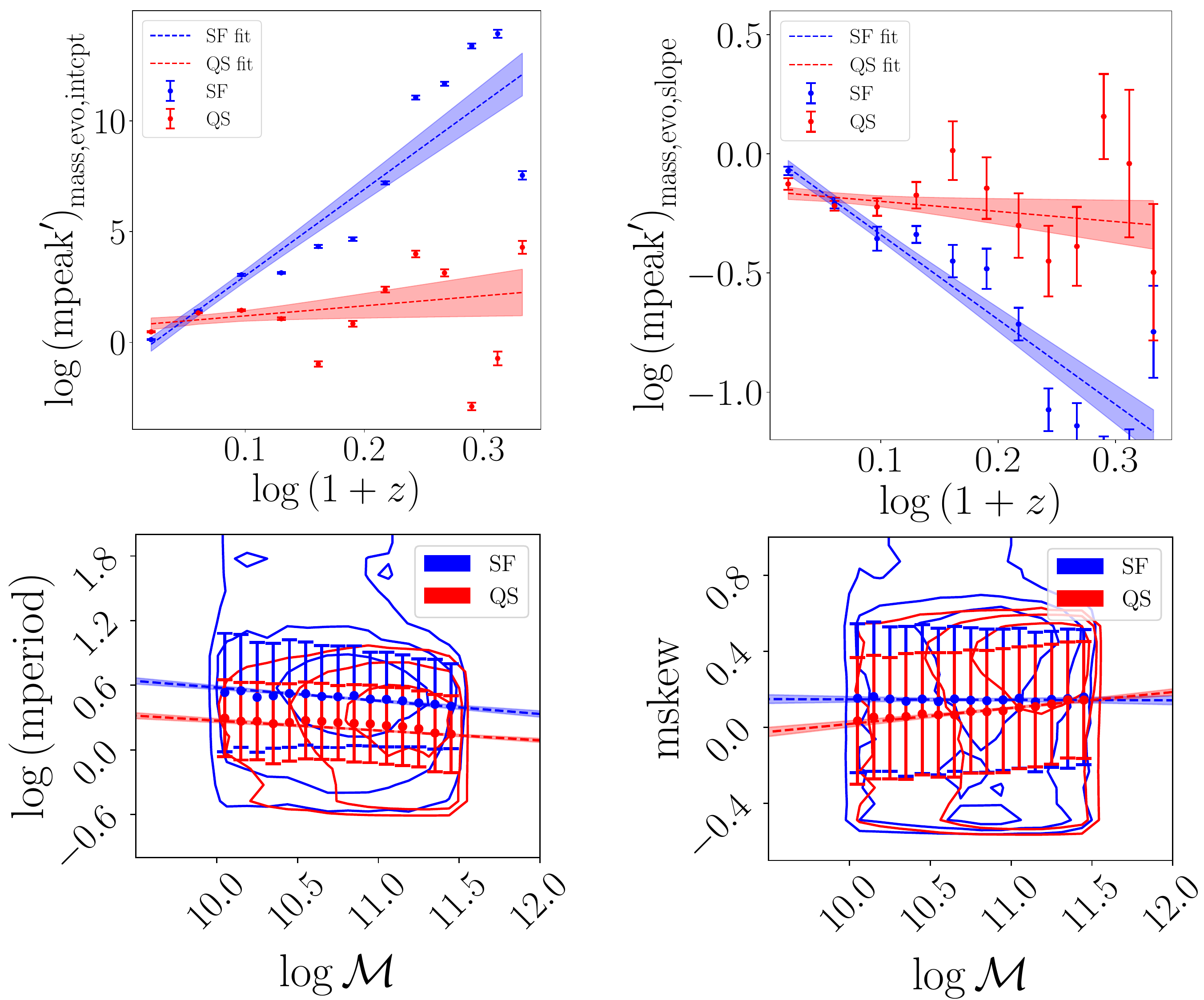} 
      \caption{Dependence of the SFH shape parameters on galaxy redshifts and stellar mass.  $\log{(\mathrm{mpeak}')}$ linearly depends on the galaxy stellar mass and its coefficients in turns depend on the galaxy redshift.  The upper panels show the dependence of the slope $\log{(\mathrm{mpeak}')}_{\mathrm{mass,evo,slope}}$ and the intercept $\log{(\mathrm{mpeak}')}_{\mathrm{mass,evo,intcpt}}$ on redshift.  Red error bars refer to quiescent galaxies from the mass-complete GAMA and DEVILS sample, while blue error bars to star-forming galaxies.  The dashed lines represent the best-fitting linear relations for the redshift evolution of these parameters, while the red and blue bands represent the $1 \sigma$ uncertainty on the fit.  Lower panels show instead the dependence of $\log{(\mathrm{mperiod})}$ and $\mathrm{mskew}$ on the galaxy stellar mass. The error bars refer to the median and standard deviation of these quantities computed in bins of $0.1 \ \mathrm{dex}$ in stellar mass,  overlayed on top of the overall distribution.  Dashed lines and colour bands represent best-fitting linear relations and $1 \sigma$ uncertainty on the fit for the populations of red and blue galaxies. }
    \label{fig:sfh_redshift_evolution}
\end{figure*}

\textsc{GalSBI-SPS} models individual galaxy SFHs in units of $M_{\odot}/\mathrm{yr}$ using a truncated skewed Normal distribution \citep{Robotham2020}. This form ensures that SFHs are anchored at zero star formation at the beginning of the Universe, naturally rising during the first billion years with a smooth, physically motivated truncation at early times. Unlike other parametric forms that assume an unphysical finite SFR at formation, this model forces the SFR to start from zero, aligning with expectations from galaxy formation theory. This approach has proven highly successful in fitting observed galaxy populations \citep{Bellstedt2020,Bellstedt2021,Thorne2021,Thorne2022,Thorne2022b,Bellstedt2024}. The functional form of the SFH reads:
\begin{equation}
    \mathrm{SFR}(t) = \mathrm{SFR}(t)_{\mathrm{norm}} \times \left [ 1 - \frac{1}{2} \left[ 1 + \mathrm{erf}\left( \frac{t - \mu}{\sigma \sqrt{2}} \right) \right] \right] \ ,
    \label{eq:sfh_trunc}
\end{equation}
where 
\begin{equation}
    \begin{split}
        \mu &= \mathrm{mpeak} \times \frac{|(\mathrm{magemax - mpeak})|}{\mathrm{mtrunc}} \ , \\
        \sigma &= \frac{|(\mathrm{magemax - mpeak})|}{2 \times \mathrm{mtrunc}} \ .
    \end{split}
\end{equation}
$\mathrm{mtrunc}= 2 \ \mathrm{Gyr}$ sets the sharpness of the early-time truncation (with 0 being no truncation), while $\mathrm{magemax} = 13.4\ \mathrm{Gyr}$ represents the maximum age of star formation, chosen to correspond to the existence of some of the earliest observed galaxies ($z \sim 11$, \citealt{Oesch2016}) within our adopted cosmology. The naming convention for the SFH parameters follows that introduced in \citet{Robotham2020}. The second term of equation \ref{eq:sfh_trunc} represents the truncation factor, while the the first term is a skewed Normal distribution expressed as:
\begin{equation}
    \mathrm{SFR}(t)_{\mathrm{norm}} = \mathrm{mSFR} \times e^{\frac{X(t)^2}{2}} \ ,
\end{equation}
where
\begin{equation}
    X(t) = \left (  \frac{t - \mathrm{mpeak}}{\mathrm{mperiod}}  \right) \left( e^{\mathrm{mskew}} \right)^{\mathrm{asinh}\left( \frac{t - \mathrm{mpeak}}{\mathrm{mperiod}} \right)} \ .
\end{equation}
The skewed Normal distribution is governed by four parameters. $\mathrm{mSFR}$ represents the peak SFR in units of $M_{\odot}/\mathrm{yr}$. $\mathrm{mpeak}$ is the age in $\mathrm{Gyr}$ at which the peak SFR occurs. $\mathrm{mperiod}$ is the width of the star formation period in $\mathrm{Gyr}$, while $\mathrm{mskew}$ is the skewness controlling the asymmetry of the SFH. The chosen parametrisation achieves the best results when 
galaxies have experienced a single dominant epoch of star formation,  while it might be less accurate if the galaxy has undergone a very recent small burst of star-formation,  which is however a notoriously hard-to-model SFH component.  As shown in \citealt{Robotham2020},  this parametrisation works at the galaxy population level as it is able to correctly reproduce the diversity of SFHs seen in the SHARK semi-analytic model \citep{Lagos2019}. Furthermore,  in \cite{Bravo2022},  the authors found that the vast majority of galaxies have a single dominant epoch of star-formation,  thereby making this parametrisation reliable even in the case of burstiness in the SFH.

To meaningfully sample the four free SFH parameters ($\mathrm{mSFR}$, $\mathrm{mpeak}$, $\mathrm{mperiod}$, and $\mathrm{mskew}$) and generate realistic galaxy SFHs for blue and red galaxies, we model the distributions of best-fitting SFH parameter values obtained from GAMA \citep{Bellstedt2020,Bellstedt2021,Bellstedt2024} and DEVILS \citep{Thorne2021,Thorne2022,Thorne2022b} data. We restrict the modelling to stellar mass-complete samples and remove objects whose best-fitting SFH parameters lie at the edges of their priors, as defined in Table 2 of \citet{Bellstedt2020}. We verify that this selection does not bias the overall parameter distributions of the mass-complete samples. Because we aim at using the SFH parametrisation in a generative mode, we cannot directly model the observed distribution of best-fitting parameters. If we were to do so,  we would sample an exact replica of the galaxy population observed by GAMA and DEVILS.  In particular, the sampled parameters would lead to a distribution of SFHs whose formed stellar masses do not coincide with the distribution of stellar masses drawn from the GSMF.  In our model,  we condition the sampling of SFH parameters such that the integrated stellar mass over each individual galaxy SFH equals the formed stellar mass sampled from the GSMF for that galaxy (see Sect.  \ref{sect:smf} for the discussion on the kind of stellar masses sampled from the GSMF).

To ensure that the latter is true,  we need to remodel the SFHs from GAMA and DEVILS and separate the sampling process of the SFH parameters between those that define the SFH shape ($\mathrm{mpeak}$,  $\mathrm{mperiod}$ and $\mathrm{mskew}$) and the SFH normalisation ($\mathrm{mSFR}$). First,  we define a common time grid in observer-frame look-back time, spanning the range from $t=10^5\ \mathrm{yr}$ to $t=13.7 \times 10^9\ \mathrm{yr}$, corresponding to the age of the Universe at $z=0$ in our adopted cosmology. Since the original SFHs in \textsc{ProSpect} are defined in galaxy-frame look-back time, in order to correctly compute the stellar mass formed, we set the SFH to $0$ beyond $\mathrm{magemax}' = 13.4 - t_{\mathrm{lb},z} \ \mathrm{Gyr}$, i.e.  beyond the age of the Universe at the galaxy redshift.  We then redefine the age of the peak star formation ($\mathrm{mpeak}$).  Instead of using an absolute value in $\mathrm{Gyr}$ and sampling it from GAMA and DEVILS data,  we define $\mathrm{mpeak}'$ to be the fraction of the galaxy age at which the SFH peaks, 
\begin{equation}
    \mathrm{mpeak}' = \frac{\mathrm{magemax}' - \mathrm{mpeak}}{\mathrm{magemax}'} \ .
\end{equation}
Values of $\mathrm{mpeak}' > 1$ correspond to rising SFHs.  $\mathrm{mperiod}$ and $\mathrm{mskew}$ are instead unchanged from their original definition. 

Observationally,  it has been shown that more massive galaxies assemble the majority of their stellar mass at earlier cosmic epochs and over a shorter period than less massive galaxies that have more prolonged periods of star-formation and assemble most of their mass at later times (see \citealt{DeLucia2015} and reference therein for an up-to-date review).  This leads us to expect a dependence of the SFH shape parameters on galaxy redshift and stellar mass.  Furthermore, we expect this trend to have different strength for blue and red galaxy populations.  To model this trend,  we parametrise $\log{(\mathrm{mpeak}')}$ for blue and red galaxies as function of stellar mass as
\begin{equation}
\begin{split}
\log{(\mathrm{mpeak}')} =& \log{(\mathrm{mpeak}')}_{\mathrm{mass,evo,intcpt}} + \\
 &\log{(\mathrm{mpeak}')}_{\mathrm{mass,evo,slope}} \times \log{\mathcal{M}} \ ,
\end{split}
\label{mpeak_evo}
\end{equation}
where the subscript `mass,evo' stands for mass evolution,  and with slope and intercept  linearly evolving with redshift as
\begin{equation}
\begin{split}
\log{(\mathrm{mpeak}')}_{\mathrm{mass,evo,intcpt}} =& \  \mathrm{a}_{\mathrm{mass,evo,intcpt}} + \\
 & \mathrm{b}_{\mathrm{mass,evo,intcpt}} \times \log{(1+z)} \ ,  \\
\log{(\mathrm{mpeak}')}_{\mathrm{mass,evo,slope}} =& \   \mathrm{a}_{\mathrm{mass,evo,slope}} + \\
& \mathrm{b}_{\mathrm{mass,evo,slope}} \times \log{(1+z)} \ .
\end{split}
\label{mpeak_z_evo}
\end{equation}
The upper panels of  Fig. \ref{fig:sfh_redshift_evolution} show the evolution of the slope $\log{(\mathrm{mpeak}')}_{\mathrm{mass,evo,slope}}$ and intercept $\log{(\mathrm{mpeak}')}_{\mathrm{mass,evo,intcpt}}$ as function of redshift for the combination of mass-complete GAMA and DEVILS quiescent and star-forming galaxies,  selected using the specific star-formation rate (sSFR) criterion of $\log{(sSFR)} \ge -10.5$, for star-forming objects, and $\log{(sSFR)} \le -10.5$, for quiescent objects \citep{Leja2017}.  The lower panels of Fig. \ref{fig:sfh_redshift_evolution} show instead the dependence of the median of $\log{(\mathrm{mperiod})}$ and $\mathrm{mskew}$ in bins of $0.1 \ \mathrm{dex}$ on stellar mass.  To keep the model simple yet realistic,  we also model this dependence as a linear relation:
\begin{equation}
\begin{split}
\log{(\mathrm{mperiod})} =& \log{(\mathrm{mperiod})}_{\mathrm{mass,evo,intcpt}} + \\
& \log{(\mathrm{mperiod})}_{\mathrm{mass,evo,slope}} \times \log{\mathcal{M}} \ , \\
\mathrm{mskew} =& \mathrm{mskew}_{\mathrm{mass,evo,intcpt}} + \\ 
& \mathrm{mskew}_{\mathrm{mass,evo,slope}} \times \log{\mathcal{M}} \ .
\end{split}
\label{mperiod_mskew_evo}
\end{equation}
Based on the data at our disposal,  we do not find a strong trend of the slope and intercept of these linear relations with redshift.  Best-fitting parameters for these relations obtained from GAMA and DEVILS data are reported in Table \ref{table:sfh_funct_parameters}.  We sample the SFH shape parameters for each individual galaxy from a Multivariate Truncated Normal distribution.  The mean of this multivariate distribution is different for each galaxy as it is given by equations \ref{mpeak_evo},  \ref{mpeak_z_evo} and \ref{mperiod_mskew_evo} conditioned on the galaxy stellar mass and redshift. The covariance is instead fixed and estimated for blue and red galaxies from the residuals between the GAMA and DEVILS original distribution of SFH shape parameters and that obtained from sampling with the best-fitting parameters in Table \ref{table:sfh_funct_parameters}.  The sampling is truncated to the observational limits defined in Table 2 of \citet{Bellstedt2020}. We then rescale $\log{(\mathrm{mpeak}')}$ back to the absolute value in $\mathrm{Gyr}$.

The sampled parameters define the SFH shape for each galaxy.  To obtain the value of the normalisation $\mathrm{mSFR}$,  we impose for each galaxy that the integral of the SFH over the galaxy lifetime returns the formed stellar $\mathcal{M}$ drawn from the GSMF:
\begin{equation}
\begin{split}
\mathrm{mSFR} \ \times \ \int_0^{\mathrm{magemax}} & \ sfh_{\mathrm{shape}}(\mathrm{mpeak,mperiod,mskew}) \ \times \\
 &\left [ 1 - \frac{1}{2} \left[ 1 + \mathrm{erf}\left( \frac{t - \mu}{\sigma \sqrt{2}} \right) \right] \right] = \mathcal{M} \ ,
\end{split}
\end{equation}
yielding the final correct SFH for each individual galaxy. The four sampled parameters are passed to the \textsc{massfunc\_snorm\_trunc} function in \textsc{ProSpect} when generating galaxy SEDs.  Additionally, we compute the average SFR over the last $100 \ \mathrm{Myr}$ for each galaxy. Redshifts, stellar masses and SFRs represent the most important physical properties for the modelling of the galaxy population. Most of the other physical properties that are described in the following sections are conditioned on these three fundamental quantities.

\begin{table}
\caption{Parameter values of the SFH shape parameters evolution with stellar mass and redshift for the red and blue galaxy populations.}
\centering
\begin{tabular}{lcc}
\hline\hline
Parameter  & Red & Blue\\
\hline
$\mathrm{a}_{\mathrm{mass,evo,intcpt}}$ & 0.223 & -0.655 \\
$\mathrm{b}_{\mathrm{mass,evo,intcpt}}$ & 15.143 & 35.018 \\
$\mathrm{a}_{\mathrm{mass,evo,slope}}$ & -0.104 & 0.006 \\
$\mathrm{b}_{\mathrm{mass,evo,slope}}$ & -1.524 & -3.544 \\
\hline
$\log{(\mathrm{mperiod})}_{\mathrm{mass,evo,intcpt}}$ & 1.198 & 2.727 \\
$\log{(\mathrm{mperiod})}_{\mathrm{mass,evo,slope}}$ & -0.094 & -0.210 \\
\hline
$ \mathrm{mskew}_{\mathrm{mass,evo,intcpt}}$ & -0.554 & 0.027 \\
$ \mathrm{mskew}_{\mathrm{mass,evo,slope}}$ & 0.063 & 0.016 \\
\hline
\end{tabular}
\tablefoot{Parameters used for the SFH modelling of red and blue galaxies. Galaxy SFHs are described by a truncated skewed Normal distribution \citep{Robotham2020}.  Shape parameters of this SFH parametrisation ($\mathrm{mpeak}$,  $\mathrm{mperiod}$ and $\mathrm{mskew}$) are sampled from a Multivariate Truncated Normal distribution,  whose mean is different for each galaxy as the shape parameters depend on its stellar mass and redshift.  The parameter distribution is modelled using the observed data from GAMA and DEVILS \citep{Bellstedt2021, Bellstedt2024, Thorne2021, Thorne2022}.  The covariance is instead fixed and estimated for blue and red galaxies from the residuals between the GAMA and DEVILS original distribution of SFH shape parameters and that obtained from sampling with the best-fitting parameters (we do not report the covariance to avoid an overly long table).   Details of the sampling are provided in Sect. \ref{sect:sfh_history}. }
\label{table:sfh_funct_parameters}
\end{table}

\subsection{Metallicity history}
\label{sect:metal_history}

Most studies in the literature, either aimed at modelling or fitting the galaxy population,  treat either the gas or the stellar metallicity (or both) as fixed quantities throughout the galaxy lifetime, often assuming the two having the same value,  or as a variable but constant value (see \citealt{Curti2025} for description of chemical evolution models). However, it is well known that the gas, and by reflection the stellar metallicity, evolves with the galaxy lifetime. Every new generation of stars forms from an interstellar medium (ISM) that gets enriched in metals through supernovae and stellar winds. The subsequent generation of stars will thus form from an ISM that contains a larger fraction of metals with respect to the one from which the previous generation of stars formed. This `closed-box' model is however complicated by the possible infall of pristine, lower metallicity gas from the inter-galactic medium and by the loss of metal-rich gas into said medium. The metallicity assumption has a great impact on galaxy SEDs and consequently on the measured galaxy properties. Indeed the well-known age-metallicity degeneracy leads higher metallicity younger SSPs to appear similar to lower metallicity but older SSPs \citep{Worthey1994}. \cite{Bellstedt2025} also found that an evolving metallicity, despite having little impact on the estimation of stellar mass and SFR, allows for the removal of systematic biases in the recovered cosmic star formation history (CSFH) and produces more realistic uncertainties in the recovered SFHs. Therefore, it is of great importance to introduce a physically motivated prescription for the gas metallicity evolution and how this reflects on the stellar metallicity, since new generation of stars form from the metal-enriched star-forming gas. 

\textsc{ProSpect} models the star-forming gas metallicity evolution of an individual galaxy by mixing discrete SSPs with varying ages and metallicities to achieve a target stellar population age and metallicity,  potentially using all metallicities available in the provided SSPs. This represents an improvement over approaches that fix the metallicity to a fiducial value, since it allows for the metallicity to be free to vary at either discrete or in between library interpolated values\footnote{\textsc{ProSpect}, at fixed age of a given SSP,  approximates the target metallicity by logarithmically weighting the coarse available grid of SSP metallicities.}.  The functional form of the metallicity evolution we adopt is the same one used in \cite{Bellstedt2021,Bellstedt2024,Alsing2023}.  The basic idea is to linearly map the stellar mass evolution of each galaxy on to the shape of the gas-phase metallicity evolution \citep{Driver2013}.  The metal enrichment is thus following a 1-to-1 relation with the mass-build, such that e.g.  when half of the stellar mass of a galaxy has been assembled,  half of the metal enrichment has occurred.  In this prescription,  the yield,  defined as the fraction in mass of a given element produced per unit mass of stars formed and returned to the ISM, is not fixed, but rather evolving, declining with time. This prescription represents what we would expect in a closed-box mode of star formation, but with some pristine inflow and a constant ejecta metallicity regardless of the metallicity of the gas that formed the stars \citep{Robotham2020}. 

This prescription has proven to reproduce the CSFH \citep{Bellstedt2021} and provide robust measurements of galaxy properties when applied to GAMA and DEVILS data \citep{Bellstedt2021,Bellstedt2024,Thorne2021,Thorne2022}. As pointed out in \cite{Robotham2020}, this type of metal evolution is appropriate to describe realistic galaxy formation as it looks physically more like the galaxy metallicity histories in semi-analytic models (e.g. SHARK \citep{Lagos2019}) than an analytic closed box. The linear mapping breaks down only in case of extreme inflow of pristine low metallicity gas, which is likely to be rare \citep{Ubler2014}.

The analytical prescription of the gas-phase metallicity evolution is as follows:
\begin{equation}
    Z_{\mathrm{gas}} (t) = Z_{\mathrm{gas,init}} + (Z_{\mathrm{gas,final}} - Z_{\mathrm{gas,init}}) \frac{1}{\mathcal{M}} \int_{0}^{t'} \mathrm{SFR}(t) \mathrm{d}t \ ,
    \label{eq:metal_history}
\end{equation}
where $Z_{\mathrm{gas,init}}$ is the initial metallicity for the earliest phases of star-formation, $\mathcal{M}$ is stellar mass in units of $M_{\odot}$ drawn from the GSMF, $\mathrm{SFR}(t)$ is the galaxy SFH, and $Z_{\mathrm{gas,final}}$ is the present-day gas-phase metallicity of the object. This prescription links the chemical enrichment of the galaxy to the SFR, such that increased star formation is associated with an increased rate of metal production. $Z_{\mathrm{gas,init}}$ is taken to be equal to the lowest metallicity of the \textsc{MIST} \citep{Dotter2016,Choi2016} isochrones used in \textsc{ProSpect},  while the maximum value that $Z_{\mathrm{gas,final}}$ can assume depends on the highest metallicity of the \textsc{MIST} isochrones. The gas-phase metallicity is important not only to establish the metallicity of the next generation of stars, but also to model the gas emission contribution to the galaxy SED, which depends on the gas-phase metallicity itself and on the gas ionisation (see Sect.  \ref{sect:gas_ionization}).

The relation between the gas-phase metallicity, often expressed in terms of the gas-phase oxygen abundance $\log{(O/H)}$, and the stellar mass is known as the mass-metallicity relation (MZR, \citealt{Tremonti2004,Kewley2008,Foster2012,Maiolino2019} and reference therein). This relation points to a positive correlation between the two quantities in such a way that the more massive (or luminous) the galaxy is, the higher the fraction of metals is going to be in comparison to its lower-mass counterparts. This is due to the increased gravitational potential well depth that allows massive galaxies to retain a large fraction of their metals within their halo. The relation flattens at high stellar masses where the mass growth mechanism is not related to star-formation anymore, but rather to gas-poor mergers. Red galaxies constitute the majority of objects at the high mass, high metallicity end of the relation, whereas the low-metallicity part is dominated by blue galaxies \citep{Bellstedt2021}. This distribution points to the fact that in reality the MZR represents only a projection of a more fundamental relation, known as Fundamental metallicity relation (FMR, \citealt{Mannucci2010,Lara-Lopez2013}), which links together the stellar mass, the gas-phase metallicity and the SFR of a galaxy.

In \textsc{GalSBI-SPS}, the present-day gas-phase metallicity $Z_{\mathrm{gas,final}}$ of an individual galaxy is obtained from the FMR which is a redshift-dependent relation that involves the galaxy stellar mass and the SFR. The present-day gas-phase metallicity of a galaxy, expressed as gas-phase oxygen abundance $\log{(O/H)}$, is drawn from a Normal distribution whose mean is described by the parametrisation introduced in \cite{Bellstedt2021}:
\begin{equation}
\begin{split}
    \left[ 12 + \log{(O/H)} \right] =& \ \alpha(t_{\mathrm{lb}}) \log{\left( \frac{\mathrm{SFR}}{M_{\odot} yr^{-1} } \right)} + \\ &\beta(t_{\mathrm{lb}}) \log{\left(\frac{\mathcal{M}}{M_{\odot}} \right)} + \gamma(t_{\mathrm{lb}}) \ ,
    \end{split}
\end{equation}
where $t_{\mathrm{lb}}$ is the observer look-back time to the galaxy redshift in $\mathrm{Gyr}$ and $\alpha(t_{\mathrm{lb}}), \beta(t_{\mathrm{lb}}), \gamma(t_{\mathrm{lb}})$ evolve quadratically with look-back time:
\begin{equation}
    \begin{split}
        \alpha(t_{\mathrm{lb}}) =& \alpha_0 + \alpha_1 t_{\mathrm{lb}} + \alpha_2 t_{\mathrm{lb}}^2\\
        \beta(t_{\mathrm{lb}}) =& \beta_0 + \beta_1 t_{\mathrm{lb}} + \beta_2 t_{\mathrm{lb}}^2\\
        \gamma(t_{\mathrm{lb}}) =& \gamma_0 + \gamma_1 t_{\mathrm{lb}} + \gamma_2 t_{\mathrm{lb}}^2 \ .
    \end{split}
\end{equation}
The scatter around this mean relation also evolves quadratically with look-back time:
\begin{equation}
    \sigma_{\mathrm{FMR}}(t_{\mathrm{lb}}) = \sigma_{\mathrm{FMR},0} + \sigma_{\mathrm{FMR},1} t_{\mathrm{lb}} + \sigma_{\mathrm{FMR},2} t_{\mathrm{lb}}^2 \ .
\end{equation}
We use for these parameters the values quoted in \cite{Bellstedt2021} and report them in Table \ref{table:mzr_funct_params}. These values have been obtained by fitting a sample of star-forming galaxies selected via sSFR cut, $\log{(sSFR/\mathrm{yr}^{-1})} > -11.5$. As a first approximation for this work, we extend the validity of the relation also to red galaxies, using the same parameter values for both population of objects.  The present-day gas-phase metallicity in solar units can be obtained from the one expressed in terms of oxygen abundance using the relation
\begin{equation}
    \frac{Z_{\mathrm{gas,final}}}{Z_{\odot}} = 10^{\left[ 12 + \log{(O/H)} \right] - \left[ 12 + \log{(O/H)} \right]_{\odot}} \ ,
\end{equation}
where $Z_{\odot}$ is the solar metallicity in absolute units and $\left[ 12 + \log{(O/H)} \right]_{\odot}$ is the solar oxygen abundance. This relation allows us to sample a relative-to-solar metallicity value, such that, depending on the specific absolute value for the gas-phase solar metallicity (e.g. $Z_{\odot}=0.020$ and  $\left[ 12 + \log{(O/H)} \right]_{\odot} = 8.69$ in \citealt{Asplund2009}), one can accordingly scale the absolute value of $Z_{\mathrm{gas,final}}$. The linear metallicity evolution is implemented in \textsc{ProSpect} through the \textsc{Zfunc\_massmap\_lin} function, while the present-day gas-phase metallicity $Z_{\mathrm{gas,final}}$ is passed to the \textsc{Zfinal} \textsc{ProSpect} parameter.

\begin{table}
\caption{Parameter values of the FMR parameter evolution with look-back time for the red and blue galaxy populations.}
\centering
\begin{tabular}{lcc}
\hline\hline
Parameter  & Red & Blue\\
\hline
$\alpha_{0}$ & -0.9202 & -0.9202 \\
$\alpha_{1}$ & 0.2916 & 0.2916 \\
$\alpha_{2} $ & -0.0508 & -0.0508 \\
\hline
$\beta_{0}$ & 1.198 & 1.198 \\
$\beta_{1}$ & -0.2658 & -0.2658 \\
$\beta_{2}$ & 0.0488 & 0.0488 \\
\hline
$\gamma_{0}$ & -2.785 & -2.785 \\ 
$\gamma_{1}$ & 2.268 & 2.268 \\
$\gamma_{2}$ & -0.4267 & -0.4267 \\
\hline
$\sigma_{\mathrm{FMR},0}$ & 0.4270 & 0.4270 \\
$\sigma_{\mathrm{FMR},1}$ & -0.0298 & -0.0298 \\
$\sigma_{\mathrm{FMR},2}$ & 0.0026 & 0.0026 \\
\hline
\end{tabular}
\tablefoot{The FMR parameters used in \textsc{GalSBI-SPS} are taken from the best-fitting values on GAMA galaxies in \citealt{Bellstedt2021}.  We sample the gas-phase oxygen abundance $\log{(O/H)}$ from a Normal distribution whose mean is given by the FMR.  The coefficients of the SFR and stellar mass dependence in the FMR evolve with redshift (expressed in terms of look-back time), as well as the scatter $\sigma_{\mathrm{FMR}}(t_{\mathrm{lb}})$ around the mean relation.}
\label{table:mzr_funct_params}
\end{table}

As for stellar metallicity $Z_*$, i.e. the fraction of metals lock in stars, a similar relation with the stellar mass exists \citep{Gallazzi2005,Zahid2017,Lian2018,Looser2024}. Rather than sampling from another parametric relation, we follow the prescription also adopted in other studies (e.g. \citealt{Alsing2023, Leja2017}) that the stellar metallicity is equal to that of the gas-phase metallicity at all times. This assumption, known as instantaneous enrichment, postulates that the stars have the same metallicity as the ISM from which they had formed. It is important to point out, as detailed in e.g. \cite{Fraser-McKelvie2022}, that the integrated stellar populations are generally more metal-poor than the ISM, especially in low-mass galaxies, even when the same metallicity indicator is adopted. This is also found in analytical models, such as the gas regulator or `bathtub' model \citep{Lilly2013,Peng2014}, which predicts a difference of roughly $\sim 0.2 \ \mathrm{dex}$. We leave the implementation of a more complex prescription that takes into account these differences to future work.

\subsection{Gas ionisation}
\label{sect:gas_ionization}

The origin of emission lines in optical galaxy spectra stems from nebular gas ionised either by stellar radiation or by radiation from the accretion disk of an AGN, with the latter confined to the central regions of galaxies.  Nebular gas ionised by radiation from massive stars is characterised by a set of physical properties that include the electron density $n_e$, gas-phase metallicity $Z_{\mathrm{gas}}$ (discussed in Sect. \ref{sect:metal_history}) and ionisation parameter $\log{U}$,  defined as the ratio of ionising photons to the total hydrogen density.  A careful modelling of the gas emission contribution to the galaxy SED therefore requires understanding the relation of these physical properties with the global properties of the host galaxies, such as stellar mass and SFR.

The emission line fluxes are typically computed by means of photoionisation codes such as \textsc{CLOUDY} \citep{Ferland2017} or \textsc{MAPPINGS} \citep{Dopita1996}.  These codes generate grids of emission line ratios by solving radiative transfer equations for multiple ionic species. They adopt different types of input ionising radiation with a dependence on gas properties, such as ionisation parameter and gas-phase metallicity, that can be fully modelled and explored \citep{Kewley2002, Dopita2013,Gutkin2016}.  Running photoionisation codes for large samples of galaxies is computationally expensive,  therefore,  in almost all of the SPS codes available in the literature (see Fig. 1 in \citealt{Thorne2021} for an overview),  the emission line fluxes come from pre-computed tables that are generated by making strong simplifying assumptions about the geometry,  usually a point source at the center of a spherical shell,  and the gas composition to reduce the number of free parameters \citep{Byler2017}\footnote{Photoionisation models might also suffer from limitations due to e.g.  \ion{H}{II} regions modelled as plane-parallel slabs of gas and relative chemical abundances assumed to scale proportionally to the solar pattern (see \citealt{Curti2025} review).}. 

\textsc{ProSpect} makes no exception,  adopting photoionisation tables from \cite{Levesque2010} to produce line emission features for a range of gas-phase metallicities and ionisation parameters.  The star formation nebular emission features are computed in \textsc{ProSpect} using a simple energy balance scheme.  Since efficient Hydrogen ionisation is largely caused by young O and B stars continuum flux short of the Lyman limit ($911.8 \ \AA$),  \textsc{ProSpect} assumes that the integrated intrinsic stellar flux at $\lambda < 911.8 \ \AA$ of stars younger than $10 \ \mathrm{Myr}$ determines the ionised flux of the nebulae \citep{Orsi2014,Robotham2020},  with no UV photon escape fraction in the default settings.  This ionising flux is then redistributed across known significant emission features interpolating between the gridded values of $Z_{\mathrm{gas}}$ and $\log{U}$ available from the \cite{Levesque2010} tables. The latter do not include the contribution from the nebular continuum,  which has been recently shown by \cite{Miranda2024} to provide a significant contribution to the galaxy broad-bands only for the extreme emission line galaxies, which are increasingly important at higher redshifts.

Another strong simplifying assumption that is adopted in pre-computed tables is to assume a fixed electron density of $n_e = 100 \ \mathrm{cm}^{-3}$ \citep{Byler2017},  which is also the assumption we use in \textsc{GalSBI-SPS} as we are adopting the \cite{Levesque2010} fluxes implemented in \textsc{ProSpect}.  This assumption is usually motivated by the interest in targeting the gas emission from \ion{H}{II} regions due to star formation, ignoring the contribution of AGN broad line regions where the electron density is higher.  However, it is well known that the typical electron density of \ion{H}{II} regions in low redshift galaxies is of the order of $n_e \sim 30-120 \ \mathrm{cm}^{-3}$ \citep{Brinchmann2008,Davies2021}, while, for high redshift galaxies, studies found values up to a few $\times 10^3 \ \mathrm{cm}^{-3}$ \citep{Masters2014,Shirazi2014,Kaasinen2017,Kashino2017}, with an average evolution at a rate of $\propto (1+z)^{1-2}$ \citep{Isobe2023,Abdurrouf2024}. 

The ionisation parameter values change as function of redshift and galaxy physical properties. Spectroscopic surveys of objects at $1 < z < 7$ have shown that there is a general increase of the ionisation parameter with redshift at fixed stellar mass \citep{Brinchmann2008,Shirazi2014,Kashino2017,Kaasinen2018,Papovich2022}. This is typically attributed to the presence of lower $\log{(O/H)}$, harder ionising spectra, and higher electron densities, characteristic of the high-redshift galaxy population. Indeed, there have been reports in the literature of an inverse correlation between the gas-phase metallicity and the ionisation parameter, since the increasing absorption due to a higher opacity in the stellar winds leads to a reduction at high metallicity of the stellar ionising photon flux illuminating the gas \citep{Orsi2014,Kashino2019}. Other studies \citep{Kaasinen2018} have instead found correlations between $\log{U}$ and $sSFR$, rather than with metallicity, therefore it is still unclear how exactly these parameters are connected and whether $\log{U}$ and $Z_{\mathrm{gas}}$ are intrinsically correlated or not.

\begin{table}
\caption{Gas ionisation relation parameter values for the red and blue galaxy populations.}
\centering
\begin{tabular}{lcc}
\hline\hline
Parameter  & Red & Blue\\
\hline
$\upsilon_0$ & -2.316 & -2.316 \\
$\upsilon_1$ & -0.360 & -0.360 \\
$\upsilon_2 $ & -0.292 & -0.292 \\
$\upsilon_3 $ & 0.428 & 0.428 \\
\hline
$\sigma_{\log{U}}$ & 0.1 & 0.1 \\
\hline
\end{tabular}
\tablefoot{The gas ionisation parameter $\log{U}$ is sampled from a Truncated Normal distribution whose mean depends on metallicity, $sSFR$ and electron density. The coefficients of this relation are taken from \cite{Kashino2019}.}
\label{table:logU_funct_params}
\end{table}

\begin{figure*}
   \centering
   \resizebox{\hsize}{!}{\includegraphics[width=\hsize]{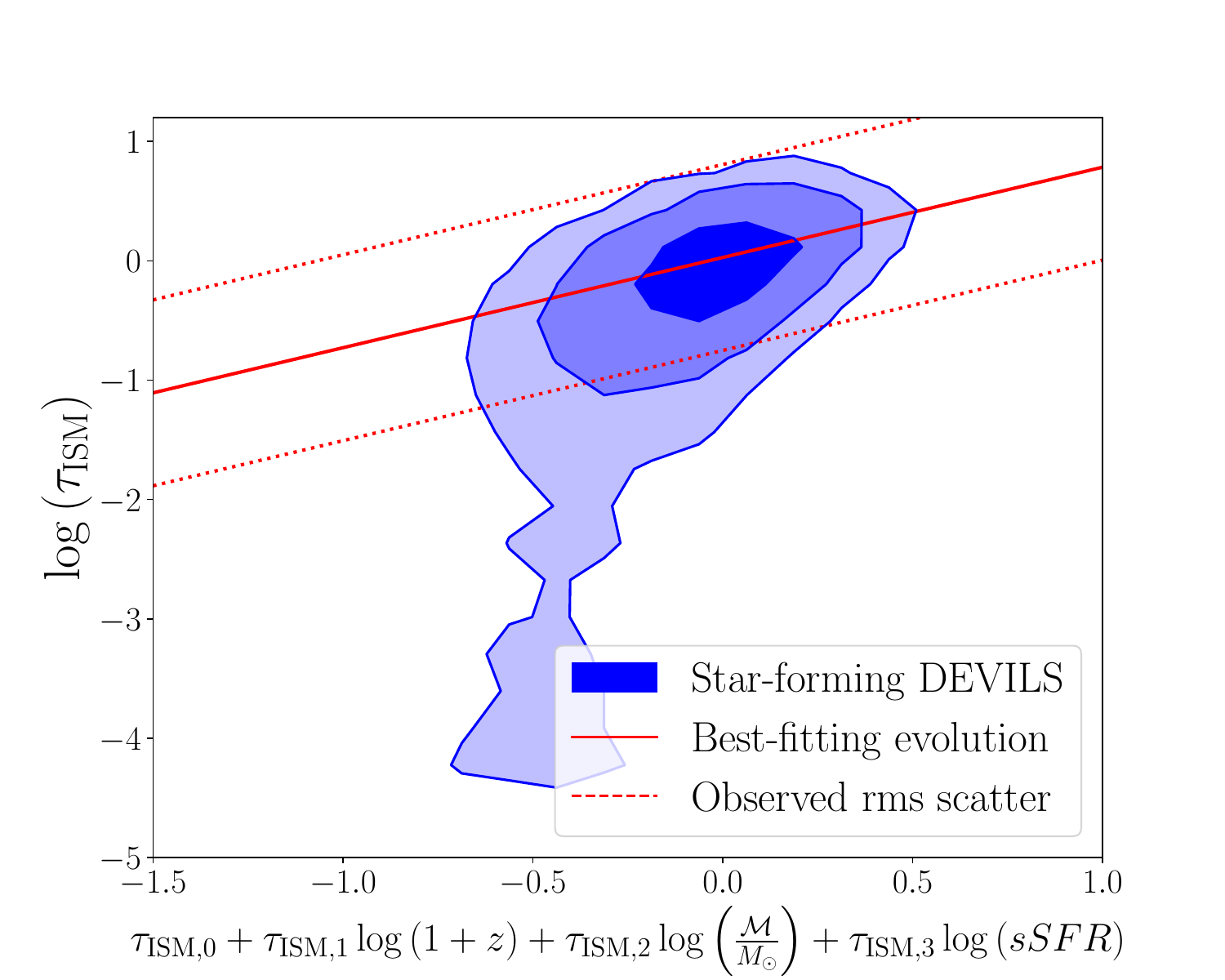}
   \includegraphics[width=\hsize]{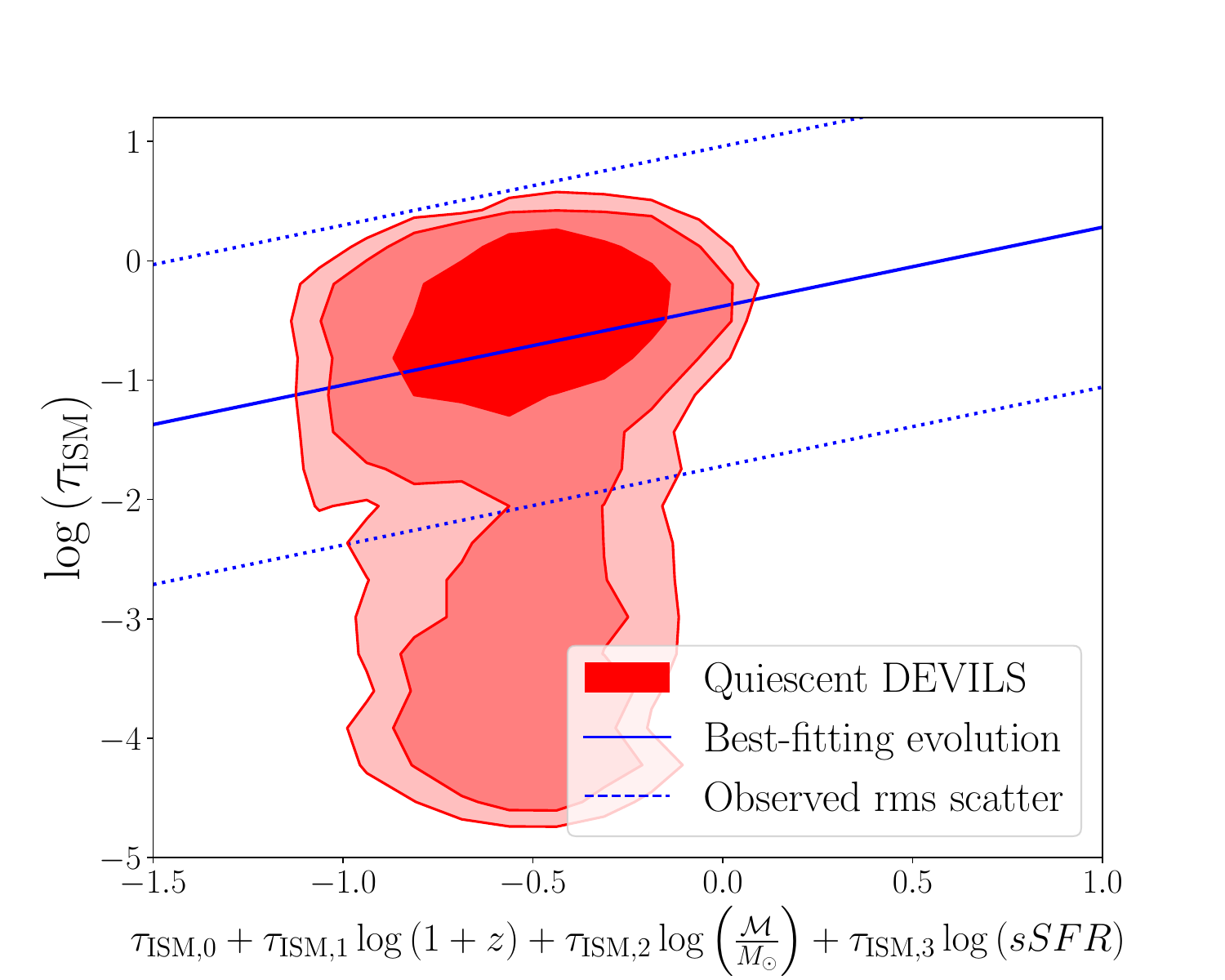}}
      \caption{Correlation between $\log{(\tau_\mathrm{ISM})}$ and redshift, stellar mass and $sSFR$ for star-forming (left panel) and quiescent (right panel) galaxies from DEVILS data. The solid red (blue) line shows the best-fitting relation expressing this dependence for star-forming (quiescent) galaxies. The dotted lines show the observed rms scatter of the relation. Each sample contour encloses $50\%,84\%,99\%$ of the values.
      }
    \label{fig:tau_ism_redshift_evolution}
\end{figure*}

In order to take into account the possible correlation among these parameters and let the data constrain their dependency, we model the gas ionisation $\log{U}$ following the prescription in equation 12 of \cite{Kashino2019},  where we assume that the galaxy can be modelled as a single \ion{H}{II} region described by a single ionisation parameter \footnote{\cite{Jamet2008} showed that the ionisation parameter varies according to the distribution of stars in individual \ion{H}{II} regions of the same galaxy, thereby impacting their nebular diagnostics.}.  We sample $\log{U}$ for individual galaxies from a truncated Normal distribution with mean
\begin{equation}
    \log{U} = \upsilon_0 + \upsilon_1 (\log{(O/H)} + 4) + \upsilon_2 \log{n_e} + \upsilon_3 (\log{(sSFR)} + 9) \ ,
\label{eq:gas_ionisation}
\end{equation}
and scatter around the mean relation equal to $\sigma_{\log{U}} = 0.1 \ \mathrm{dex}$, as quoted in \cite{Kashino2019}. The electron density is fixed to $n_e = 100 \ \mathrm{cm}^{-3}$, while $\log{(O/H)}$ is computed from $Z_{\mathrm{gas,final}}$ as follows
\begin{equation}
    \log{(O/H)} = \log{\left( \frac{Z_{\mathrm{gas,final}}}{Z_{\odot}} \right) } + \left[ 12 + \log{(O/H)} \right]_{\odot} -12 \ .
\end{equation}
$\log{sSFR}$ is instead computed from the stellar mass and the SFR drawn from the relations in Sects. \ref{sect:smf} and \ref{sect:sfh_history}, $\log{(sSFR \ \mathrm{yr}^{-1})} = \log{(\mathrm{SFR} / \mathcal{M})}$. We limit the sampled parameters in the range $\log{U}= [-4, -1]$, which is the typical range used in \textsc{CLOUDY} and \textsc{MAPPINGS}. The parameters for this relation are set to the best-fitting values in \cite{Kashino2019} and reported in Table \ref{table:logU_funct_params}. All the parameters are well constrained in \cite{Kashino2019}, fully rejecting the null hypothesis of no dependence on said parameters. \cite{Kashino2019} derived this relation by measuring the nebular conditions from the emission-line flux ratios of a sample of low redshift $0.027 \leq z \leq 0.25$ galaxies from the Sloan Digital Sky Survey (SDSS) that have been stacked into bins of stellar mass and $sSFR$. Although in this sample the gas-phase metallicity has been estimated for objects having $\log{(sSFR \ \mathrm{yr}^{-1})} \geq -10.5$, therefore star-forming objects, we apply this relation also to the red population of galaxies. This is motivated by the fact that the impact of the gas emission on the red galaxies spectra (and therefore magnitudes) is in principle negligible.

$\log{U}$ is a dimensionless quantity that expresses the ratio of hydrogen ionising photons to total hydrogen density. In the literature, some studies refer to the ionisation parameter as $\log{U}$ \citep{Byler2017}, while others use the quantity $q \ [\mathrm{cm} \ \mathrm{s}^{-1}]$ \citep{Levesque2010}. The two definitions simply differs by the speed of light in units of $[\mathrm{cm} \ \mathrm{s}^{-1}]$, $q = U \times c$. \textsc{ProSpect} expresses the ionisation parameter of the gas in terms of $q$ and allows the user to relate this parameter to the provided gas-phase metallicity using equation 5 in \cite{Orsi2014}. To account for the possible dependence on other physical quantities, we provide the code with our value of the ionisation parameter obtained by converting the quantity sampled from equation \ref{eq:gas_ionisation} into the appropriate units recognised by our generative SED engine.

\subsection{Dust component}
\label{sect:dust_component}

In \cite{Tortorelli2024}, we showed that the dust attenuation was one of the SED modelling components that had the strongest impact on the galaxy colours (up to $\sim 1 \ \mathrm{mag}$ in $u-g$) and on the mean redshift of the tomographic bins ($\Delta z \sim 0.1$), therefore a realistic modelling of dust and its relation with the galaxy physical properties is a key ingredient for a SPS-based galaxy population model. Dust forms in the atmospheres of evolved stars or in remnants of supernovae \citep{Draine2003} and released in the ISM where, observationally, modifies the stellar continuum through the absorption and the scattering of starlight and the re-emission of the absorbed energy in the infrared. The dust attenuation involves not only absorption and scattering out of the line of sight, but also the complex star-dust geometry in a galaxy and the scattering back into the line of sight.

Observations from large samples of galaxies at low redshift \citep{Salim2018} show that the dust attenuation laws vary as function of galaxy types even among similar mass galaxies at different redshifts \citep{Salim2020}. They also change between lines of sight within the same galaxy \citep{Maiz2024}. This points to the fact that it does not exist a single dust attenuation law with fixed parameters that is generic enough that it can be applied to the whole galaxy population, but rather that the use of a dust attenuation prescription should be tightly linked to the spectral type and physical properties of the individual galaxies \citep{Nagaraj2022}.

We model the effect of dust in \textsc{GalSBI-SPS} using the \cite{Charlot2000} two-phase dust attenuation model, which modifies the intrinsic flux $f_{\mathrm{intr}}$ into the observed attenuated one $f_{\mathrm{att}}$ at a given wavelength $\lambda$ through the attenuation factor $A(\lambda)$, 
\begin{equation}
     f_{\mathrm{att}} (\lambda) = f_{\mathrm{intr}} (\lambda) \times 10^{-A(\lambda)/2.5},
\end{equation}
where $A(\lambda)$ is related to the effective optical depth of attenuation $\tau$ by
\begin{equation}
     A(\lambda) = e^{-\tau (\lambda / \lambda_0)^\nu} .
\end{equation}
$\tau$ changes the column density of dust, $\lambda_0 = 5500 \AA$ is the pivot wavelength and $\nu$ is the modifying power-law index, which we set to $-0.7$. This model splits the galaxy starlight in a component coming from stars younger than $10 \ \mathrm{Myr}$ and another from stars older than $10 \ \mathrm{Myr}$, which is the typical timescale for the disruption of a molecular cloud \citep{Blitz1980}. Older stars are solely attenuated by a dust `screen', regulated by the $\tau_{\mathrm{ISM}}$ parameter, which represents the dust in the diffuse ISM. Young stars are instead attenuated by the `dust screen' and by an additional `birth cloud' component that adds extra attenuation towards young stars, simulating their embedding in molecular clouds and \ion{H}{II} regions. The `birth cloud' component is regulated by the $\tau_\mathrm{BC}$ parameter. From the definition, it is clear that $\tau_{\mathrm{ISM}}$ is the parameter that has the strongest effect on the galaxy SED, while $\tau_\mathrm{BC}$ mostly impacts the UV flux (see Fig. 3 in \citealt{Thorne2021}). An additional factor that impact the dust attenuation in the UV is the $2175 \AA$ bump. This a prominent absorption feature that is postulated to arise from polycyclic aromatic hydrocarbon (PAH) dust grains, even though the actual carrier has not been definitively established \citep{Battisti2025}. The bump seems to be redshift-independent \citep{Battisti2022} and in general its strength does not correlated with any galaxy property, with a median value of roughly $1/3$ of the MW value \citep{Battisti2020} among different type of galaxies. A recent work by \cite{Battisti2025} reported though an anti-correlation with surface SFR, in such a way that the bump is supposed to be applied only to old stars, since the carrier seems to be easily destroyed by UV photons. Given the size of the sample and the uncertainty concerning the dependency of the bump strength on galaxy properties, we use the default value for the bump strength in the \textsc{ProSpect} SED computation, which is set to this roughly universal value of $1/3$ of the MW value.

\begin{table}
\caption{Dust attenuation parameter values for the red and blue galaxy populations.}
\centering
\begin{tabular}{lcc}
\hline\hline
Parameter  & Red & Blue\\
\hline
$\tau_{\mathrm{ISM},0}$ & 2.213 & -1.421 \\
$\tau_{\mathrm{ISM},1}$ & 2.923 & 0.533 \\
$\tau_{\mathrm{ISM},2} $ & -0.319 & 0.235 \\
$\tau_{\mathrm{ISM},3} $ & 0.003 & 0.131 \\
$\sigma_{\mathrm{obs},\log{(\tau_\mathrm{ISM})}}$ & 1.3 & 0.78 \\
\hline
$\mu_{\log{(\tau_{\mathrm{BC,\{ b,r \}}})}}$ & -0.18 & -0.10 \\
$\sigma_{\log{(\tau_{\mathrm{BC,\{b,r\}}})}}$ & 0.21 & 0.24 \\
\hline
\end{tabular}
\tablefoot{Dust attenuation parameters used to sample $\tau_\mathrm{ISM}$ and $\tau_\mathrm{BC}$. The first five rows are obtained by fitting the hyper-plane that express the dependence of $\tau_\mathrm{ISM}$ on redshift, stellar mass and $sSFR$ modelled from DEVILS data. The last two rows show mean and scatter of the Normal distribution from which we sample $\tau_\mathrm{BC}$.}
\label{table:tauism_funct_params}
\end{table}

In order to sample the dust parameters for our generative model in a physically meaningful way, we use the DEVILS data to model the dependence of $\tau_\mathrm{ISM}$ and $\tau_\mathrm{BC}$ on the integrated galaxy properties. We do not find any dependence of $\log{(\tau_\mathrm{BC})}$ on redshift, stellar mass or $sSFR$. Therefore, this parameter is simply sampled from a Normal distribution that has different means $\mu_{\log{(\tau_{\mathrm{BC}})}}$ and scatters $\sigma_{\log{(\tau_{\mathrm{BC}})}}$ for the population of blue and red galaxies, truncated at the limits in Table 2 of \cite{Bellstedt2020}. On the contrary, we find that $\log{(\tau_\mathrm{ISM})}$ does show a dependence on redshift, stellar mass and $sSFR$. We decide to model this dependence as:
\begin{equation}
\begin{split}
    \log{(\tau_\mathrm{ISM})} =& \tau_{\mathrm{ISM},0} + \tau_{\mathrm{ISM},1} \log{(1+z)} +\\ &\tau_{\mathrm{ISM},2} \log{\left(\frac{\mathcal{M}}{M_{\odot}} \right)} + \tau_{\mathrm{ISM},3} \log{(sSFR)} .
\end{split} 
\label{eq:tau_ism}
\end{equation}
In order to obtain the best-fitting parameters of this relation, we fit an hyper-plane to the DEVILS catalogue data using the \textsc{ltsfit} library \citep{Cappellari2013}. Figure \ref{fig:tau_ism_redshift_evolution} shows the best-fitting relation expressing the dependence of $\log{(\tau_\mathrm{ISM})}$ on redshift, stellar mass and $sSFR$. $\log{(\tau_\mathrm{ISM})}$ is then sampled from a Normal distribution with mean following \ref{eq:tau_ism} and scatter $\sigma_{\mathrm{obs},\log{(\tau_\mathrm{ISM})}}$ around this relation measured from the data, truncated at the limits in Table 2 of \cite{Bellstedt2020}. The parameter values of the relations from which we sample $\log{(\tau_\mathrm{BC})}$ and $\log{(\tau_\mathrm{ISM})}$ for the population of red and blue galaxies are reported in Table \ref{table:tauism_funct_params}. The two-phase dust attenuation is directly implemented in \textsc{ProSpect} by simply providing the values of $\tau_\mathrm{BC}$ and $\tau_\mathrm{ISM}$ as input to the parameters \textsc{tau\_birth} and \textsc{tau\_screen}, respectively. 

\textsc{ProSpect} contains also an energy balance prescription that re-emits the bolometric sum of attenuated light in the FIR using the \cite{Dale2014} templates. As discussed in \cite{Tortorelli2024} and also visible in Fig. 3 in \citealt{Thorne2021}, this emission impacts the galaxy SED mostly beyond the $20000 \AA$ rest-frame, which is in the $Ks$ band, the reddest band of interest, thereby providing an impact on galaxy colours that is negligible for our science case. Therefore, we set the \textsc{alpha\_SF\_birth} and \textsc{alpha\_SF\_screen} parameters in \textsc{ProSpect} to their default values ($1$ and $3$, respectively) and we do not model the dust emission parameter distributions, leaving a more thorough modelling to a future analysis.

\subsection{Velocity dispersion}
\label{sect:vel_disp}

The velocity dispersion in an integrated spectrum is a light-weighted superposition of all the velocity dispersions of the individual stellar populations. Hence, the observed value is going to be similar to that of the stars dominating the light from the galaxy. From an SPS point of view, the velocity dispersion manifests itself as a broadening of the spectral lines that is usually modelled as a Gaussian smoothing in velocity space in $\mathrm{km} \  \mathrm{s}^{-1}$. As shown in \cite{Tortorelli2024}, varying the velocity dispersion of a galaxy does not make an appreciable contribution to the broad-band fluxes. However, we include a model for sampling the velocity dispersion in \textsc{GalSBI-SPS} as our aim is to build a model that is realistic enough to be used to also forward-model spectroscopic surveys. 

\begin{table}
\caption{Velocity dispersion parameter values for the red and blue galaxy populations.}
\centering
\begin{tabular}{lcc}
\hline\hline
Parameter  & Red & Blue\\
\hline
$\log{(M_b/M_{\odot})}$ & 10.26 & 10.26 \\
$\log{(\sigma_b/km \ s^{-1})}$ & 2.073 & 2.073 \\
$s_1 $ & 0.403 & 0.403 \\
$s_2 $ & 0.293 & 0.293 \\
\hline
\end{tabular}
\tablefoot{Velocity dispersions are sampled from a Truncated Normal distribution whose mean follows the relation with stellar mass in \cite{Zahid2016}.  The parameter values of this relation are taken from the first row of Table 1 in \cite{Zahid2016}.}
\label{table:veldisp_funct_params}
\end{table}

As in \cite{Tortorelli2024}, we model the velocity dispersion $\sigma_{\mathrm{disp}}$ for our population of galaxies following the stellar mass-velocity dispersion relation introduced in \cite{Zahid2016}:
\begin{equation}
\begin{split}
    \sigma_{\mathrm{disp}}(\mathcal{M}) &= \sigma_b \left ( \frac{\mathcal{M}}{M_b} \right)^{s_1} \ \mathrm{for} \ \mathcal{M} \le M_b \\
    \sigma_{\mathrm{disp}}(\mathcal{M}) &= \sigma_b \left ( \frac{\mathcal{M}}{M_b} \right)^{s_2} \ \mathrm{for} \ \mathcal{M} > M_b
\end{split}
\label{eq:vel_disp}
\end{equation}
where the values of $M_b,\sigma_b,s_1,s_2$ are reported in Table \ref{table:veldisp_funct_params} and are taken from the first row of Table 1 in \cite{Zahid2016}. These values are redshift independent, as \cite{Zahid2016} showed that the Sloan Digital Sky Survey and Smithsonian Hectospec Lensing Survey data at $z<0.7$ are consistent with no evolution in redshift. We sample the velocity dispersion from a Normal distribution whose mean follows the relation in equation \ref{eq:vel_disp} and whose scatter is taken from the values quoted in \cite{Zahid2016}. The minimum velocity dispersion is set to $10 \ \mathrm{km \ s^{-1}}$. We smooth the galaxy spectrum as a Gaussian broadening in velocity space, where the Gaussian FWHM is $FWHM = 2.35482 \times \sigma_{\mathrm{disp}}$. The smoothing is applied prior to adding the AGN flux contribution.

\subsection{Active galactic nuclei}
\label{sect:agn_model}

AGNs are powered by supermassive black-holes ($M_{\mathrm{BH}} \gtrsim 10^6 M_{\odot}$) in the centre of galaxies that are actively accreting mater. They have high luminosities ($L_{\mathrm{bol,AGN}} \lesssim 10^{48} \ \mathrm{erg \ s^{-1}}$) and strong evolution of their luminosity function with redshift \citep{Kulkarni2019,Shen2020,Thorne2022}. Their emission covers the whole electromagnetic spectrum \citep{Padovani2017}, with our wavelength range of interest, from the UV to the infrared, being powered by the accretion disk, mostly in the form of broad and narrow emission lines strongly ionised by the central engine, and by the thermal emission of the dust heated by the central engine. When actively accreting matter and contributing more than $10\%$ to the galaxy bolometric luminosity, an AGN can dramatically modify galaxy colours, inducing strong changes in the redshift distribution of galaxies detected in a survey \citep{Tortorelli2024}.

\begin{figure}
   \centering
   \includegraphics[width=\hsize]{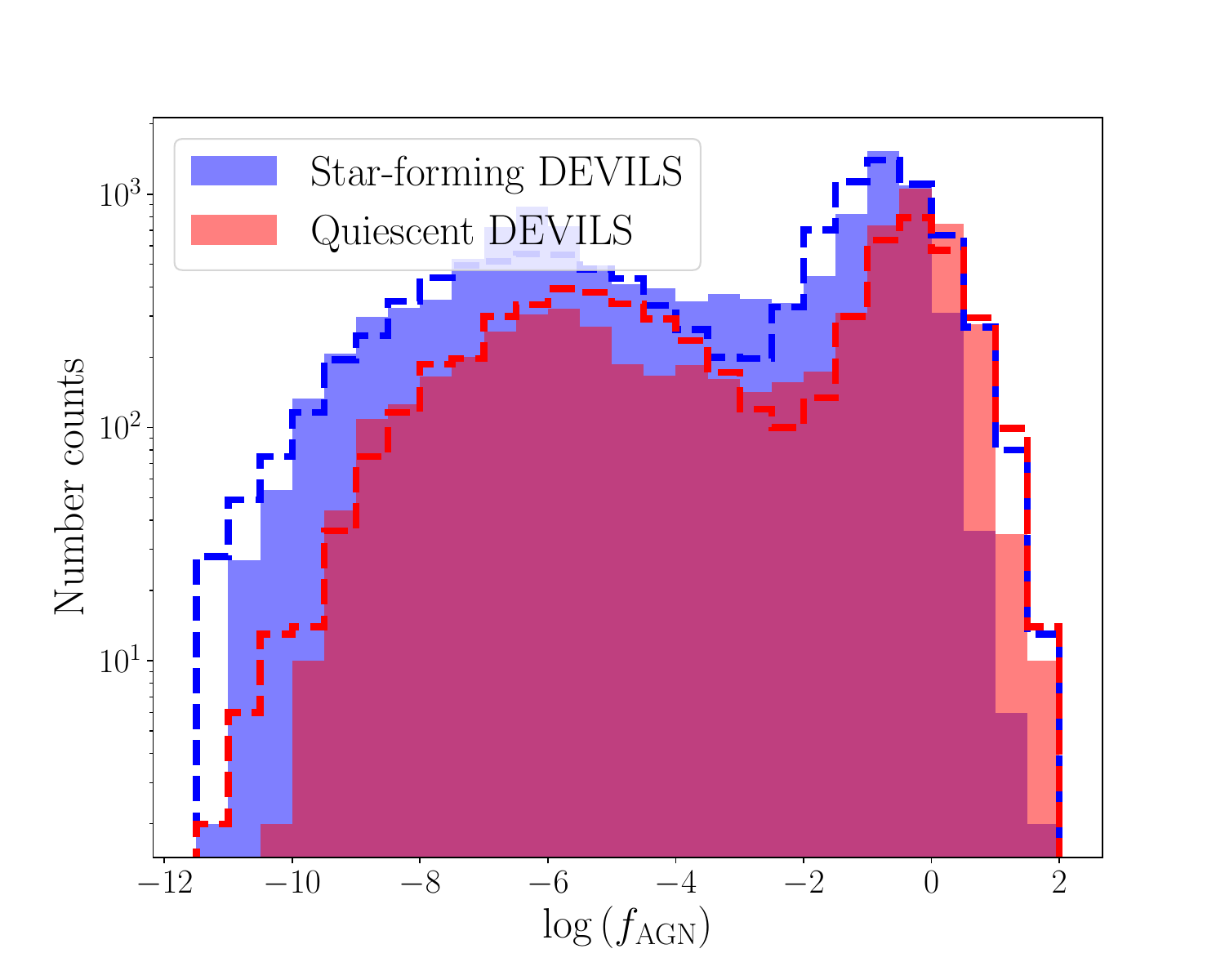}
      \caption{Histogram of the ratio between the AGN bolometric luminosity and the galaxy luminosity in log-space, $\log{(f_{\mathrm{AGN}})}$, for star-forming (blue histogram) and quiescent (red histogram) galaxies. The dashed lines represent the samples drawn from the sum of two Normal distributions with which we model the $\log{(f_{\mathrm{AGN}})}$ values from DEVILS data.
              }
    \label{fig:log_fagn}
\end{figure}

To model AGN hosts in \textsc{GalSBI-SPS},  we adopt the prescription of \cite{Lopez-Lopez2024}.  We first compute the probability that a galaxy hosts an AGN as function of redshift and stellar mass by taking the ratio between the redshift-evolving GSMFs of blue and red galaxies and the AGN host galaxy mass function (HGMF) derived in \cite{Bongiorno2016}.  Then,  we statistically assign an AGN to a host galaxy with a Bernoulli trial proportional to the computed probability.  As in \cite{Lopez-Lopez2024},  we assume fixed AGN fractions outside the redshift ranges specified by \cite{Bongiorno2016}. In particular,  the probability of hosting an AGN is $1\%$ at all stellar masses below $z=0.55$, while at $z>2$ we maintain a constant fraction taken from the last redshift bin in \cite{Bongiorno2016}.  These assumptions are support by recent observational studies \citep{Birchall2022,Williams2022,Osorio-Clavijo2023}.  

To assign an AGN spectral contribution,  we adopt a different set of AGN templates with respect to those implemented in \textsc{ProSpect}, which are the \cite{Dale2014} model, the extended AGN template from \cite{Andrews2018} and the \cite{Fritz2006} model. These templates model the emission from the AGN as composition of power-laws with different spectral indices, however AGN spectra in the optical/UV are also characterised by the presence of broad and/or narrow emission lines coming from the strongly ionised material in the accretion disc. Similarly to what we did in \cite{Tortorelli2024}, we replace the AGN model in \textsc{ProSpect} with the parametric quasar SED model presented in \cite{Temple2021}, calibrated to match the observed $u,g,r,i,z,Y,J,H,K,W1,W2$ colours of the SDSS DR16 quasar population \citep{Lyke2020}. This model covers the full AGN contribution in the $912 \AA \lesssim \lambda  \lesssim 30000 \AA $ rest-frame wavelength range by modelling the the low-frequency tail of the direct emission from the accretion disc ($900 \AA \lesssim \lambda \lesssim 10000 \AA$), the emission from the hot dust torus ($10000 \AA \lesssim \lambda \lesssim 30000 \AA$), the Balmer continuum ($\sim 3000 \AA$), the broad and narrow emission lines in the optical/UV region with the Baldwin eﬀect \citep{Baldwin1977} applied, and the Lyman-absorption suppression at $\lambda < 912 \AA$. The \cite{Temple2021} model will be implemented directly into \textsc{ProSpect} in a future release of the code\footnote{Private communication with the code author.}.

The \cite{Temple2021} model requires as input the the monochromatic $3000 \AA$ continuum luminosity and the absolute i-band magnitude at $z = 2$, as defined by \cite{Richards2006}, to produce the AGN spectrum with the right units and normalisation. We relate the bolometric luminosity of the AGN $L_{\mathrm{bol,AGN}}$ to these two quantities by following the prescription in the \cite{Temple2021} templates code repository\footnote{We follow the author guidelines provided to us in a private communication.}. The bolometric AGN luminosity, in turns, is related to the galaxy bolometric luminosity $L_{\mathrm{bol,galaxy}}$ through the $f_{\mathrm{AGN}}$ parameter, $L_{\mathrm{bol,AGN}} = f_{\mathrm{AGN}} \times L_{\mathrm{bol,galaxy}}$, which is the only parameter we are sampling in \textsc{GalSBI-SPS} to add the contribution from the AGN to the galaxy SED. We model the $f_{\mathrm{AGN}}$ distribution from the DEVILS data used in \cite{Thorne2022}. We do not use the values of $f_{\mathrm{AGN}}$ reported in this paper though, as its definition is different from the one highlighted above. We instead recompute $f_{\mathrm{AGN}}$ to match our definition by re-running \textsc{ProSpect} with the exact same setup as in \cite{Thorne2022}.  We model $\log{(f_{\mathrm{AGN}})}$ as a two component mixture of Normal distributions. Figure \ref{fig:log_fagn} shows the best-fitting distributions for red and blue galaxies in DEVILS separated by $sSFR$ superimposed on the newly computed $f_{\mathrm{AGN}}$ values. The best-fitting mean and standard deviation of the Normal distributions are reported in Table \ref{table:fagn_funct_params}.  We then assign AGN flux contributions to all galaxies, regardless of whether they are classified as AGN hosts. However,  $f_{\mathrm{AGN}}$ is sampled from different distributions depending on the AGN classification.  Galaxies hosting AGNs draw from the high-$\log{(f_{\mathrm{AGN}})}$ distribution, while non-AGN galaxies draw from the low-$\log{(f_{\mathrm{AGN}})}$ one.  We adopt a threshold of $\log(f_{\mathrm{AGN}}) = -1$ to separate these two regimes \citep{Thorne2022}.

\begin{table}
\caption{Ratio of AGN to galaxy luminosity parameter values for the red and blue galaxy populations.}
\centering
\begin{tabular}{lcc}
\hline\hline
Parameter & Red & Blue\\
\hline
$\mu_{f_{\mathrm{AGN,1}}}$ & -5.5 & -6.25 \\
$\sigma_{f_{\mathrm{AGN,1}}}$ & 1.75 & 2 \\
\hline
$\mu_{f_{\mathrm{AGN,2}}}$ & -0.25 & -0.75 \\
$\sigma_{f_{\mathrm{AGN,2}}}$ & 0.7 & 0.8 \\
\hline
\end{tabular}
\tablefoot{The ratio of AGN to galaxy luminosity $\log{f_{\mathrm{AGN}}}$ is sampled from either one of two different Normal distributions depending on whether the galaxy is an AGN host.Galaxies hosting AGNs draw from the high-$\log{f_{\mathrm{AGN}}}$ distribution, while non-AGN galaxies draw from the low-$\log{f_{\mathrm{AGN}}}$ one.  We report mean and standard deviation of the two Normal distributions.}
\label{table:fagn_funct_params}
\end{table}

\begin{figure*}
   \centering
   \resizebox{\hsize}{!}{\includegraphics[width=\hsize]{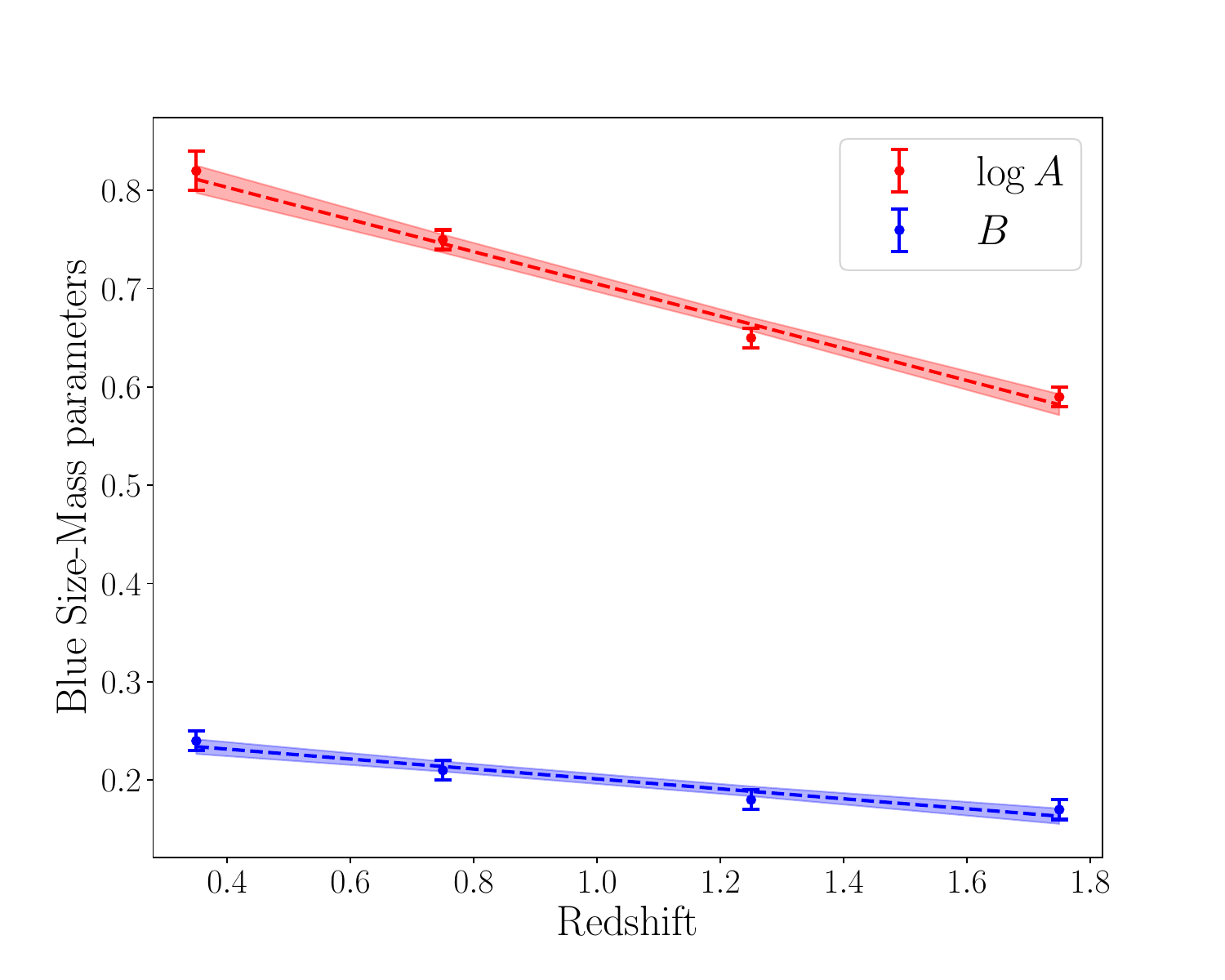}
   \includegraphics[width=\hsize]{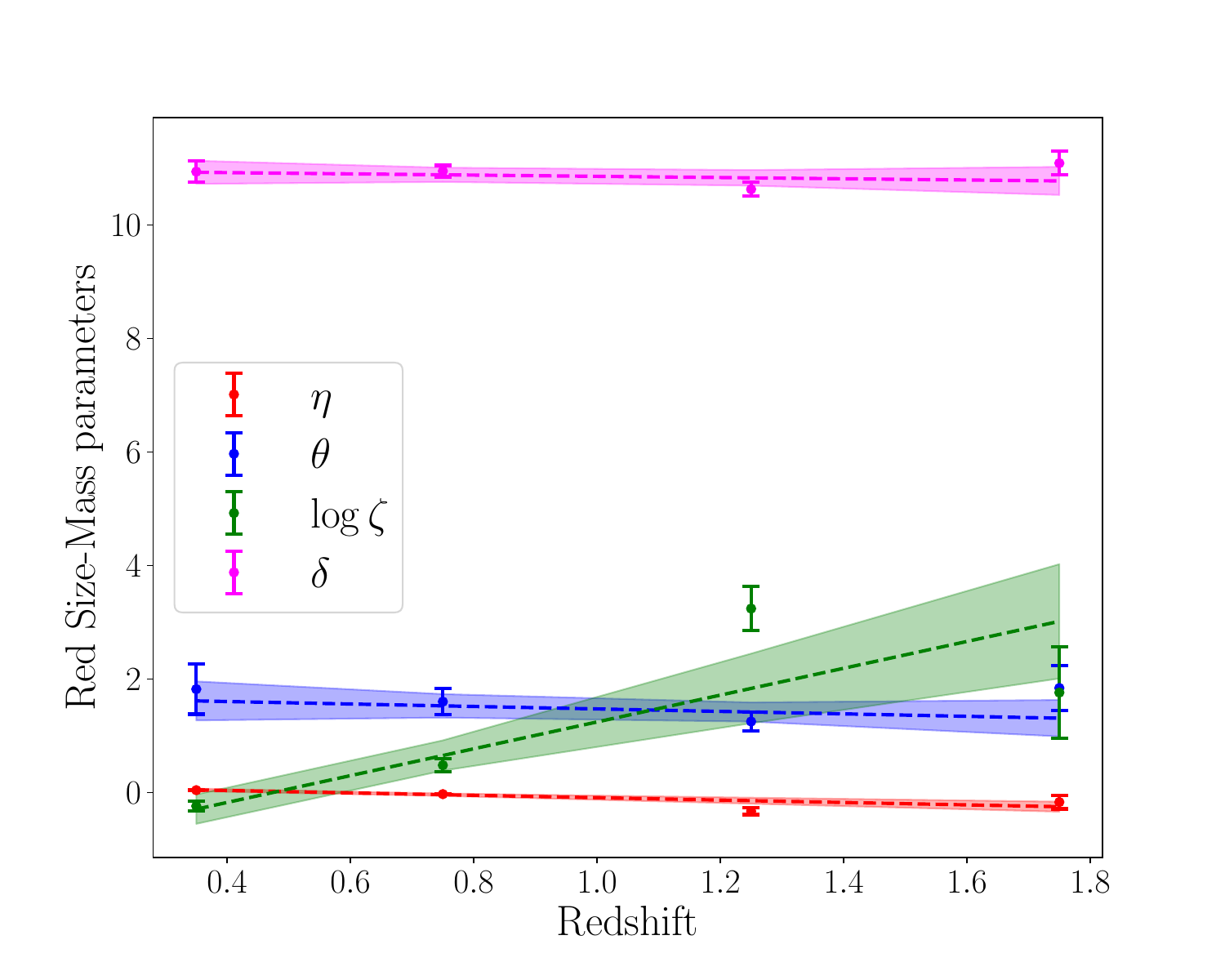}}
      \caption{The best-fitting linear evolution in redshift of the size-stellar mass parameters for star-forming (left panel) and quiescent (right panel) galaxies. The scatter points represent the observed values of the parameters and are taken from Tables 2 and 3 in \cite{Nedkova2021}. The dashed lines show the best-fitting relations, while the bands represent the $1\sigma$ uncertainty on the fit.}
    \label{fig:size_mass_redshift_evolution}
\end{figure*}

\subsection{Magnitude computation}
\label{sect:mag_computation}

The galaxy physical properties sampled in the previous sections are essential to create galaxy SEDs with \textsc{ProSpect}. The parameters we provide as input to the code are $\mathrm{mSFR}$, $\mathrm{mpeak}$, $\mathrm{mperiod}$ and $\mathrm{mskew}$ for the SFH, $\tau_\mathrm{BC}$ and $\tau_\mathrm{ISM}$ for the two component dust attenuation, $Z_{\mathrm{gas,final}}$ for the present day value of the linearly evolving gas-phase metallicity, the ionisation parameter $q$, and the SSP library described in Sect. \ref{sect:stellar_pop_model}. We set the redshift to $z=0$ in order to create a rest-frame SED for each galaxy in units of $L_{\odot}/\AA$. We then apply a Gaussian broadening to the rest-frame spectra to model the effect of the galaxy velocity dispersions (see Sect. \ref{sect:vel_disp}). We compute the observed-frame spectra in units of $\mathrm{erg} \ \mathrm{s}^{-1} \ \mathrm{cm}^{-2} \ \AA^{-1}$ using the luminosity distance $D_L$ for each object obtained with \textsc{PyCosmo} \citep{Refregier2018}. The addition of an AGN contribution to the flux is left as a choice to the user through a specific flag in the code. The AGN contribution is computed with the \cite{Temple2021} templates as detailed in Sect. \ref{sect:agn_model}. The output AGN SED is also in the observed-frame and it can directly be added to the galaxy flux contribution. We then apply reddening due to galactic extinction to the observed-frame spectra using the \cite{Fitzpatrick1999} extinction law. The extinction map of the Milky-Way is the one presented in \cite{Schlegel1998}, where the $E(B-V)$ values are obtained based on the galaxy $RA$ and $DEC$ coordinates. The final spectra are then integrated in the $g,r,i,z,y$ HSC wavebands to obtain the observed-frame AB magnitudes. We also compute the rest-frame magnitudes by de-redshifting the galaxy spectra before applying the galactic extinction.  The Python package \textsc{galsbi} further allows the user to skip the rest-frame SED computation and save computational time by obtaining directly the observed and rest-frame magnitudes from the sampled physical quantities using the \textsc{ProMage} magnitude emulator presented in \cite{Tortorelli2025}. 

\begin{table}
\caption{Galaxy size-stellar mass relation parameter values for the red and blue galaxy populations.}
\centering
\begin{tabular}{lcc}
\hline\hline
Parameter  & Red & Blue\\
\hline
$\log{A}_{\mathrm{intcpt}}$ &  & 0.869 \\
$\log{A}_{\mathrm{slope}}$ &  & -0.164 \\
$B_{\mathrm{intcpt}}$ &  & 0.252 \\
$B_{\mathrm{slope}}$ &  & -0.051 \\
$\eta_{\mathrm{intcpt}}$ & 0.119 &  \\
$\eta_{\mathrm{slope}}$ &  -0.211  &  \\
$\theta_{\mathrm{intcpt}}$ & 1.689 &  \\
$\theta_{\mathrm{slope}}$ &  -0.217  &  \\
$\log{\zeta}_{\mathrm{intcpt}}$ & -1.122 &  \\
$\log{\zeta}_{\mathrm{slope}}$ &  2.364  &  \\
$\delta_{\mathrm{intcpt}}$ & 10.966 &  \\
$\delta_{\mathrm{slope}}$ &  -0.108  &  \\
\hline
\end{tabular}
\tablefoot{Galaxy size parameter values obtained from fitting the redshift evolution of the values in Tables 2 and 3 in \cite{Nedkova2021}.  The galaxy sizes are drawn from a log-Normal distribution whose mean is given by the galaxy size-stellar mass relation. This relation is parametrised as a single power-law for blue galaxies and as a double power-law for red galaxies.}
\label{table:re_funct_params}
\end{table}

\subsection{Morphological parameters distribution}
\label{sect:morpho_params}

The modelling components described in the previous sections have been used to generate realistic galaxy SEDs and, consequently, magnitudes. A crucial aspect of our forward-modelling approach is the ability to simulate realistic survey images with \textsc{UFig}. This requires not only magnitudes, but also assigning each galaxy a light profile, which we assume to be a S\'ersic light profile \citep{Sersic1963}, and sampling additional morphological parameters, such as sizes, S\'ersic indices, ellipticities and positions. We leave the use of more complex morphologies, such as bulge and disk, to a future work, as their relation with galaxy SFHs and physical properties require more careful modelling as shown in \cite{Robotham2022,Bellstedt2024}.

Galaxy sizes are key observational quantities to understand the formation pathways of galaxies since they imprint on the galaxy structure (e.g. \citealt{Mo1998}). While it is well established that galaxy sizes evolve with redshift (e.g. \citealt{Daddi2005,Trujillo2007,vanDokkum2008}), the degree with which they do so and whether this size growth is due to major mergers (e.g. \citealt{Naab2007}), minor mergers (e.g. \citealt{Naab2009}), progenitor bias (e.g. \citealt{Fagioli2016}), feedbacks (e.g. \citealt{Damjanov2009}), or other processes, is still under debate. Galaxy physical sizes in $\mathrm{kpc}$ exhibit a tight scaling relation with the stellar mass known as size-stellar mass relation (e.g. \citealt{vanderwel2014}). This is well characterised at low redshift and at high mass end (e.g. \citealt{Shen2003}), while the behaviour at higher redshift (e.g. \citealt{Yang2021}) and the low-mass end (e.g. \citealt{Lange2015}) is still subject of intense research. 

\begin{figure}
   \centering
   \includegraphics[width=\hsize]{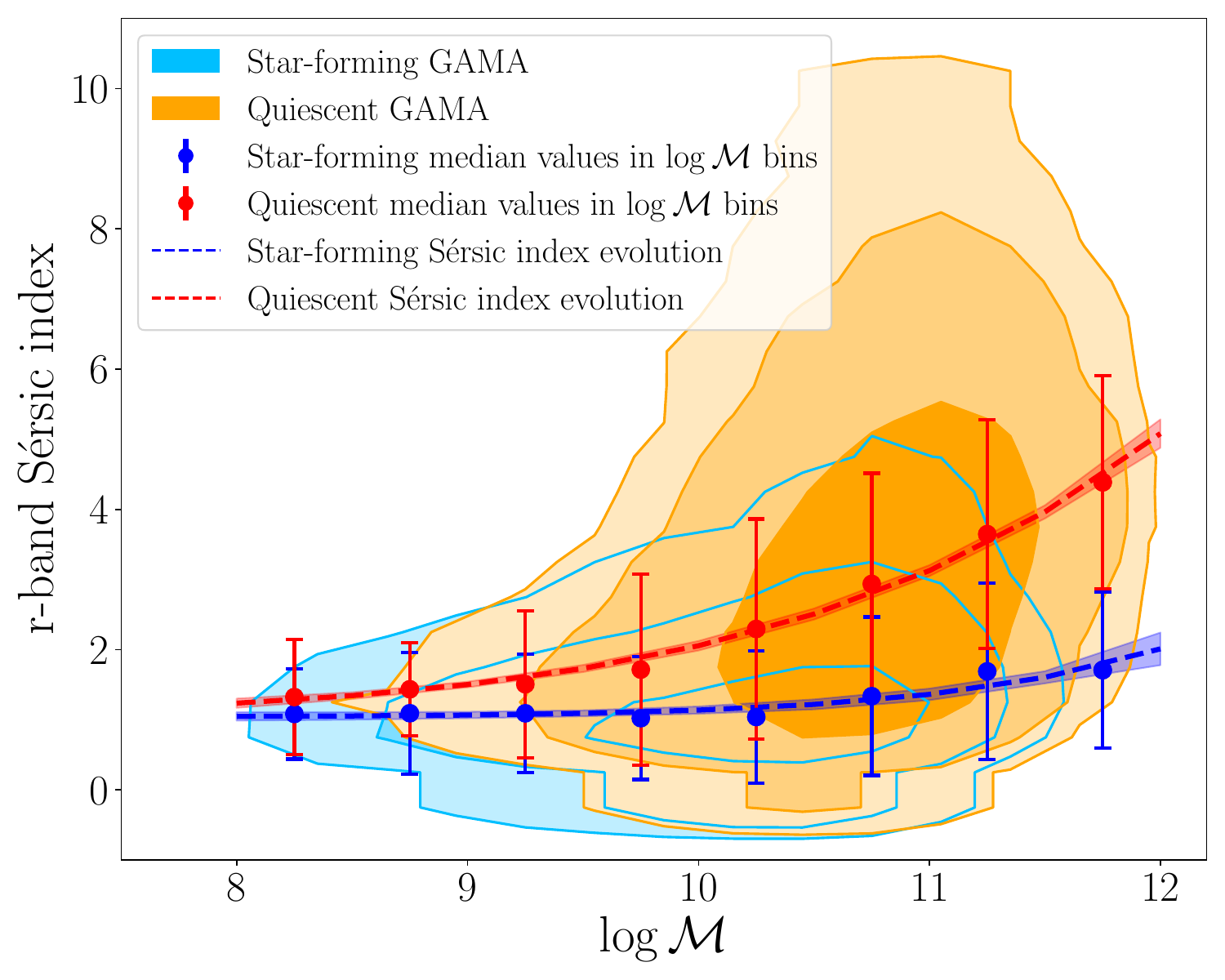}
      \caption{Stellar mass dependence of the $r$-band S\'ersic index $n$. The deep sky blue and orange contours are obtained by matching the star-forming and quiescent galaxies selected via $sSFR$ of the GAMA mass complete sample with the morphological catalogue in \cite{Kelvin2012}. The blue and red error bars represent the $r$-band S\'ersic index median trend in stellar mass bins for star-forming and quiescent galaxies, respectively, with the size of the error bars representing the standard deviation in said bin. The dashed lines represent the evolution with stellar mass obtained from the best-fitting parameters of Equation \ref{eq:sersic_index_stellar_mass} on the observed data. The bands represent the $1\sigma$ uncertainty on the best-fitting relations. Each sample contour encloses $50\%, 84\%, 99\%$ of the values.
              }
    \label{fig:sersic_index_stellar_mass}
\end{figure}

In \cite{Fischbacher2025a}, the authors model galaxy sizes as a log-normal distribution whose mean and scatter were related to the galaxy absolute magnitude \citep{Shen2003}. In our work, galaxy sizes, expressed in terms of the half-light radii in $\mathrm{kpc}$, are also sampled from a log-normal distribution, but the mean of this distribution is given by size-stellar mass relation, while the standard deviation is taken from \cite{Shen2003}. The size-mass relation has a different parametrisation for the population of blue and red galaxies. We adopt the parametrisations introduced in \cite{Nedkova2021}, where the authors measured the size-stellar mass relation for a large sample of blue and red galaxies in clusters and fields with HST photometry in the redshift range $0.2 < z < 2$, extending the sample down to stellar masses of $\sim 10^{7.5} \ M_{\odot}$. The dependence of the galaxy size with stellar mass is parametrised for blue galaxies as a single power-law,
\begin{equation}
    R_e = A \left( \frac{\mathcal{M}}{5 \times 10^{10} M_{\odot}} \right)^B \ ,
\end{equation}
where $\log{A}$ and $B$ are assumed to evolve linearly with redshift:
\begin{equation}
    \begin{split}
        \log{A} &= \log{A}_{\mathrm{intcpt}} + \log{A}_{\mathrm{slope}} \times z \ , \\
        B &= B_{\mathrm{intcpt}} + B_{\mathrm{slope}} \times z.
    \end{split}
\end{equation}
This assumption is motivated by fitting the redshift evolution of the $\log{A}$ and $B$ values from Table 3 in \cite{Nedkova2021} for the sample of blue galaxies covering the whole mass range probed (left panel of Fig. \ref{fig:size_mass_redshift_evolution}). We report the best-fitting parameters of the linear fit of $\log{A}$ and $B$ in Table \ref{table:re_funct_params}. The size-mass relation for red galaxies tends instead to flatten at low stellar masses, therefore it is parametrised as a double power-law,
\begin{equation}
    R_e = \zeta (\mathcal{M})^{\eta} \left( 1 + \frac{\mathcal{M}}{10^{\delta}} \right)^{\theta - \eta} \ ,
\end{equation}
where $\eta$ and $\theta$ are the slope at the low and high mass end, $\zeta$ is the normalisation, while $10^{\delta}$ is the stellar mass at which the second derivative of the function is at a maximum. All these parameters also evolve linearly with redshift:
\begin{equation}
    \begin{split}
        \eta &= \eta_{\mathrm{intcpt}} + \eta_{\mathrm{slope}} \times z \ , \\
        \theta &= \theta_{\mathrm{intcpt}} + \theta_{\mathrm{slope}} \times z \ , \\
        \log{\zeta} &= \log{\zeta}_{\mathrm{intcpt}} + \log{\zeta}_{\mathrm{slope}} \times z \ , \\
        \delta & = \delta_{\mathrm{intcpt}} + \delta_{\mathrm{slope}} \times z \ .
    \end{split}
\end{equation}
The linear evolution fit has been obtained from the values in Table 2 of \cite{Nedkova2021} for the whole mass range of the red sample (right panel of Fig. \ref{fig:size_mass_redshift_evolution}), while the obtained best-fitting parameters are reported in Table \ref{table:re_funct_params}. In this work, we assume the size in $\mathrm{kpc}$ to be wavelength independent, although observations do show an half-light radii dependence on wavelength (e.g. \citealt{Tortorelli2023}) due to the different stellar populations probed. The assumption of a single S\'ersic profile with no radial dependence of the stellar population justify the choice of a wavelength independent size for this work.

\begin{table}
\caption{ S\'ersic index-stellar mass relation parameter values for the red and blue galaxy populations.}
\centering
\begin{tabular}{lcc}
\hline\hline
Parameter  & Red & Blue\\
\hline
$n_0$ & 1.054 & 1.042 \\
$n_1$ & 1.453  & 0.176 \\
$n_2$ & 7.634 & 12.750 \\
$\sigma_n$ & 1.439 & 0.909 \\
\hline
\end{tabular}
\tablefoot{Best-fitting parameters obtained by modelling the S\'ersic index dependence on stellar mass with the GAMA data of \cite{Kelvin2012,Bellstedt2020}.}
\label{table:sersic_index_stellar_mass}
\end{table}

In \cite{Fischbacher2025a}, galaxy S\'ersic indices are drawn from a redshift-independent beta prime distribution, since extending the parametrisation of the S\'ersic index by allowing its evolution with redshift did not improve the distance metrics between the simulations and the observations. This is consistent with findings in the literature \citep{Martorano2025} that show little to no evolution of the median S\'ersic index in the redshift range $0.5 < z < 2.5$ . \cite{Martorano2025} also found that high-mass galaxies have higher S\'ersic indices than lower-mass galaxies at all redshift and that star-forming galaxies have lower S\'ersic indices than quiescent galaxies. Therefore, in our work, we introduce a stellar mass dependence of the S\'ersic index for both blue and red galaxies. To model this dependence, we match the GAMA mass complete sample with the single-component S\'ersic fit catalogue presented in \cite{Kelvin2012} and downloaded from the GAMA II database, separating blue and red galaxies via the $sSFR$. We use the $r$-band S\'ersic index $n$ to model its dependence on stellar mass and we remove from the catalogue bad fits utilising the primary $\chi^2$ of both the galaxy and the PSF, as suggested by the authors. We model the relation as
\begin{equation}
n = n_0 + n_1 \times \left( \frac{\log{\mathcal{M}}}{\log{(10^{10.5} M_{\odot})}} \right)^{n_2} \,
    \label{eq:sersic_index_stellar_mass}
\end{equation}
where $10^{10.5} M_{\odot}$ is the median stellar mass of the sample.  Figure \ref{fig:sersic_index_stellar_mass} shows the best-fitting evolution of $n$ as function of stellar mass. As in \cite{Martorano2025}, we also find that high-mass galaxies tend to have higher $n$ than lower mass ones and that star-forming galaxies have lower $n$ than quiescent galaxies at all masses. We sample $n$ from a truncated Normal distribution, with mean given by Equation \ref{eq:sersic_index_stellar_mass} and standard deviation $\sigma_n$ given by the median scatter across the stellar mass range. The sampled values are within the range $0.2 < n < 10$. The best-fitting parameters of Equation \ref{eq:sersic_index_stellar_mass} and the scatter across stellar mass for blue and red galaxies are reported in Table \ref{table:sersic_index_stellar_mass}.

Galaxy positions are randomly assigned within the \textsc{UFig} simulated image. Future works are aiming at introducing a realistic clustering prescription, as developed in \cite{Berner2022,Berner2024,Fischbacher2025}. This is an important point to keep in mind for the use of forward-modelling for cosmological analysis. Indeed, \cite{Fischbacher2025a} found that by not including clustering, and therefore not being affected by cosmic variance, the phenomenological model returns tomographic redshift distributions that are much smoother than those obtained from COSMOS2020 photo-zs \citep{Weaver2022}. The authors attributed this difference to the fact that the reported uncertainty from \textsc{GalSBI} is underestimating the real one since it is only including the uncertainty of the galaxy population model. 

Finally, we sample both blue and red galaxies absolute ellipticities from the same modified Beta distribution as in \cite{Moser2024,Fischbacher2025a}. The $\alpha_{\mathrm{ell}}$ and $\beta_{\mathrm{ell}}$ parameters of the standard Beta distribution are related to the mode $e_{\mathrm{mode}}$ and spread $e_{\mathrm{spread}}$ parameters of the modified Beta via $e_{\mathrm{mode}} = (\alpha_{\mathrm{ell}} - 1) / (e_{\mathrm{spread}} - 2)$ and $e_{\mathrm{spread}} = \alpha_{\mathrm{ell}} + \beta_{\mathrm{ell}}$. We set $e_{\mathrm{mode}}=0.2$ and $e_{\mathrm{spread}}=2.9$. The two ellipticity components $e_1$ and $e_2$ are computed by assigning a random phase to the absolute ellipticity, i.e. no shear or intrinsic alignment prescription applied. 

\section{HSC image simulation}
\label{sect:hsc_image_sims}

HSC is an excellent precursor dataset for Stage IV surveys and it has been used in \cite{Moser2024,Fischbacher2025a} to both constrain and validate the phenomenological version of \textsc{GalSBI}. These two motivations lead us to  consistently use this dataset to validate the performance of \textsc{GalSBI-SPS} against its phenomenological version and the observed data.

The catalogue of intrinsic galaxy properties sampled from the galaxy population model described in Sect. \ref{sect:gal_pop_model} is used to create simulated HSC DUD images with the image simulator \textsc{UFig} \citep{Berge2013,Fischbacher2024ufig}. \textsc{UFig} needs as input the catalogue of galaxy properties and the instrumental characteristics of the HSC images, namely image sizes, pixel scales, zero-points, and central pixel coordinates. We take the instrumental properties from the metadata provided by the HSC database\footnote{https://hsc-release.mtk.nao.ac.jp/datasearch/}. We simulate the HSC images in the $g,r,i,z,y$ broad-bands. The $r$ and $i$ filters have been replaced by more uniform $r2$ and $i2$ filters, whose throughputs are the ones we use in our simulations. We do not simulate single exposures, but rather the image co-added versions through the use of exposure time maps, derived from the metadata, and 2D background maps, derived from each individual observed HSC image. The background noise is added to the simulated images in the form of a Gaussian noise plus a Lanczos resampling \citep{Duchon1979} of order 3 to create correlated noise, mimicking the contribution from sky brightness, readout and errors during data processing. The Gaussian noise mean is read off the image header, while the standard deviation is estimated by generating a 2D background image from the observations and then applying a $3\sigma$ sigma-clipping with \textsc{photutils} \citep{bradley2024}, ensuring a different standard deviation per pixel. The 2D background image is estimated from background-subtracted observed images, so no background subtraction is performed on simulations. The magnitude zeropoint is set to $27 \ \mathrm{mag/ADU}$ for the HSC co-adds, while the CCD gain of a single exposure multiplied by the number of exposures per pixel gives us a rough estimate of the effective gain to convert between ADUs and number of photons. Galaxies are randomly distributed on the image and rendered via photon shooting according to the galaxy S\'ersic profile, naturally including Poisson noise \citep{Berge2013,Bruderer2016}. Galaxy profiles are then convolved with a PSF whose model is generated with a Convolutional Neural Network, presented in \cite{Herbel2018} and updated in \cite{Kacprzak2020}. The model PSF is evaluated at the position of bright unsaturated stars ($18 \le i \le 22$) and then interpolated across the co-add using  Chebyshev polynomials as detailed in Appendix C of \cite{Kacprzak2020}. The observed star positions are taken from GAIA DR3 \citep{Gaia2016,Vallenari2023}. Simulated bright stars ($i<18$) are not randomly drawn, but placed at the positions where bright stars are in real HSC DUD images. This allows us to apply the same bright object masks of HSC to our simulated images. Fainter stars are instead drawn from the Besan\c{c}on model of the Milky Way \citep{Robin2003}. Lastly, we convert photons to ADU by dividing out the effective gain and saturate pixels that are above the maximum value of the observed data. More details on the HSC DUD image simulation procedure are provided in section 3.2 of \citealt{Moser2024}, while Figs. 2 and 3 of the same work shows the good agreement between the simulated and observed image background pixel distribution and co-added image.

\begin{figure}
   \centering
   \includegraphics[width=\hsize]{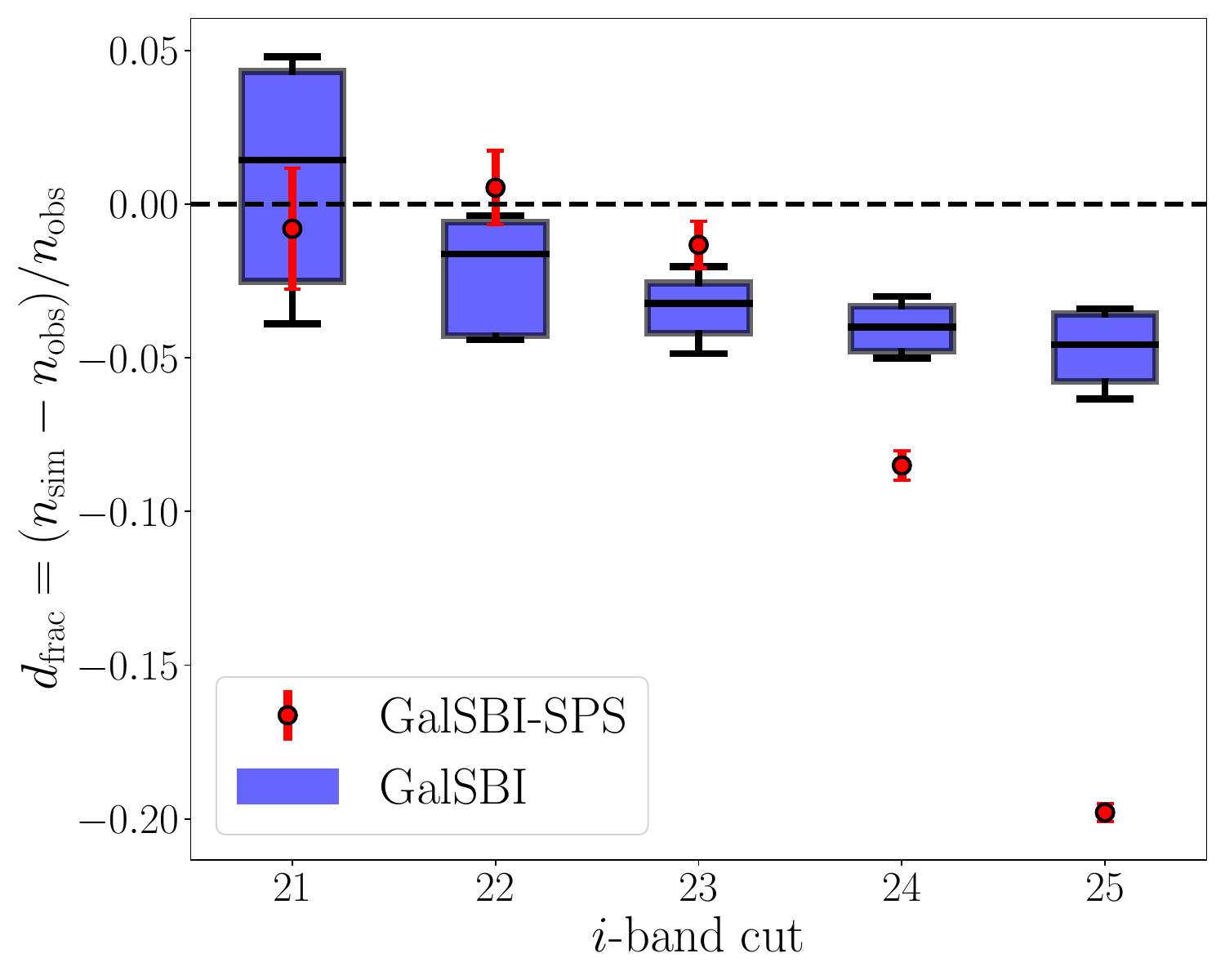}
      \caption{Fractional difference between the number of detected galaxies in the simulations $n_{\mathrm{sim}}$ and in the observations $n_{\mathrm{obs}}$ as a function of the $i$-band magnitude cut. Red scatter points refer to the \textsc{GalSBI-SPS} model, while blue box plots to the phenomenological version in \cite{Fischbacher2025a}. The \textsc{GalSBI-SPS} error bars consider Poisson noise on the counts, while \textsc{GalSBI} box plots have been constructed using the posterior samples in \cite{Fischbacher2025a}. The black dashed line marks the perfect agreement in the number counts with observations.
              }
    \label{fig:dfrac_ibandthresh}
\end{figure}

\begin{table*}[h]
\caption{Median, $84$th-$50$th, and $50$th-$16$th percentile values for the galaxy magnitudes in the $g, r, i, z, y$ bands and half-light radii $r_{50}$ in the $i$-band.}
    \centering
    \small
    \renewcommand{\arraystretch}{1.2}
    \setlength{\tabcolsep}{2pt}
    \begin{adjustbox}{max width=\textwidth}
    \begin{tabular}{c|cccccc|cccccc|cccccc}
        & \multicolumn{6}{c|}{HSC data} & \multicolumn{6}{c|}{\textsc{GalSBI-SPS}} & \multicolumn{6}{c}{\textsc{GalSBI}} \\
        \hline
        $i_{\mathrm{cut}}$ & $g$ & $r$ & $i$ & $z$ & $y$ & $r_{50}$ & $g$ & $r$ & $i$ & $z$ & $y$ & $r_{50}$ & $g$ & $r$ & $i$ & $z$ & $y$ & $r_{50}$\\
        \hline
        21 & $21.73^{+0.91}_{-1.10}$ & $20.79^{+0.63}_{-1.07}$ & $20.29^{+0.52}_{-1.05}$ & $19.99^{+0.53}_{-1.02}$ & $19.81^{+0.55}_{-1.00}$ & $0.90^{+0.35}_{-0.21}$ & 
        $21.59^{+0.80}_{-1.20}$ & $20.76^{+0.58}_{-1.12}$ & $20.28^{+0.53}_{-1.07}$ & $20.04^{+0.53}_{-1.05}$ & $19.93^{+0.53}_{-1.04}$ & $1.14^{+0.50}_{-0.32}$ & 
        $21.63^{+0.99}_{-1.14}$ & $20.81^{+0.61}_{-1.16}$ & $20.26^{+0.55}_{-1.14}$ & $19.97^{+0.56}_{-1.13}$ & $19.81^{+0.57}_{-1.13}$ & $0.87^{+0.39}_{-0.23}$ \\
        22 & $22.64^{+0.88}_{-1.16}$ & $21.80^{+0.68}_{-1.23}$ & $21.23^{+0.57}_{-1.15}$ & $20.90^{+0.59}_{-1.11}$ & $20.72^{+0.64}_{-1.11}$ & $0.80^{+0.28}_{-0.17}$ & 
        $22.53^{+0.90}_{-1.27}$ & $21.71^{+0.67}_{-1.23}$ & $21.20^{+0.59}_{-1.19}$ & $20.94^{+0.59}_{-1.16}$ & $20.81^{+0.60}_{-1.15}$ & $0.99^{+0.40}_{-0.24}$ & 
        $22.50^{+0.88}_{-1.14}$ & $21.78^{+0.64}_{-1.25}$ & $21.21^{+0.59}_{-1.22}$ & $20.91^{+0.60}_{-1.21}$ & $20.75^{+0.62}_{-1.20}$ & $0.78^{+0.29}_{-0.17}$ \\
        23 & $23.44^{+0.77}_{-1.12}$ & $22.76^{+0.64}_{-1.31}$ & $22.16^{+0.62}_{-1.26}$ & $21.82^{+0.65}_{-1.23}$ & $21.63^{+0.71}_{-1.23}$ & $0.72^{+0.24}_{-0.15}$ & 
        $23.46^{+0.94}_{-1.39}$ & $22.65^{+0.74}_{-1.37}$ & $22.08^{+0.67}_{-1.31}$ & $21.81^{+0.68}_{-1.27}$ & $21.68^{+0.69}_{-1.26}$ & $0.88^{+0.33}_{-0.19}$ & 
        $23.31^{+0.82}_{-1.14}$ & $22.72^{+0.64}_{-1.30}$ & $22.15^{+0.63}_{-1.30}$ & $21.83^{+0.65}_{-1.27}$ & $21.67^{+0.67}_{-1.27}$ & $0.71^{+0.23}_{0.13}$ \\
    \end{tabular}
    \end{adjustbox}
    \vspace{0.1cm}
    \tablefoot{Median magnitudes in five bands ($g, r, i, z, y$) and half-light radii $r_{50}$ (size estimate in $\mathrm{arcsec}$) in the $i$-band for different $i$-band magnitude cuts across observations (HSC data) and simulated galaxies (\textsc{GalSBI-SPS} and \textsc{GalSBI}). The quoted upper and lower errors refer to the $84$th-$50$th and $50$th-$16$th percentile values of the distributions.}
    \label{tab:median_magnitudes}
\end{table*}

One crucial aspect of forward-modelling is the ability to apply the same measurement steps and selection cuts on both data and simulations. Therefore, we do not use the photometric catalogue from COSMOS2020, but we measure galaxy photometric properties with \textsc{Source Extractor} using the same settings for observed and simulated images. The use of the same settings ensures that all uncertainties in the extraction process are consistently modelled between data and simulations, resulting in the constraints on the galaxy population model being independent of the particular choice of the extraction software. We run \textsc{Source Extractor} in dual-image mode, with the $i$-band image as detection image, using the hyper-parameters settings reported in Appendix C of \cite{Moser2024}. While galaxies detected on the real images are matched to the COSMOS2020 catalogue by positions and magnitudes, galaxies detected on the simulated images are matched to the true injected galaxies by means of the \textsc{Source Extractor} segmentation map. For each detection, we find the overlapping simulated object that minimises the sum of the differences between the measured magnitude ($\mathrm{MAG\_AUTO\_b}$) and the true magnitude ($\mathrm{mag_b}$) in all bands,
\begin{equation}
 \Delta \mathrm{mag} = \sum_{b \in g,r,i,z,y}  | \mathrm{mag_b} - \mathrm{MAG\_AUTO\_b}  |\ .
\end{equation}
This procedure improves the matching especially in the case of a crowded field. The matching procedure allows us to obtain catalogues of detected objects with measured quantities, such as magnitudes and sizes, for both data and simulations. These properties are then accompanied by high quality photo-z information, in the case of data, and true galaxy physical properties, including redshifts, in the case of simulations.

\section{Results}
\label{sect:results}

In this section, we compare the photometric properties (magnitudes, colours and sizes), the redshift distributions and the scaling relations between physical properties of the galaxies that have been detected on the simulated and the observed images. The photometric measurements have been performed on the $56$ HSC DUD patches fully overlapping with the COSMOS2020 photo-z catalogue, with the exact same \textsc{Source Extractor} setup between observations and simulations. The simulated HSC images created in this work have the same instrumental characteristics as those in \cite{Fischbacher2025a}. Therefore, we compare the detected galaxies drawn from \textsc{GalSBI-SPS} not only against observations, but also against those drawn from the phenomenological version of \textsc{GalSBI}. It is important to point out that in this section we are comparing the observed galaxy properties against those sampled from a model, \textsc{GalSBI-SPS}, whose parameter have not been explicitly constrained using SBI, as it is the case, instead, for the \textsc{GalSBI} phenomenological model. 

\subsection{Photometric properties comparison}
\label{sect:photometric_comparison}

The \textsc{Source Extractor} configuration file we use in our work is described in appendix C of \cite{Moser2024}. The same setup was also used in \cite{Fischbacher2025a}. We apply selection cuts on the catalogues of detected galaxies to avoid spurious detections and contaminations, separate galaxies from stars, and select objects with reliable photometry in every waveband. To ensure a meaningful comparison with \textsc{GalSBI}, the cuts we apply are the same ones that have been used in \cite{Fischbacher2025a} to compare the photometric properties and the redshift distributions between observations and simulations. The cuts are applied on the \textsc{Source Extractor} measurements for every waveband with strict $\mathrm{AND}$ conditions:
\begin{equation}
    \begin{split}
    0 \le \ &\mathrm{FLAGS} \le 3 \, \\
    14 < \ &\mathrm{MAG\_APER3} \ < 30 \ ,\\
    16 < \ &\mathrm{MAG\_APER3\_i} \ < 25 \ ,\\
      0  < \ &\mathrm{MAG\_AUTO} \ < 99 \ ,\\
      0  < \ &\mathrm{z} \ < 6 \ ,\\
         0.1 < \ &\mathrm{FLUX\_RADIUS} \ < 30 \, \\
         -3  < \ &\log{(\mathrm{SNR})} \equiv \log{(\frac{\mathrm{FLUX\_AUTO}}{\mathrm{FLUXERR\_AUTO}})} < 4 \ ,\\
         0 < \ &\mathrm{ELL} < 1 \ , \\
         0 < \ &\mathrm{N\_EXPOSURES} \, \\
         0.5 < \ &r_{50}/\mathrm{PSF\_FWHM} \ , \\
         0 \le \ &\mathrm{CLASS\_STAR\_i} < 0.95 \ .
    \end{split}
\end{equation}
$\mathrm{FLAGS}$ is an indicator for the photometric extraction quality, $\mathrm{MAG\_APER3}$ is the magnitude in a $3\arcsec$ aperture, $\mathrm{MAG\_AUTO}$ is the total magnitude (which is closest to the \textsc{UFig} true magnitude), $z$ is the redshift, $\mathrm{FLUX\_RADIUS}$ is the radius containing $50 \%$ of the total light, $\mathrm{SNR}$ is the signal-to-noise ratio computed using the total flux $\mathrm{FLUX\_AUTO}$ and its error $\mathrm{FLUXERR\_AUTO}$, $\mathrm{ELL}$ is the absolute ellipticity computed using the \textsc{Source Extractor} windowed moments, $\mathrm{N\_EXPOSURES}$ is the number of exposures in the co-added images at the position of the object, $r_{50}$ is the object size computed using the \textsc{Source Extractor} windowed moments as in \cite{Kacprzak2020}, $\mathrm{PSF\_FWHM}$ is the point-spread-function full width at half-maximum, and $\mathrm{CLASS\_STAR}$ is the \textsc{Source Extractor} stellarity index. Since in the image simulations we assume that bright saturated stars are in the same position as those from the observations, we apply the HSC bright object mask to both observed and simulated images. The mask allows us to avoid sources whose photometry is impacted by the higher background value caused by nearby bright objects. We visually inspect galaxies close to bright objects in observed images, finding that the HSC bright object mask efficiently removes galaxies contaminated by starlight. The bright object induces in its proximity 
a higher background value that leads the extraction software to artificially increase the measured size of the detected galaxy. 

\begin{figure*}
   \centering
   \resizebox{\hsize}{!}{\includegraphics[width=\hsize]{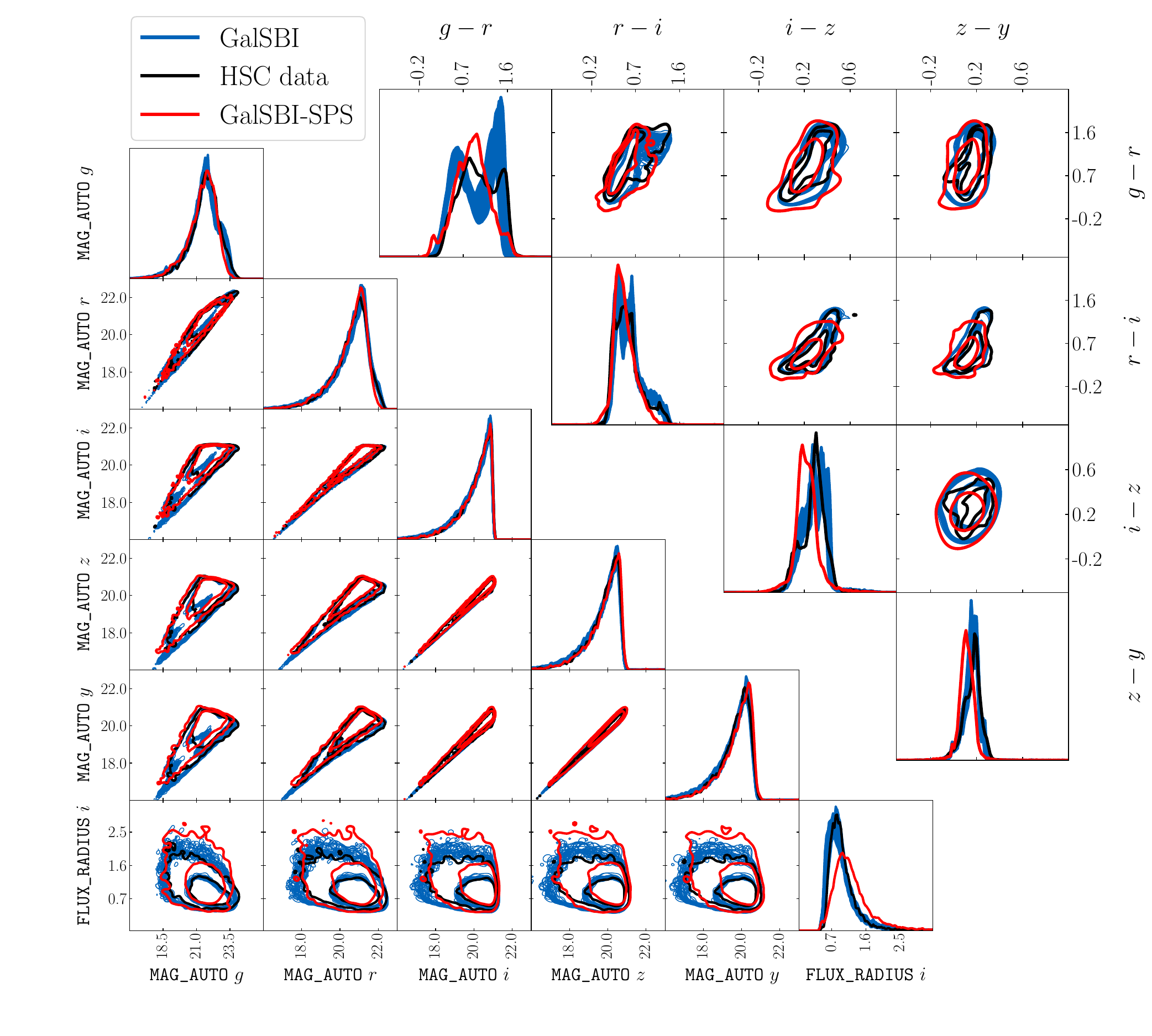}}
      \caption{Comparison of magnitude, size, and colour distributions between the observed HSC galaxies (black contours) and those from \textsc{GalSBI} (blue contours) and \textsc{GalSBI-SPS} (red contours) models for the $i \le 21$ cut. The lower left contours refer to the magnitude and size distributions, while the upper right to the colour distributions. $\mathrm{MAG\_AUTO}$ refers to the galaxy magnitudes as measured by \textsc{Source Extractor}, $\mathrm{FLUX\_RADIUS}$ to the galaxy effective radii, while the colours are obtained as difference between the \textsc{Source Extractor} magnitudes. We report in the figure the \textsc{GalSBI} contours obtained from each posterior distribution sample in \cite{Fischbacher2025a}.}
    \label{fig:magnitude_size_colour_contours_21}
\end{figure*}

The first metric we compare is the number of detected galaxies, expressed as fractional difference between the number of detected galaxies in the simulations $n_{\mathrm{sim}}$ and in the observations $n_{\mathrm{obs}}$:
\begin{equation}
    d_{\mathrm{frac}} = \frac{n_{\mathrm{sim}}-n_{\mathrm{obs}}}{n_{\mathrm{obs}}} \ .
\end{equation}
Since we are using \textsc{Source Extractor} in forced photometry mode and we are considering only those objects that have photometric measurements in all bands (we do not consider possible UV dropouts), the number of detected galaxies is band-independent. Figure \ref{fig:dfrac_ibandthresh} shows the fractional difference of detected galaxies as function of the $i$-band magnitude cut ($i \le j \ , \ j=\{ 21,22,23,24,25 \}$) for both versions of \textsc{GalSBI}. The error bars for \textsc{GalSBI-SPS} consider Poisson noise on the counts, while the boxplots for \textsc{GalSBI} represent the distribution of counts estimated for each posterior sample. Since we determine the model parameters for \textsc{GalSBI-SPS} using mostly bright galaxy samples (see e.g.  Fig. \ref{fig:GAMA_DEVILS_iband_hist} in Appendix \ref{appendix:devils_completeness} for the relations involving GAMA and DEVILS data), it is not surprising to see that the fractional difference of detected galaxies is at the percent level at the bright-end cut, progressively increasing at the faint-end cut ($i \le 24 ,  \ i \le 25$) where the model is mostly uninformed. By being constrained against HSC DUD deep data using SBI, the fractional difference of detected galaxies for \textsc{GalSBI} is always at the percent level. 

\begin{figure*}
   \centering
   \resizebox{\hsize}{!}{\includegraphics[width=\hsize]{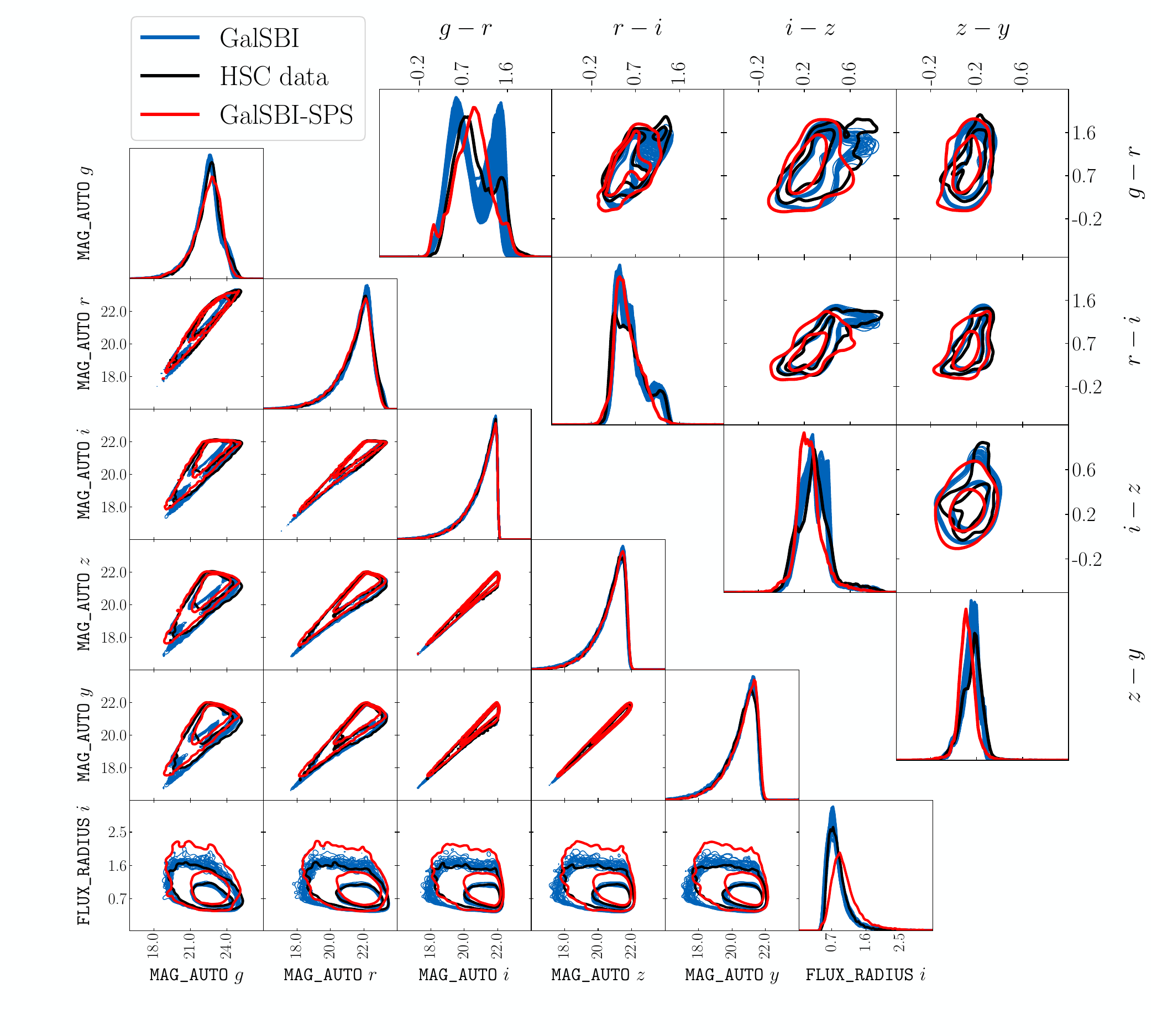}}
      \caption{Comparison of magnitude, size, and colour distributions between the observed HSC galaxies (black contours) and those from \textsc{GalSBI} (blue contours) and \textsc{GalSBI-SPS} (red contours) models for the $i \le 22$ cut. The lower left contours refer to the magnitude and size distributions, while the upper right to the colour distributions. $\mathrm{MAG\_AUTO}$ refers to the galaxy magnitudes as measured by \textsc{Source Extractor}, $\mathrm{FLUX\_RADIUS}$ to the galaxy effective radii, while the colours are obtained as difference between the \textsc{Source Extractor} magnitudes. We report in the figure the \textsc{GalSBI} contours obtained from each posterior distribution sample in \cite{Fischbacher2025a}.}
    \label{fig:magnitude_size_colour_contours_22}
\end{figure*}

\begin{figure*}
   \centering
   \resizebox{\hsize}{!}{\includegraphics[width=\hsize]{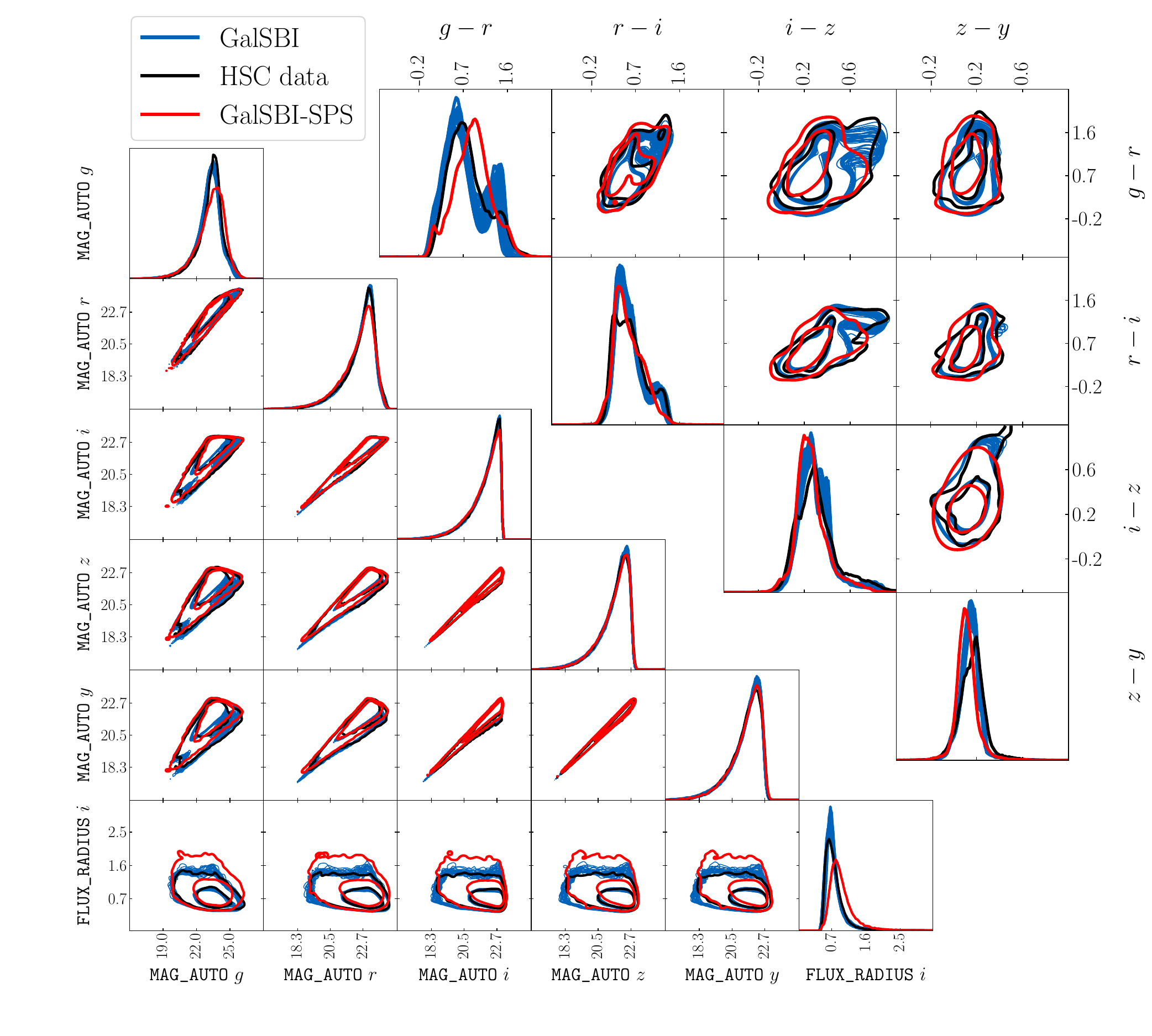}}
      \caption{Comparison of magnitude, size, and colour distributions between the observed HSC galaxies (black contours) and those from \textsc{GalSBI} (blue contours) and \textsc{GalSBI-SPS} (red contours) models for the $i \le 23$ cut. The lower left contours refer to the magnitude and size distributions, while the upper right to the colour distributions. $\mathrm{MAG\_AUTO}$ refers to the galaxy magnitudes as measured by \textsc{Source Extractor}, $\mathrm{FLUX\_RADIUS}$ to the galaxy effective radii, while the colours are obtained as difference between the \textsc{Source Extractor} magnitudes. We report in the figure the \textsc{GalSBI} contours obtained from each posterior distribution sample in \cite{Fischbacher2025a}.}
    \label{fig:magnitude_size_colour_contours_23}
\end{figure*}

The second metric we compare is the distribution of magnitudes in the five HSC filters $g,r,i,z,y$ and sizes measured in the $i$-band. We focus on the $i$-band sizes because our current single S\'ersic modelling does not account for radial gradients in the galaxy size.  Galaxy magnitudes are measured on the HSC images using the \textsc{Source Extractor} $\mathrm{MAG\_AUTO}$ parameter, while galaxy sizes, defined as half-light radii in $\mathrm{arcsec}$, are derived from the \textsc{Source Extractor} $\mathrm{FLUX\_RADIUS}$ parameter. Figures \ref{fig:magnitude_size_colour_contours_21}, \ref{fig:magnitude_size_colour_contours_22}, and \ref{fig:magnitude_size_colour_contours_23} present comparisons of magnitude, size, and colour distributions among  \textsc{GalSBI-SPS} (red contours), \textsc{GalSBI} (blue contours) and observed measurements (black contours) for various $i$-band magnitude cuts. We focus on the $i\le 21,22,23$ samples, as this is the magnitude range spanned by the data we use to create our forward model. For completeness, comparisons for the fainter $i\le 24,25$ samples are included in Appendix \ref{appendix:faint_sample}.

Figure \ref{fig:magnitude_size_colour_contours_21} shows that, for the brightest galaxy sample ($i \le 21$), the $g,r,i,z,y$ magnitude distributions exhibit very good agreement among the three samples. This is quantified in Table \ref{tab:median_magnitudes}, which reports the median, $50$th-$16$th, and $84$th-$50$th percentile values for each distributions. At this magnitude cut, the \textsc{GalSBI-SPS} median magnitudes differ from the observed data by at most $\sim 0.1 \ \mathrm{mag}$ in the $g$ and $y$-bands, and remain below this threshold in all other wavebands. \textsc{GalSBI} shows sub-$0.1 \ \mathrm{mag}$ median magnitude agreement across all bands, except for the $g$-band, although this difference is not visually noticeable in the magnitude histogram. In the magnitude-size planes (bottom-left row of Fig. \ref{fig:magnitude_size_colour_contours_21}), both \textsc{GalSBI-SPS} and \textsc{GalSBI} captures the bulk of the observed magnitude-size distribution. However, \textsc{GalSBI-SPS} tends to generate a population of galaxies with larger median apparent size, whereas \textsc{GalSBI} achieves a notably better match, likely due to being directly constrained by HSC data through SBI. This improved agreement is reflected in the median size differences, which is of only $0.03\arcsec$ for \textsc{GalSBI}, compared to $0.24\arcsec$ for \textsc{GalSBI-SPS}. 

The $i \le 22$ galaxy sample (Fig. \ref{fig:magnitude_size_colour_contours_22}) displays similar trends. Magnitude distributions remain consistent between observations and simulations, with all median differences between \textsc{GalSBI-SPS} and the data remaining below $0.1 \ \mathrm{mag}$. The median size discrepancy for \textsc{GalSBI-SPS} is slightly reduced to $0.19\arcsec$, but still significant.  In contrast, \textsc{GalSBI} shows a $\sim 0.03$ mag agreement with the observations in the median $r,i,z,y$ bands, while the $g$-band median magnitude differs from the observed value by an amount ($0.14$ mag) that is larger than the \textsc{GalSBI-SPS} case. Also at this magnitude cut, \textsc{GalSBI} size distribution perfectly agrees with the one from the data.

For the fainter $i \leq 23$ sample (Fig.  \ref{fig:magnitude_size_colour_contours_23}), the $g$, $i$, $z$, and $y$-band median magnitudes from \textsc{GalSBI-SPS} continue to show sub-$0.1$ mag agreement with the data. The $r$-band, instead, increased its median difference, with respect to the previous $i$-band magnitude cuts, to roughly $0.11$ mag. As for \textsc{GalSBI}, the median magnitudes in the $r$, $i$, $z$, and $y$ bands differ less than $0.04$ mag, while the $g$-band median magnitude differs by $0.13$ mag with respect to that from the data. It is also worth noticing that while the median magnitudes in the $g$-band show a better agreement with data for \textsc{GalSBI-SPS}, the $g$-band histograms in the top-left box of Fig.  \ref{fig:magnitude_size_colour_contours_23} show a distribution for \textsc{GalSBI-SPS} galaxies that is skewed towards fainter values with respect to that of \textsc{GalSBI} and the data.The inclusion of fainter galaxies, which dominate the number counts and tend to have smaller apparent sizes, helps reduce the median size mismatch between \textsc{GalSBI-SPS} and the data down to $0.16\arcsec$. \textsc{GalSBI} shows instead a very good agreement with data since the model has been constrained against observations by minimising distance metrics on the $ \le 23$ galaxy sample.

\begin{table*}[h]
\caption{Median, $84$th-$50$th, and $50$th-$16$th percentile values for the galaxy colours in the $g, r, i, z, y$ bands.}
    \centering
    \renewcommand{\arraystretch}{1.2} 
    \setlength{\tabcolsep}{2pt}  
    \begin{adjustbox}{max width=\textwidth}
    \begin{tabular}{c|cccc|cccc|cccc}
        & \multicolumn{4}{c|}{HSC data} & \multicolumn{4}{c|}{\textsc{GalSBI-SPS}} & \multicolumn{4}{c}{\textsc{GalSBI}} \\
        \hline
        $i_{\mathrm{cut}}$ & $g-r$ & $r-i$ & $i-z$ & $z-y$ & $g-r$ & $r-i$ & $i-z$ & $z-y$ & $g-r$ & $r-i$ & $i-z$ & $z-y$ \\
        21 & $1.00^{+0.47}_{-0.39}$ & $0.50^{+0.30}_{-0.22}$ & $0.29^{+0.08}_{-0.13}$ & $0.18^{+0.05}_{-0.08}$ & $0.89^{+0.34}_{-0.35}$ & $0.46^{+0.25}_{-0.16}$ & $0.22^{+0.08}_{-0.07}$ & $0.11^{+0.06}_{-0.06}$ & $1.00^{+0.42}_{-0.47}$ & $0.56^{+0.30}_{-0.22}$ & $0.29^{+0.09}_{-0.11}$ & $0.16^{+0.05}_{-0.07}$ \\
        22 & $0.88^{+0.54}_{-0.33}$ & $0.55^{+0.41}_{-0.27}$ & $0.29^{+0.12}_{-0.14}$ & $0.18^{+0.07}_{-0.10}$ & $0.90^{+0.35}_{-0.37}$ & $0.49^{+0.32}_{-0.19}$ & $0.23^{+0.11}_{-0.08}$ & $0.11^{+0.07}_{-0.06}$ & $0.80^{+0.57}_{-0.34}$ & $0.55^{+0.39}_{-0.21}$ & $0.28^{+0.11}_{-0.11}$ & $0.16^{+0.06}_{-0.08}$  \\
        23 & $0.75^{+0.51}_{-0.32}$ & $0.55^{+0.40}_{-0.28}$ & $0.29^{+0.18}_{-0.15}$ & $0.17^{+0.09}_{-0.11}$ & $0.89^{+0.36}_{-0.38}$ & $0.53^{+0.36}_{-0.22}$ & $0.24^{+0.14}_{-0.10}$ & $0.12^{+0.08}_{-0.07}$ & 
        $0.67^{+0.60}_{-0.31}$ & $0.54^{+0.38}_{-0.21}$ & $0.28^{+0.13}_{-0.11}$ & $0.15^{+0.07}_{-0.08}$ \\
    \end{tabular}
    \end{adjustbox}
    \vspace{0.1cm}
    \tablefoot{Median colours in the five HSC bands ($g, r, i, z, y$) for different $i$-band magnitude cuts across observations (HSC data) and simulated galaxies (\textsc{GalSBI-SPS} and \textsc{GalSBI}). The quoted upper and lower errors refer to the $84$th-$50$th and $50$th-$16$th percentile values of the distributions.}
    \label{tab:median_colours}
\end{table*}

We complete the photometric comparison among \textsc{GalSBI-SPS}, \textsc{GalSBI} and the observations by considering the $g-r,r-i,i-z,z-y$ colour distributions. This comparison is very informative as colours are a proxy of galaxy SEDs and of the modelling choices that go into creating them. The comparisons for the $i \le 21,22,23$ galaxy samples are reported in the upper right panels of Figs. \ref{fig:magnitude_size_colour_contours_21}, \ref{fig:magnitude_size_colour_contours_22}, and \ref{fig:magnitude_size_colour_contours_23}, while the comparison for the fainter samples are reported in Appendix \ref{appendix:faint_sample}. The median,  $50$th-$16$th, and $84$th-$50$th percentile values for each colour distribution, are reported in Table \ref{tab:median_colours}. The three samples span a very similar broad-band colour space at all $i$-band magnitude cuts. For the $i \le 21$ sample (Fig. \ref{fig:magnitude_size_colour_contours_21}), the difference between the \textsc{GalSBI-SPS} and the observed median colours is $\lesssim 0.7$ mag for the $r-i,i-z,z-y$ colours, while the \textsc{GalSBI-SPS} $g-r$ median
colour differs by $0.11$ mag with respect to the observed value. As for \textsc{GalSBI}, the median values of the colours are more consistent with the observed ones than what \textsc{GalSBI-SPS} shows. While the 2D distributions show a similar colour space coverage among the three samples, the 1D colour distributions show a more marked visual agreement for \textsc{GalSBI} rather than for \textsc{GalSBI-SPS} with the data. Both \textsc{GalSBI} and the data show a more pronounced bimodal $g-r$ colour distribution than \textsc{GalSBI-SPS}. While \textsc{GalSBI-SPS} tends to produce a larger fraction of bluer galaxies with respect to the observations, thereby leading to a lower $g-r$ median colour, \textsc{GalSBI} has the opposite trend, producing an excess of redder galaxies with respect to the observations. The $r-i$ colour distribution shows a similar trend for \textsc{GalSBI-SPS}, \textsc{GalSBI} and the data, while for the $i-z$ colour, \textsc{GalSBI-SPS} tends to produce a population of bluer objects compared to \textsc{GalSBI} and the data. The $z-y$ colour distribution is similar among the three samples instead. 

For the $i \le 22$ sample (Fig. \ref{fig:magnitude_size_colour_contours_22}), we observe little evolution in the median colours for \textsc{GalSBI-SPS} with respect to the $i \le 21$ case. For \textsc{GalSBI} and the data, the $r-i,i-z,z-y$ colours also show little evolution in their median values, while the $g-r$ median colour becomes bluer than the $i \le 21$ sample. Notably the difference in the $g-r$ median colour for \textsc{GalSBI-SPS} and the data is of the order of $0.02$ mag, while for \textsc{GalSBI} this difference is $0.08$ mag. The other three colours show differences in the median values with respect to the observations of $\sim 0.06$ mag for \textsc{GalSBI-SPS} and $0.02$ mag for \textsc{GalSBI}. The colour space spanned by three samples is similar albeit with some differences with respect to the $i \le 21$ case. In the $g-r$ colour, observed data are dominated by bluer colours, as we would expect for a sample containing higher redshift objects. The bimodal distribution predicted by \textsc{GalSBI} shows instead a similar number of objects with bluer or redder $g-r$ colours, while \textsc{GalSBI-SPS} predicts a distribution that is very similar to the previous case. Similar considerations to the $i\le 21$ case apply for the $r-i$ and $z-y$ colours, while the $i-z$ colour distribution for \textsc{GalSBI} and the data shows an excess of redder objects with respect to \textsc{GalSBI-SPS}.

In the $i \le 23$ sample (Fig. \ref{fig:magnitude_size_colour_contours_23}), the observed $g - r$ colour distribution becomes increasingly bluer, whereas \textsc{GalSBI-SPS} does not show the same trend, maintaining a roughly constant median of $\sim 0.89 \ \mathrm{mag}$. This leads to a discrepancy of $0.14$ mag between \textsc{GalSBI-SPS} and the data, while for the same colour \textsc{GalSBI} median differs by $0.08$ mag. In this magnitude cut, \textsc{GalSBI} has an excess of galaxies with red $g-r$ colours with respect to the observations, while the distribution of blue $g-r$ colours is similar. The difference of the median $r-i,i-z,z-y$ colours between \textsc{GalSBI-SPS} and the data is smaller than the previous case, of the order of $0.05$ mag, while for \textsc{GalSBI} is of the order of $0.02$ mag. The colour space spanned by three samples is similar, with the $r-i,i-z,z-y$ 1D colour distributions show a better agreement among the three samples than the previous magnitude cut.

To summarise, \textsc{GalSBI-SPS} produces a fairly realistic galaxy population up to $i \le 23$. It yields a percent-level agreement with the HSC data in the number of detected galaxies up to a $i$-band cut of $i \le 23$. Beyond this threshold, the model is uninformed and the number counts differ up to $20 \%$ at $i \le 25$. \textsc{GalSBI-SPS} magnitude distributions are in agreement with those from \textsc{GalSBI} and the observed HSC data at all magnitude cuts. The size distribution, instead, is skewed towards larger apparent size galaxies for \textsc{GalSBI-SPS} compared to \textsc{GalSBI} and the data, with the tension decreasing from $0.24$ to $0.16\arcsec$ when including fainter galaxies. \textsc{GalSBI-SPS} galaxy colours span a very similar colour space as their observed counterparts. However, the fraction of bluer to redder $g-r$ colours differ both from the data and \textsc{GalSBI}. At increasingly fainter magnitude cuts, the $g-r$ colours of HSC data tend to become bluer, while \textsc{GalSBI-SPS} $g-r$ colours do not evolve appreciably. While \textsc{GalSBI} produces a more clear bimodal $g-r$ colour distribution, the ratio of bluer to redder objects is different from that of the observations. Overall, the better agreement that \textsc{GalSBI} has with the observations compared to \textsc{GalSBI-SPS} is motivated by \textsc{GalSBI} model parameters being constrained against HSC data with SBI. \textsc{GalSBI-SPS} model parameters are instead a combination of prescriptions fitted to low-redshift observations and values taken from literature studies that use very differently selected samples of objects. A SBI run with appropriate distance metrics can jointly constrain the model parameters, mitigating the residual discrepancies we see in the current setup.

\begin{figure*}
   \centering
   \resizebox{\hsize}{!}{\includegraphics[width=\hsize]{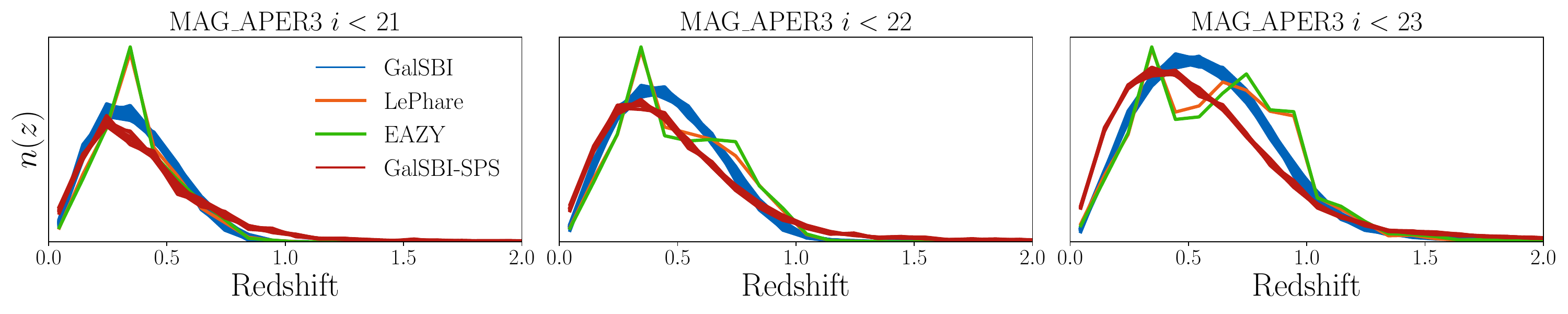}}
      \caption{Comparison of the redshift distributions for different magnitude cuts between \textsc{GalSBI-SPS} (red lines), \textsc{GalSBI} (blue lines), and the photo-z estimates with \textsc{EAZY} (green line) and \textsc{LePhare} (orange line) of the observed galaxies in the COSMOS field. We report in the figure the \textsc{GalSBI} redshift distributions obtained from each posterior sample in \cite{Fischbacher2025a} and the \textsc{GalSBI-SPS} redshift distributions obtained from each bootstrap sample.}
    \label{fig:redshift_distributions}
\end{figure*}

\begin{table*}[h]
\caption{Mean,  median, $84$th-$50$th, and $50$th-$16$th percentile values of the galaxy redshift distributions from the \textsc{GalSBI-SPS}, \textsc{GalSBI}, \textsc{EAZY}, and \textsc{LePhare} estimates.}
    \centering
    \renewcommand{\arraystretch}{1.2}
    \setlength{\tabcolsep}{2pt} 
    \begin{adjustbox}{max width=\textwidth}
    \begin{tabular}{c|cccc|cc|cc}
        & \multicolumn{4}{c|}{HSC data} & \multicolumn{2}{c|}{\textsc{GalSBI-SPS}} & \multicolumn{2}{c}{\textsc{GalSBI}} \\
        \hline
        $i_{\mathrm{cut}}$ & $\bar{z}_{\mathrm{EAZY}}$ & $Me(z)_{\mathrm{EAZY}}$ & $\bar{z}_{\mathrm{LePhare}}$ & $Me(z)_{\mathrm{LePhare}}$ & $\bar{z}$ & $Me(z)$ & $\bar{z}$ & $Me(z)$ \\
        21 & $0.375$ & $0.354^{+0.531}_{-0.219}$ & $0.368$ & $0.349^{+0.530}_{-0.210}$ & $0.452 \pm 0.006$ & $0.356^{+0.663}_{-0.174}$ & $0.359 \pm 0.006$ & $0.343^{+0.537}_{-0.184}$ \\
        
        22 & $0.493$ & $0.464^{+0.738}_{-0.255}$ & $0.484$ & $0.454^{+0.726}_{-0.250}$ & $0.542 \pm 0.003$ & $0.431^{+0.778}_{-0.208}$ & $0.460 \pm 0.005$ & $0.441^{+0.683}_{-0.238}$  \\
        
        23 & $0.632$ & $0.624^{+0.930}_{-0.316}$ & $0.618$ & $0.607^{+0.925}_{-0.307}$ & $0.644 \pm 0.002$ & $0.514^{+0.921}_{-0.242}$ & $0.596 \pm 0.006$ & $0.563^{+0.870}_{-0.301}$ \\
    \end{tabular}
    \end{adjustbox}
    \vspace{0.1cm}
    \tablefoot{Mean and median redshifts for \textsc{GalSBI-SPS}, \textsc{GalSBI}, \textsc{EAZY}, and \textsc{LePhare} at different $i$-band magnitude cuts. The mean redshift and its error for \textsc{GalSBI} is estimated from the posterior samples, while for \textsc{GalSBI-SPS} via bootstrap resampling.The quoted upper and lower errors refer to the $84$th-$50$th and $50$th-$16$th percentile values of the distributions.}
    \label{table:redshift_means}
\end{table*}

\subsection{Redshift distributions comparison}

In this section we compare the redshift distributions derived for simulations and observations. While redshifts in simulations are drawn from the input galaxy population model, and are by definition without error, apart from matching errors, the redshift estimates for observations are taken from the state-of-the-art COSMOS2020 photo-z catalogue (see Sect. \ref{sect:cosmos}). These estimates have been obtained using two different template fitting codes, \textsc{EAZY} and \textsc{LePhare}. The comparisons between the simulated and observed redshift distributions are reported in Fig. \ref{fig:redshift_distributions} for different $i$-band cuts (using \textsc{Source Extractor} $\mathrm{MAG\_APER3}$ magnitudes as in \citealt{Fischbacher2025a}), while the mean redshifts, and their uncertainties, median, $50$th-$16$th, and $84$th-$50$th percentile values are reported in Table \ref{table:redshift_means}. Given that we generate a single realisation of the COSMOS field with \textsc{GalSBI-SPS}, the errors on the mean redshift estimates are obtained using bootstrap resampling. We repeatedly sample with replacement from the original set of galaxies to create multiple bootstrap samples, we apply the magnitude selection, and then we compute the mean for each bootstrap sample and the standard deviation of these bootstrap means. We generate a total of $10^3$ bootstrap samples. 

For all the magnitude-limited samples, we note that \textsc{GalSBI-SPS} redshift distributions visually peak at a lower redshift compared to the estimates from \textsc{GalSBI}, \textsc{EAZY} and \textsc{LePhare}. For the $i \le 21$ sample (left-most panel of Fig. \ref{fig:redshift_distributions}), the median redshift of \textsc{GalSBI-SPS} differs by just $0.002$ and $0.007$ with respect to those of \textsc{EAZY} and \textsc{LePhare}, respectively. However, the presence of high-redshift bright objects at $i \le 21$ in \textsc{GalSBI-SPS} leads to a noticeable shift in the mean redshift, which is higher by $0.077$ and $0.084$ when compared to the \textsc{EAZY} and \textsc{LePhare} estimates, respectively. \textsc{GalSBI} shows instead a better agreement with the observations in the mean redshift, with a $0.016$ and $0.009$ difference for \textsc{EAZY} and \textsc{LePhare}, respectively.

As we move to the $i \le 22$ sample (central panel of Fig. \ref{fig:redshift_distributions}), the redshift distributions of \textsc{EAZY}, \textsc{LePhare}, and \textsc{GalSBI} shifts towards higher redshifts, an effect that is less pronounced for \textsc{GalSBI-SPS}. \textsc{GalSBI-SPS} median redshift that is now $0.033$ and $0.023$ smaller than the \textsc{EAZY} and \textsc{LePhare} estimates, respectively, with the mean redshift still remaining higher than the other estimates, differing by $0.049$ from \textsc{EAZY} and $0.058$ from \textsc{LePhare}. As for \textsc{GalSBI}, the mean redshifts now differs by $0.033$ from \textsc{EAZY} and $0.024$ from \textsc{LePhare}.

In the $i \le 23$ sample (right-most panel of Fig. \ref{fig:redshift_distributions}), it is visually clear how the $i \le 23$ \textsc{GalSBI-SPS} redshift distribution is skewed towards lower redshift objects than the observations. This is quantified by the difference in the median redshift, which is now of the order of $0.11$ for \textsc{EAZY} and $0.093$ for \textsc{LePhare}, with \textsc{GalSBI-SPS} median redshift being smaller than the other two estimates. As for the mean redshift, \textsc{GalSBI-SPS} estimate is larger by a value of $0.012$ and $0.026$ with respect to \textsc{EAZY} and \textsc{LePhare} estimates, respectively. This is a consistent trend across the magnitude cuts that is due to bright high redshift galaxies passing the $i$-band magnitude cut. \textsc{GalSBI} shows instead a mean redshift that is lower by $0.036$ and $0.022$ with respect to \textsc{EAZY} and \textsc{LePhare} estimates, respectively. Table 4 in \cite{Fischbacher2025a} reports the comparison of \textsc{GalSBI} mean redshifts with \textsc{EAZY} and \textsc{LePhare} estimates for fainter magnitude cuts down to $i \le 25$. In the table, the authors include also \textsc{EAZY} and \textsc{LePhare} mean redshift estimates obtained using the re-weighted mocks from \cite{Moser2024}. Using those re-weighted values leads to \textsc{GalSBI} mean redshift being consistent within the errors with \textsc{EAZY} and \textsc{LePhare} estimates.

One notable feature is the smoothness of the simulated redshift distributions compared to real ones. As discussed in \cite{Moser2024,Fischbacher2025}, this is due to the absence of clustering in our simulations, as well as the fact that we do not run the template fitting codes on simulated galaxies, hence leading to smoother than the COSMOS2020 distributions. This also implies that the simulated uncertainties on the mean redshift underestimate the real ones since they only include the uncertainty of the galaxy population model. \cite{Moser2024} analysed the effect of sample variance on the COSMOS field, finding a mean redshift offset between the COSMOS field and the other deep fields of roughly $\sim 0.015$ when cut at $i\sim 23$. This effect is similar in magnitude to the systematic offset between \textsc{EAZY} and \textsc{LePhare}. 

\begin{figure*}
   \centering
   \includegraphics[width=8.93cm]{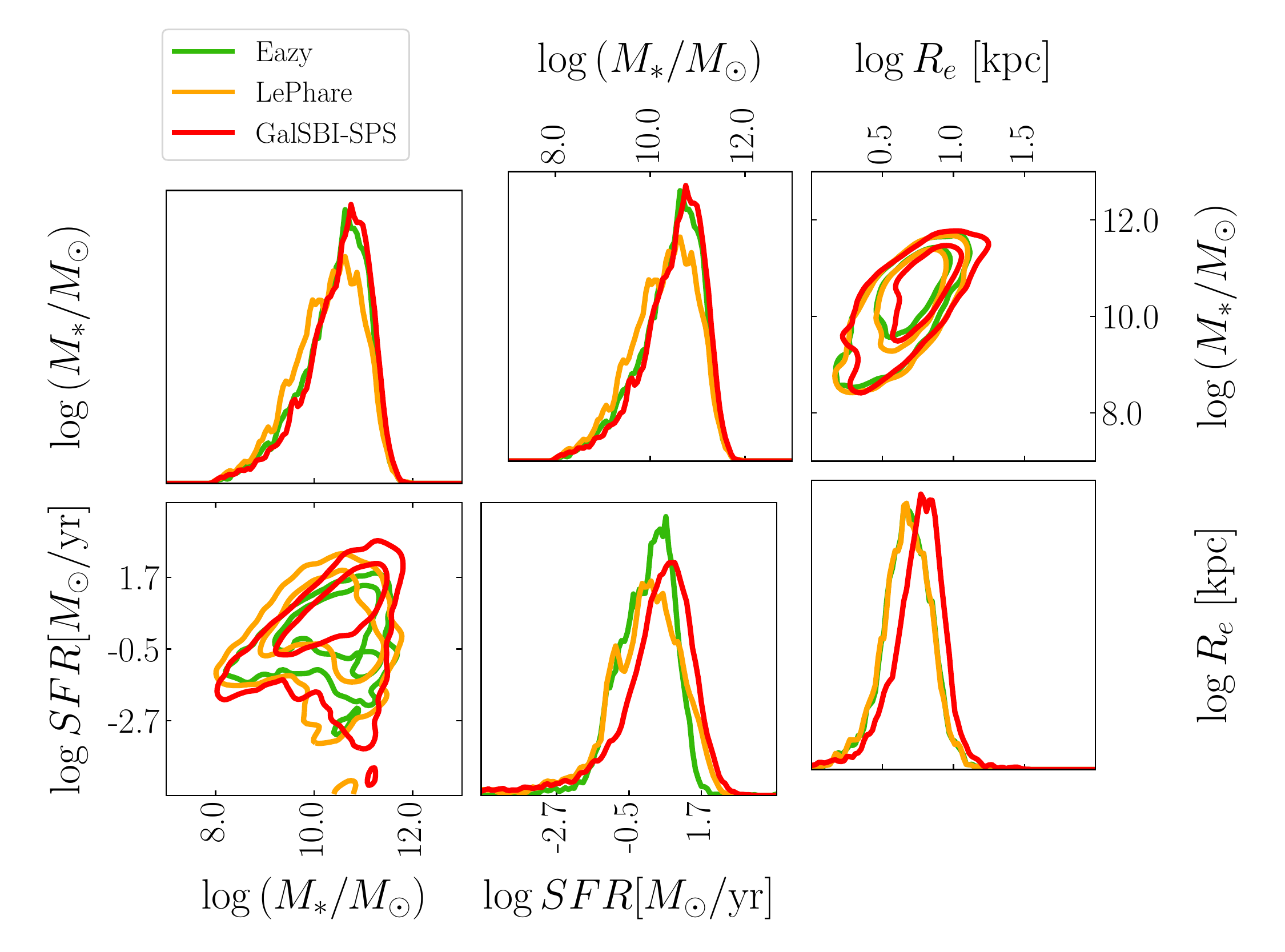} \\
   \includegraphics[width=8.93cm]{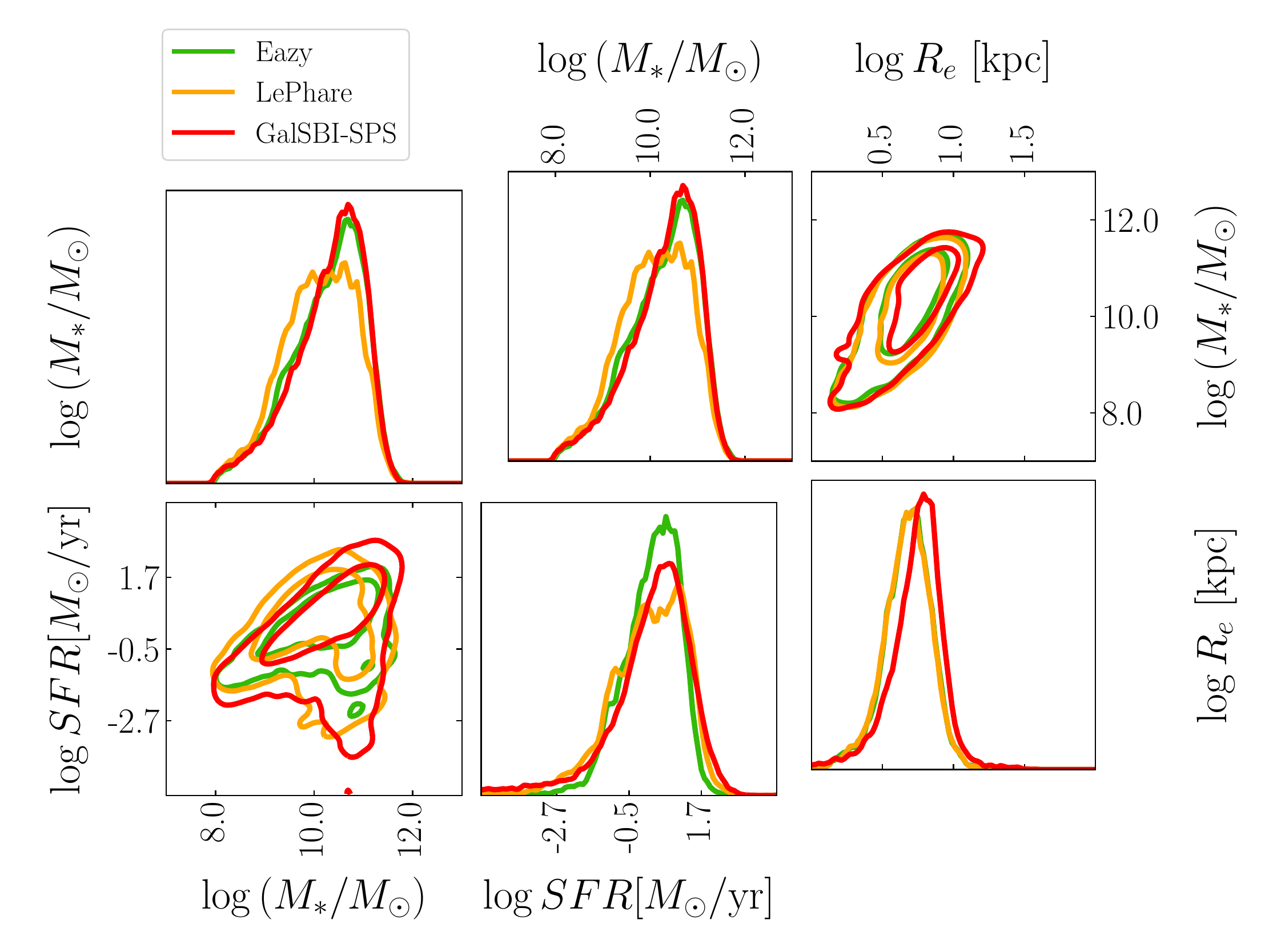} \\
   \includegraphics[width=8.93cm]{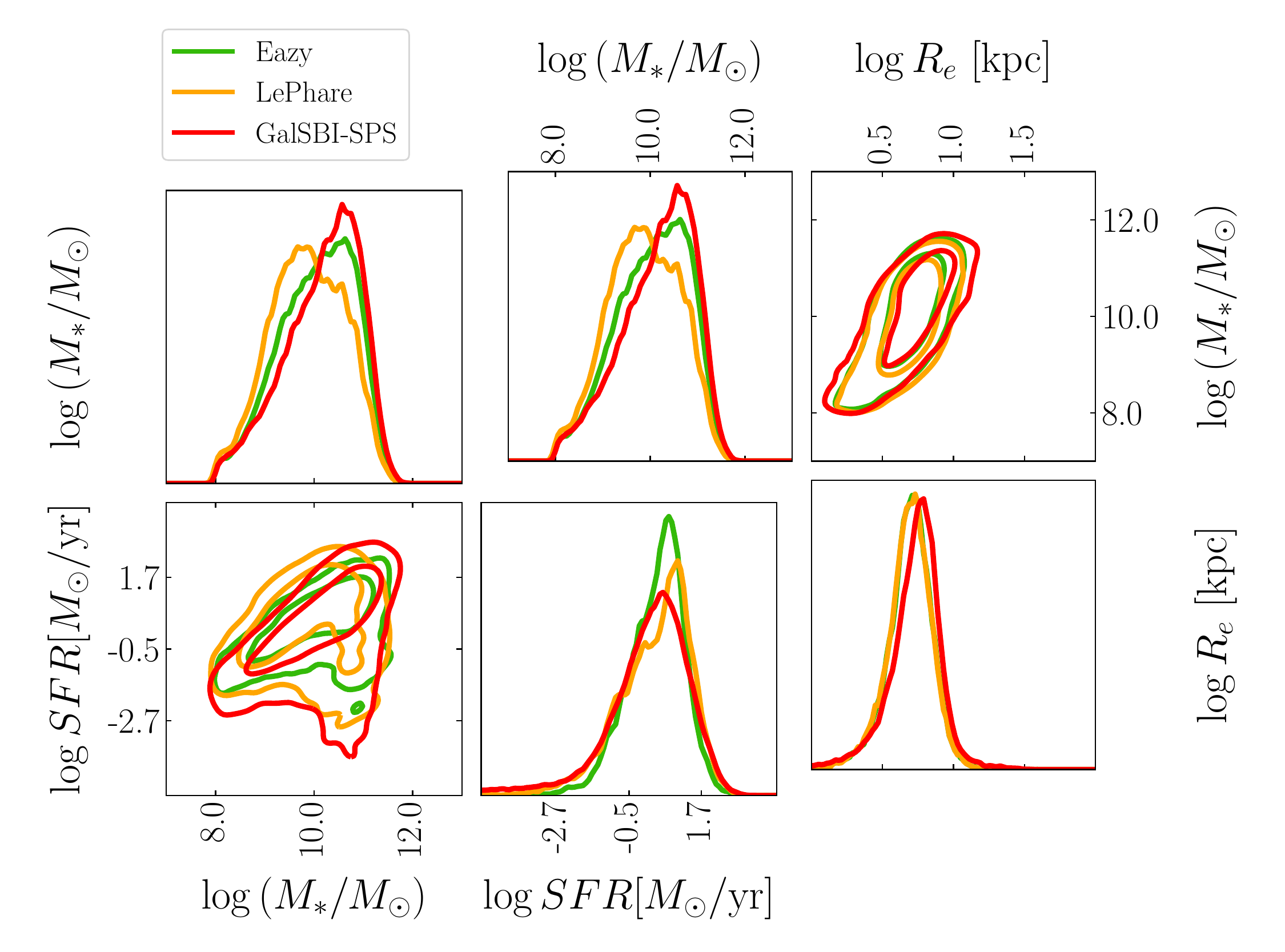}
      \caption{Comparison of the stellar mass-SFR and the size-stellar mass planes between the observed galaxies (green and orange contours) and \textsc{GalSBI-SPS} (red contours) model galaxies for the $i \le 21$ (upper left panel), $i \le 22$ (upper right panel), and $i \le 23$ (bottom panel) cuts. The lower left contours refer to the stellar mass-SFR planes, while the upper right to size-stellar mass planes. Stellar masses and SFRs for the observations are taken from the values in the COSMOS2020 catalogue, which have been estimated using \textsc{EAZY} (green contours) and \textsc{LePhare} (orange contours) photo-zs. The physical size estimates are provided by the \textsc{Source Extractor} $\mathrm{FLUX\_RADIUS}$ parameter in the $r$-band converted to $\mathrm{kpc}$ using the adopted cosmology.}
    \label{fig:stellarmass_sfr_size_plane}
\end{figure*}

\subsection{Galaxy physical properties}

One of the main features of \textsc{GalSBI-SPS} is the ability to sample galaxy physical properties from the model and provide precise estimates of the relations between these properties that can be compared against those obtained from observations. The COSMOS2020 catalogue contains stellar mass and SFR estimates obtained  using \textsc{EAZY} and \textsc{LePhare} photo-z estimates. In particular, the \textsc{LePhare} photo-zs are used in combination with \cite{Bruzual2003} templates and a delayed tau model for the SFH, while \textsc{EAZY} photo-zs are used to fit galaxy SEDs with \textsc{FSPS}. In the following, we refer to the stellar masses and the SFRs obtained from these fitting procedures as \textsc{EAZY} and \textsc{LePhare} estimates of said properties. We can compare the distribution of stellar masses and SFRs against those from the \textsc{GalSBI-SPS} model for the detected population of galaxies, allowing us to evaluate how realistic is the distribution of simulated galaxies physical properties we sample from our model. In particular, we qualitatively compare the stellar mass-SFR and the size-mass planes, two important proxies of the galaxy population in a survey. A quantitative comparison is left to a future work where the model parameters are jointly constrained against observations with appropriate distance metrics and where the stellar mass and SFR are consistently estimated across the three dataset.

Figure \ref{fig:stellarmass_sfr_size_plane} shows the stellar mass-SFR (lower left contours) and the size-stellar mass (upper right contours) planes for the same three $i$-band magnitude cuts adopted in the previous sections. The figure shows how \textsc{GalSBI-SPS} detected galaxies are able to reproduce both the star-forming main sequence \citep{Brinchmann2004,Noeske2007,Daddi2007,Salim2007,Whitaker2012,Popesso2023} and the passive cloud in the stellar mass-SFR plane, overlapping with the distributions from the observations. The stellar mass distribution of \textsc{GalSBI-SPS} galaxies seems to agree more with the \textsc{EAZY} estimates than with the \textsc{LePhare} ones, with the latter two being discrepant due to the different modelling choices.  The figure shows also that the simulations tend to produce objects with an overall higher SFR distribution compared to observations in the $i \le 21$ cut. The trend reverses instead for the fainter samples. This trend could explain what we see in the colour and redshift distribution plots. The $i \le 21$ sample contains higher SFR simulated objects, which tends to be brighter, bluer, and visible at higher redshifts, thereby producing the high redshift tail in the \textsc{GalSBI-SPS} $i \le 21$ redshift distribution. At fainter $i$-band cuts, the observations have more objects with higher SFR, leading to bluer colours than those predicted by \textsc{GalSBI-SPS}.

The upper right panels of Fig.  \ref{fig:stellarmass_sfr_size_plane} compares the size-stellar mass planes for the three $i$-band cuts. The physical size estimates are provided by the \textsc{Source Extractor} $\mathrm{FLUX\_RADIUS}$ parameter in the $r$-band converted to $\mathrm{kpc}$ using the adopted cosmology. The sizes have been consistently measured on data and simulations with the same \textsc{Source Extractor} setup. The figure shows very good qualitative agreement at all $i$-band cuts. We also observe a similar trend as in Figs. \ref{fig:magnitude_size_colour_contours_21}, \ref{fig:magnitude_size_colour_contours_22}, and \ref{fig:magnitude_size_colour_contours_23}, where \textsc{GalSBI-SPS} tends to produce a size distribution that is skewed towards larger values with respect to the observations. We also observe, as in Fig. \ref{fig:magnitude_size_colour_contours_23}, that the distribution of physical sizes tend to be more in agreement in the faint sample rather than in the bright sample.

\section{Discussion}
\label{sect:discussion}

The results of this work demonstrate the ability of \textsc{GalSBI-SPS} to provide a realistic and robust representation of the galaxy population up to a depth of $i \le 23$. Within this range, the model successfully reproduces the observed galaxy number counts and captures the multidimensional distribution of magnitudes, sizes, colours, and physical properties of observed galaxies. This level of agreement is particularly noteworthy,  considering that the model parameters were taken from different literature sources, each based on a different sample selections. 

There are however some tensions with the observations. Apparent galaxy sizes in $\mathrm{arcsec}$ are systematically larger than the observed counterparts, with median offsets of by $0.24\arcsec, 0.19\arcsec, 0.16\arcsec$ at $i \le 21, \ i \le 22, \ i\le 23$, respectively.  Furthermore, at $i \le 21$, \textsc{GalSBI-SPS} under-predicts the fraction of redder objects in the bimodal $g-r$ colour distribution with respect to observations, whereas the phenomenological model (\textsc{GalSBI}) over-predicts the number of this population.  At fainter cuts, $i \le 22, \ i\le 23$, \textsc{GalSBI-SPS} still under-predicts the fraction of redder objects in the bimodal $g-r$ colour, while also producing an overall distribution of $g-r$ colours that is redder than what the observations suggest. \textsc{GalSBI-SPS} redshift distributions tend to have a mean redshift that is higher than what the COSMOS2020 photo-zs seem to suggest.  Furthermore, the galaxy population in \textsc{GalSBI-SPS} has a higher average SFR at $i \le 21$ and a lower average SFR at $i \le 23$ compared to the values quoted in the COSMOS2020 catalogue.

These discrepancies could be due to the estimated values for the mean and scatter of the SFH shape parameter distributions and to the fact that we model these complex distributions from GAMA and DEVILS data using a single truncated Multivariate Normal distribution. This choice is motivated by the necessity to not dramatically increase the number of model parameters. We also assume linear redshift and stellar mass evolutions of the SFH shape parameters. While this modelling captures the bulk of the population,  the parameter values used thus far are not able to reproduce bimodal or strongly non-Gaussian features in the SFH parameter space.  As a result,  the predicted SFR distribution at $i \le 21$ is skewed toward higher values than what is reported in the COSMOS2020 catalogue. This excess star formation makes galaxies appear brighter and therefore detectable at higher redshifts, leading to a larger redshift distribution mean compared to the observations. At fainter limits ($i\le 22, \ i\le 23$), the redshift evolution of the SFH parameters leads to a SFR distribution that has lower values relative to the data. This under-prediction of strongly star-forming galaxies results in a lack of evolution in the $g-r$ colour distribution, inconsistent with observational trends.

A key avenue for future improvements is to enforce consistency in galaxy formation histories,  ensuring,  for example,  that the stellar mass assembled by galaxies at $z=1$ reflects the stellar mass distribution of the $z=0$ progenitors at that epoch. This is a non-trivial task, as galaxies do not evolve through smooth SFHs alone,  but are affected by mergers,  feedback,  and environmental processes.  While unimodal SFHs, like those adopted in this work, are effective at capturing the long-term star formation trends of most galaxies,  many of which are dominated by a single formation epoch,  they struggle to reproduce short-lived bursts of recent star formation. These bursts are particularly important for modelling galaxies that lie above the star-forming main sequence, but are notoriously hard to model and constrain. A more flexible SFH model that incorporates a burst component, such as those available in \textsc{ProSpect}, could help disentangle long-term trends from recent star formation episodes (e.g. \citealt{Ciesla2015,Malek2018}). Indeed, as shown in Fig.  \ref{fig:stellarmass_sfr_size_plane}, \textsc{GalSBI-SPS} does not reproduce the population of highly star-forming galaxies above the main sequence at stellar masses of $\sim 10^{9-10.5} M_{\odot}$, a shortcoming that could likely be addressed by including a burst component. In this work, however, we do not implement such a component, as the SFH parameters are calibrated on best-fitting values from GAMA and DEVILS data, both of which assume unimodal SFHs. Although the lack of a burst model may partially contribute to the observed discrepancies in magnitudes and colours, it is unlikely to be the sole driver. In particular, it cannot account for the systematic offset in apparent sizes, which are consistently overestimated across all magnitude cuts. This suggests that some of the discrepancy is due to the use of parameter values obtained from observations, while in reality the intrinsic distribution of the SFH or size parameters might be different than that observed.

A second key limitation to the predictive power of \textsc{GalSBI-SPS} arises from the use of literature-based parameter values. As the primary goal of this work is to motivate and illustrate the modelling choices for our galaxy population description, we adopt parameters from published studies, each based on galaxy samples with different selection criteria and properties, such as redshift and star formation activity. For example, we infer SFH, metallicity, and dust attenuation parameters for blue and red galaxies in GAMA and DEVILS using a $\mathrm{sSFR}$ cut of $\log{(sSFR/\mathrm{yr}^{-1})} > -10.5$. In contrast, \cite{Weaver2023} adopt a colour-based selection in the $NUVrJ$ diagram when deriving the GSMF for the two populations. Even among studies that use $sSFR$-based classifications, thresholds vary. Some studies use our same threshold to separate blue and red galaxies \citep{Kashino2019}, while others adopt a more conservative cut at $\log{(sSFR/\mathrm{yr}^{-1})} > -11.5$ \citep{Bellstedt2021}. 

Additional discrepancies may stem from the treatment of dust attenuation and the inclusion of an AGN component, both of which primarily affect rest-frame UV and blue optical magnitudes and colours \citep{Tortorelli2024}. To assess the impact of these components, we conduct two tests. First, we consider a limiting case where no dust attenuation is applied to any galaxy. If dust were the primary cause of the observed discrepancies, we would expect a notable increase in UV and blue optical flux, leading to brighter $g$-band magnitudes, bluer colours, and a larger number of detections at faint flux levels. However, we find that this limiting case is not alleviating the observed tensions. Similarly, we test removing the AGN contribution to the galaxy flux. While this alters the shape of the multidimensional colour distribution by removing AGN-like colours, the trend in the $g-r$ colour evolution remains unchanged.

Beyond refining the modelling of SFH parameter distributions, the most significant improvement in the predictive power of \textsc{GalSBI-SPS} is set to come from constraining the model parameters with informative observational data using SBI. A core strength of our forward-modelling framework is the ability to adjust model parameters, guided by physically motivated priors and carefully chosen distance metrics, to achieve closer agreement with observations. We have not performed this optimisation in the present work to focus on motivating the modelling choices for \textsc{GalSBI-SPS}.  Optimising model parameters is computationally intensive due to the run-time of generating large numbers of catalogue realisations from broad prior distributions for the model parameters.  While \textsc{ProSpect} is considerably more time-efficient at generating SEDs than other SPS codes, its execution speed is still not sufficient to function as a scalable generative model across the wide prior space required for SBI.  This calls for the inclusion in our framework of a machine-learning-based emulator to enable fast computation of synthetic magnitudes.  For this reason,  we developed \textsc{ProMage} a feed-forward neural network that emulates the computation of observer- and rest-frame magnitudes from the generative galaxy SED package ProSpect.  This emulator will be crucial in enabling a full scale SBI run to optimise the model parameters.  Generating full-survey image simulations for every prior sample also presents a major computational bottleneck. As shown in \citet{Fischbacher2025}, this can be mitigated by using an emulator that maps intrinsic galaxy properties directly to detected ones, bypassing image-level simulations. Such emulators are also critical for testing model extensions to \textsc{GalSBI-SPS}. In keeping with the forward-modelling philosophy, additional complexity should only be introduced when residual tensions with the data remain after the SBI procedure. This could include a recent burst in the SFHs, the use of a single rather than a two component (blue/red) population, or the inclusion of realistic clustering of galaxies with the SHAM-OT method \citep{Fischbacher2025}. This last aspect could further expand the scientific reach of our framework, enabling studies of the stellar-to-halo mass relation, clustering-based cosmological parameter estimation,  and the environmental dependence of galaxy physical properties.

A major strength of the forward-modelling approach lies in its ability to combine heterogeneous datasets to constrain the model parameters via SBI. Each dataset carries distinct constraining power towards different components of the model.  As shown in \cite{Tortorelli2024}, several SED modelling ingredients,  such as dust attenuation,  AGN emission,  gas properties,  IMF,  and the SFH (as shown in this work),  can significantly alter broad-band galaxy colours. These affect,  in turn,  forward-modelling-based redshift distributions,  often producing shifts larger than what is required for Stage IV surveys. To mitigate this, it is crucial to design distance metrics and use the appropriate datasets that are sensitive to these SED modelling components. For instance,  HSC DUD data provide strong constraints on components that affect broadband magnitudes and colours, such as GSMF,  SFH,  dust attenuation,  galaxy sizes and the AGN contribution. This dataset,  already used in \cite{Fischbacher2025a} to calibrate the phenomenological version of \textsc{GalSBI},  serves as an ideal precursor for Stage IV photometric surveys, thereby providing a constrained model that is able to reproduce the colour space spanned by these future data.  However,  several key components of galaxy SED models have little effect on broad-band photometry \citep{Tortorelli2024}.  These include the current gas-phase metallicity,  ionisation state of the gas,  along with other stellar population features. Constraining these features requires spectroscopic data with higher spectral resolution,  as provided by surveys like DESI and 4MOST.  These surveys enable more detailed SED modelling, making it possible to create galaxy population models applicable to both photometric and spectroscopic datasets. Examples of valuable spectroscopic constraints from DESI include the use of the quasar sample \citep{Chaussidon2023}, which would help in refining AGN modelling in the near-UV/optical, DESI BGS \citep{Hahn2023BGS} high signal-to-noise spectra, which could help in linking Balmer decrement-estimated extinction with galaxy global physical properties, and the emission line galaxy samples \citep{Raichoor2023}, which can be used to trace gas ionisation and metallicity up to high redshift and link them to the galaxy global physical properties.  On the 4MOST side,  programs such as 4MOST-STePS \citep{Iovino2023},  targeting long-integration spectra of intermediate-redshift galaxies, offer further opportunities to constrain stellar population models. In order to forward-model these spectroscopic surveys and constrain the relevant parameters, it is essential to develop realistic and efficient spectral simulators that map intrinsic galaxy SEDs to observed spectra under realistic observational conditions. The \textsc{USpec} \citep{Fagioli2018,Fagioli2020} and \textsc{USpec2} (Tortorelli et al. ,  in prep.) codes are designed for this purpose, supporting the simulation of fiber-based spectra for SDSS, DESI, and 4MOST. Finally, to make SBI over large parameter spaces feasible, it is necessary to accelerate the generation of synthetic spectra from \textsc{ProSpect}.  This will require the development of a dedicated spectral emulator of \textsc{ProSpect} outputs,  analogous what we have already achieved with \textsc{ProMage}.

The choice of the SBI scheme is critical for obtaining robust and unbiased posterior estimates of model parameters. Several recent studies \citep{Tortorelli2020,Tortorelli2021,Kacprzak2020,Moser2024,Fischbacher2025} have successfully employed ABC, one of the earliest and most widely used SBI techniques, to derive posterior estimates. The use of ABC, one of the earliest SBI methods developed, is justified by its inherent conservative nature. It mitigates the risk of convergence to local minima and is particularly well suited for models with a large number of parameters, such as \textsc{GalSBI} and \textsc{GalSBI-SPS}. ABC’s robustness stems from its reliance on physically motivated priors and well-chosen distance metrics tailored to be sensitive to specific aspects of the model. The construction of informative priors plays a key role in guiding the inference process, anchoring it to observationally supported values while allowing for uncertainty through appropriate prior variances. Despite its strengths, ABC is computationally intensive. It converges more slowly to the approximate posterior than more modern machine learning–based SBI methods, such as neural density estimators \citep{Papamakarios2016,Lueckmann2017,Papamakarios2018,Alsing2019} or full-field inference approaches \citep{Zeghal2024}. These techniques are capable of modelling complex posteriors using fewer simulations, but may yield overconfident results, particularly when sample sizes are small or priors are misspecified \citep{Hermans2021,Tam2022}. This can lead to the premature exclusion of plausible regions of parameter space, an issue ABC avoids by its incremental acceptance process. Nevertheless, ABC is not without limitations. One of the primary challenges is the sharp drop in acceptance rate as the distance threshold $\epsilon$ approaches zero, which can render the method inefficient for fine-grained posterior estimation \citep{Alsing2018}.

A well-constrained galaxy population model, coupled with a realistic forward-modelling framework for photometric and spectroscopic surveys, enables a wide range of scientific applications. From a galaxy evolution perspective, a major strength of our approach is its ability to accurately measure the luminosity and the GSMFs for large galaxy samples, as well as providing an unbiased estimate of the intrinsic scaling relations among global galaxy physical properties. This is enabled by the simulation's perfect knowledge of intrinsic quantities such as stellar mass, luminosity, and redshift, alongside a precise characterisation of survey completeness. Using emulators to test model extensions, we can evaluate which physical prescriptions best match the observed trends, e.g.  the redshift evolution of gas-phase metallicity or the dependence of dust attenuation on global galaxy properties. From a cosmological standpoint,  our framework offers a robust avenue for meeting the stringent redshift calibration requirements of Stage IV surveys, thereby enabling precise measurements of LSS cosmological parameters.  Once the model parameters are constrained via SBI, the resulting redshift distributions can be used to re-analyse existing cosmological datasets,  quantify the gains in parameter precision from improved calibration,  and support future Stage IV analyses. These capabilities are made possible by the forward-modelling framework’s core feature: the ability to apply identical selection functions and analysis pipelines to both simulations and observational data.

To conclude,  we have presented the first implementation of \textsc{GalSBI-SPS},  a comprehensive SPS–based galaxy population model designed for forward-modelling applications. This model represents a major step forward in predictive galaxy population modelling,  as it consistently generates both physical and morphological galaxy properties.  For this first implementation,  despite adopting parameter values drawn from literature sources, each based on different datasets and redshift regimes, \textsc{GalSBI-SPS} offers a realistic description of the galaxy population down to $i \le 23$. The model reproduces the observed distributions of magnitudes,  colours,  and sizes from HSC data with good fidelity,  although moderate discrepancies remain,  particularly in the $g-r$ colour and in the median apparent sizes.  We benchmark \textsc{GalSBI-SPS} against its phenomenological counterpart, \textsc{GalSBI},  which has undergone extensive calibration against both broad- and narrow-band photometric data across multiple studies.  Although \textsc{GalSBI-SPS} is still in an early development phase and has not yet been optimised via SBI,  it achieves a comparable level of realism, demonstrating the strength of its underlying physical modelling framework. The predictive power of \textsc{GalSBI-SPS} is expected to increase significantly once its parameters are jointly constrained using SBI and informative photometric and spectroscopic datasets.  A calibrated version of the model will provide forward-modelled redshift distributions with the accuracy required for Stage IV cosmological analyses.  It will also allow for robust measurements of fundamental scaling relations,  such as the luminosity function,  GSMF,  size–mass relation,  and stellar mass–SFR relation, derived from a model where the true physical parameters and their redshift evolution are precisely known.  \textsc{GalSBI-SPS} is publicly available as part of the existing \textsc{galsbi} Python package,  offering the community a unified framework for generating both physical and phenomenological galaxy population models.

\section{Conclusions}
\label{sect:conclusions}

The forward-modelling of photometric and spectroscopic galaxy surveys represents a robust approach to characterise the galaxy population observed in a survey and in doing so obtaining an accurate estimate of galaxy redshift distributions for cosmological applications.  Key to its success is the development of a realistic galaxy population model and a mapping of galaxy physical properties to their SEDs. In this work, we present the first implementation of \textsc{GalSBI-SPS},  a new SPS-based galaxy population model, designed for forward-modelling applications in both galaxy evolution and cosmology. This model represents a major step forward in predictive galaxy population modelling, as it consistently generates both physical and morphological galaxy properties. The prescriptions we implement are motivated by the sensitivity analysis conducted in \cite{Tortorelli2024}. The model generates realistic, survey-independent galaxy catalogues,  which we use to forward model HSC DUD observations in the COSMOS field.  This work aims at introducing \textsc{GalSBI-SPS},  motivating the modelling choices adopted for the physical galaxy properties and for their SEDs, and evaluating the model performance with literature values for the parameters against its phenomenological version, \textsc{GalSBI} \citep{Fischbacher2025a}, and observed data.

The model samples galaxy physical properties from analytical parametrisations. Galaxies are drawn from two overlapping population of blue and red objects that share similar parametrisations, but different parameter values. The sampling starts by jointly drawing formed stellar masses of galaxies and their redshifts from double Schechter GSMFs,  with different parameter values for blue and red objects.  SFHs are modelled using the truncated skewed Normal distribution \citep{Robotham2020}. The shape parameters for the SFH are sampled from modelling the distribution of best-fitting values in GAMA and DEVILS,  while the overall normalisation comes from enforcing the integrated SFH to produce the formed stellar mass sampled from the GSMF.   We then linearly map the stellar mass evolution of each galaxy on to the shape of the gas metallicity evolution, with the final gas-phase metallicity at the time of observation dependent on the galaxy stellar mass, redshift and SFR averaged over the last $100 \ \mathrm{Myr}$. We then sample gas ionisations, dust parameters, and velocity dispersions, and we assign each galaxy an AGN template whose flux contribution depends on a sampled $f_{\mathrm{AGN}}$ parameter. We also sample morphological properties for our galaxy samples, whose light distributions are modelled as single S\'ersic light profiles. Galaxy physical sizes and S\'ersic indices are sampled conditioned on their redshifts and stellar masses,  with different analytical prescriptions for red and blue galaxies,  while the ellipticities are sampled for both blue and red galaxies from the same modified Beta distribution as in \cite{Moser2024,Fischbacher2025}.  The galaxy physical properties are then passed to the generative galaxy SED package \textsc{ProSpect},  allowing for self-consistent modelling of stellar,  nebular,  dust,  and AGN emission. We then compute apparent magnitudes from \textsc{ProSpect} SEDs after adding the AGN flux contribution. 

The catalogue of intrinsic galaxy properties,  sampled from the galaxy population model,  is then used to create simulated HSC DUD images in the COSMOS field with the \textsc{UFig} image simulator. We simulate images in the five broad-bands $g,r,i,z,y$ following the prescriptions adopted in \cite{Moser2024,Fischbacher2025} and we run \textsc{Source Extractor} on the images to produce catalogues of detected galaxies.  We consistently apply the same \textsc{Source Extractor} setup on both real and simulated HSC images.

After applying the same selection on both data and simulations to obtain a sample with reliable photometric information, we compare \textsc{GalSBI-SPS} model galaxies against those sampled from \textsc{GalSBI} and against the observed data using a number of distance metrics. We first compare the relative number of detected galaxies between the data and the simulations for five different $i$-band cuts. We find that both \textsc{GalSBI} and \textsc{GalSBI-SPS} yield a similar number of detected galaxies to the observations, within few percent, up to $i\le 23$, a threshold that is similar to a Stage III-like experiment (e.g.  Kilo Degree Survey,  \citealt{Wright2025a}). However,  at fainter magnitudes,  the relative difference between \textsc{GalSBI-SPS} and the data increases beyond $10 \%$, while for \textsc{GalSBI} it stays at the percent level.  This different behaviour is due to two factors. The first is that \textsc{GalSBI} has been constrained against HSC DUD data,  while \textsc{GalSBI-SPS} relies entirely on literature-based parameters and is not yet tuned via SBI.  The second is that the model is uninformed beyond $i\le 23$ given the GAMA,  DEVILS,  and literature data being used. We then compare the magnitude,  size,  and colour distributions among the three samples up to the threshold where the relative number of detected objects agrees within few percent ($i\le 23$). The magnitude distributions show a very good agreement at all $i$-band thresholds, with median differences always below $0.14$ mag.  Model galaxies cover the full colour space spanned by HSC data.

However some tensions are visible in the 1D $g-r$ colour distributions. \textsc{GalSBI-SPS} does not show the same fraction of redder colour objects in the bimodal $g-r$ distribution compared to the observations and \textsc{GalSBI}. Furthermore, the $g-r$ colour distribution does not evolve with the $i$-band cut,  leading to an overall distribution of redder $g-r$ colours at $i \le 22, \ i \le 23$.  Apparent sizes are systematically overestimated by $\sim0.2\arcsec$ \ in the median across all magnitude cuts.  Furthermore, \textsc{GalSBI-SPS} predicts redshift distributions that have higher mean values than what the observations and \textsc{GalSBI} return.  This difference goes from $\sim 0.08$ at $i\le 21$ to $0.012$ at $i\le 23$. The comparison of the galaxy physical properties drawn from \textsc{GalSBI-SPS} with those from the COSMOS2020 catalogue show the ability of \textsc{GalSBI-SPS} to qualitatively reproduce the distribution of galaxies in the stellar mass-SFR and size-stellar mass planes.  \textsc{GalSBI-SPS} tends to produce a population of higher SFR galaxies at $i\le 21$ and lower SFR at $i\le 23$ compared to observed data. 

In Sect. \ref{sect:discussion} we discuss the strengths and limitation of the current version of \textsc{GalSBI-SPS}. While providing a realistic, robust, and survey-independent representation of the galaxy population up to a depth of $i \le 23$, the model predictive power can be significantly increased in the future.  We mostly use values taken from the literature for the model parameters,  which introduces inconsistencies across different physical components of the model,  as these are often derived from samples with differing redshift coverage, selection criteria, and sSFR definitions.  In this work, we choose to not constrain the model with SBI to focus on motivating our choices for the \textsc{GalSBI-SPS} modelling components.  Future work will focus on constraining the model parameters with SBI, leveraging the joint constraining power of photometric and spectroscopic datasets such as HSC, LSST, DESI, and 4MOST.  This improvement in the modelling would contribute to generate more realistic $g-r$ colours and bluer/redder fraction of objects.  We will use the machine-learning galaxy magnitude emulator \textsc{ProMage} to accelerate the generation of large numbers of catalogue realisations from broad prior distributions for the model parameters. The use of emulators coupled with physically motivated priors and carefully chosen distance metrics will allow for jointly constraining \textsc{GalSBI-SPS} model parameters with informative photometric and spectroscopic data.  These improvements will enhance the model's ability to meet the stringent requirements on the redshift distributions for a Stage IV cosmological analysis. Furthermore, they will enable robust inference of the evolution of galaxy properties across cosmic time,  such as the luminosity function,  GSMF,  size-stellar mass relation,  and the stellar mass-SFR relation,  by using a model where the true physical properties are precisely known.  \textsc{GalSBI-SPS} is publicly released as part of the \textsc{galsbi} Python package \citep{Fischbacher2024b},  providing the community with a unified framework for generating physically and phenomenologically motivated galaxy population models suitable for forward-modelling wide-field galaxy surveys.

\begin{acknowledgements}
LT is grateful to Andrew Battisti, Luke Davies, Claudia Lagos,  Adam Marshall, Risa Wechsler, Sucheta Cooray for useful discussions.This work was funded by the Deutsche Forschungsgemeinschaft (DFG, German Research Foundation) under Germany’s Excellence Strategy – EXC-2094 – 390783311. This project was supported in part by grant 200021\_192243 from the Swiss National Science Foundation. SB acknowledges funding by the Australian Research Council (ARC) Laureate Fellowship scheme (FL220100191). ASGR acknowledges funding by the Australian Research Council (ARC) Future Fellowship scheme (FT200100375). GAMA is a joint European-Australasian project based around a spectroscopic campaign using the Anglo-Australian Telescope. The GAMA input catalogue is based on data taken from the Sloan Digital Sky Survey and the UKIRT Infrared Deep Sky Survey. Complementary imaging of the GAMA regions is being obtained by a number of independent survey programmes including GALEX MIS, VST KiDS, VISTA VIKING, WISE, Herschel-ATLAS, GMRT and ASKAP providing UV to radio coverage. GAMA is funded by the STFC (UK), the ARC (Australia), the AAO, and the participating institutions. The GAMA website is https://www.gama-survey.org/.  DEVILS is an Australian project based around a spectroscopic campaign using the Anglo-Australian Telescope. The DEVILS input catalogue is generated from data taken as part of the ESO VISTA-VIDEO \citep{Jarvis2013} and UltraVISTA \citep{McCracken2012} surveys.  DEVILS is part funded via Discovery Programs by the Australian Research Council and the participating institutions. The DEVILS website is https://devilsurvey.org. The DEVILS data is hosted and provided by AAO Data Central (https://datacentral.org.au/).
\end{acknowledgements}

\bibliographystyle{aa}
\bibliography{aa55759-25}

\begin{appendix}

\section{GAMA completeness-selected sample}
\label{appendix:gama_completeness}

\begin{figure}[h]
   \centering
   \includegraphics[width=8cm]{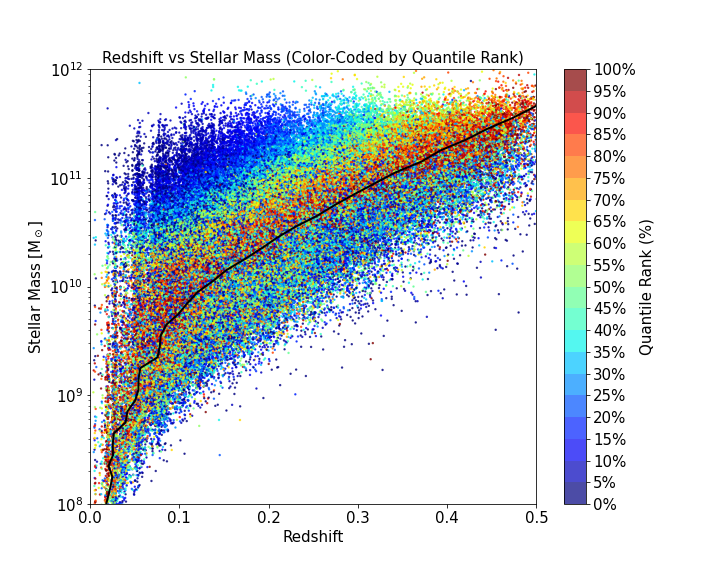}
      \caption{Stellar mass-redshift selection limit for the GAMA sample. At fixed stellar mass slice of $0.1 \ \mathrm{dex}$ width, each galaxy is colour-coded by the $g-i$ colour quantile in that slice. The black line represents the low redshift $95\%$ extreme of the $g-i$ distribution at all stellar masses.
              }
    \label{fig:GAMA_completeness_limit}
\end{figure}

\begin{figure}[h]
   \centering
   \includegraphics[width=8cm]{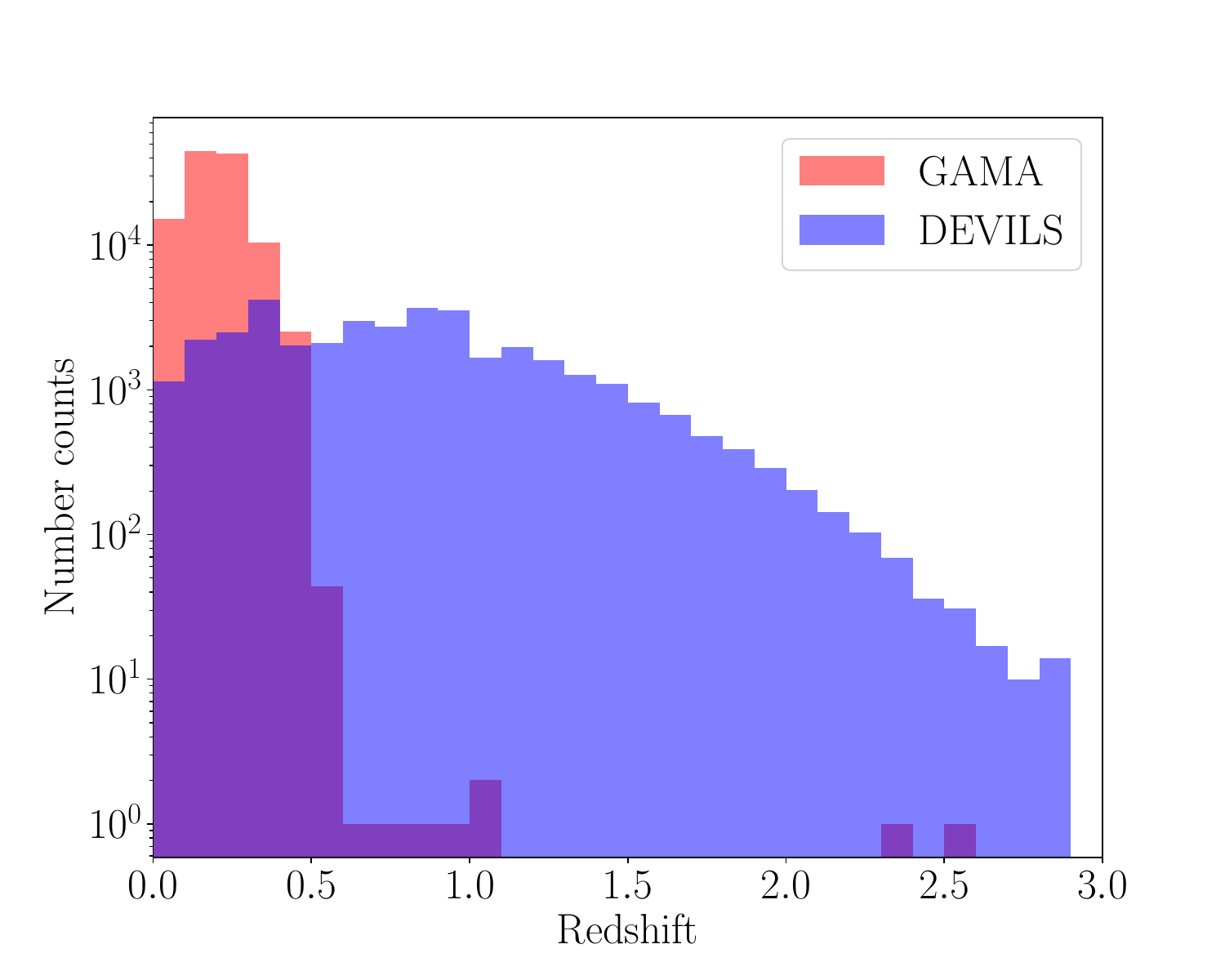}
      \caption{Redshift distributions for the mass complete samples of GAMA (red) and DEVILS (blue) galaxies.
              }
    \label{fig:GAMA_DEVILS_z_hist}
\end{figure}

The Value Added Catalogue used in \cite{Bellstedt2020,Bellstedt2021} contains gas, dust and stellar population properties measured with \textsc{ProSpect} for $233707$ galaxies. We match this catalogue to the \textsc{gkvScienceCatv02} catalogue (trimmed-down, science-ready version of table \textsc{gvkInputCat}) in the GAMA DR4 database via the \textsc{CATAID} keyword to assign the corresponding $u,g,r,i,Z,Y,J,H,Ks$ photometry to each object. Following the GAMA DR4 catalogue documentation, we select sources having $SC \ge 7$, which is the GAMA III Main Survey $95\%$ spectroscopic completeness limit. This cut reduces the number of objects to $195260$ and it is meant to include sources that are not classified as ambiguous, that lie within the survey region and outside of the star-masked regions, and that have $r < 19.65$. In order to select a sample of objects that contains all galaxy types at a given stellar mass without being biased to bright, star-forming galaxies, we define the lower stellar mass range for GAMA using the spectroscopic completeness limit defined by \cite{Robotham2014}. This completeness cut takes into account the fact that at higher redshifts we are able to see only the bluer galaxies at fixed stellar mass, thereby creating an unrepresentative sample of objects that is strongly biased towards blue galaxies. We therefore define a variable stellar mass redshift selection limit by computing, at fixed stellar mass slice of $0.1 \ \mathrm{dex}$ width, the $g-i$ colour quantile of each galaxy in that slice. Figure \ref{fig:GAMA_completeness_limit} shows the stellar mass redshift plane. Each galaxy is colour coded by the $g-i$ quantile it belongs to at fixed stellar mass slice, from $0\%$  to $100\%$, depending on whether the galaxy is the bluest or the reddest in the stellar mass bin selected. The stellar mass completeness limit is the low redshift $95\%$ extreme of the $g-i$ distribution at all stellar masses (black line in Fig.  \ref{fig:GAMA_completeness_limit}). The final mass complete sample of galaxies contains $101784$ objects. We report in Figs.  \ref{fig:GAMA_DEVILS_z_hist} and \ref{fig:GAMA_DEVILS_iband_hist} the redshift and $i$-band magnitude distributions of this sample.

\section{DEVILS completeness-selected sample}
\label{appendix:devils_completeness}

\begin{figure}[h]
   \centering
   \includegraphics[width=8cm]{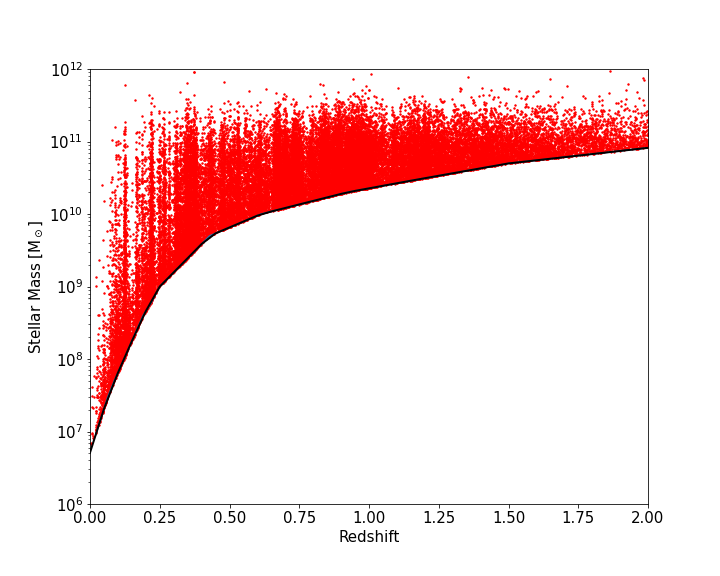}
      \caption{Stellar mass-redshift selection limit for the DEVILS sample. The black line represents stellar mass completeness selection as function of redshift defined in \cite{Thorne2022}.
              }
    \label{fig:DEVILS_completeness_limit}
\end{figure}

\begin{figure}[h]
   \centering
   \includegraphics[width=8cm]{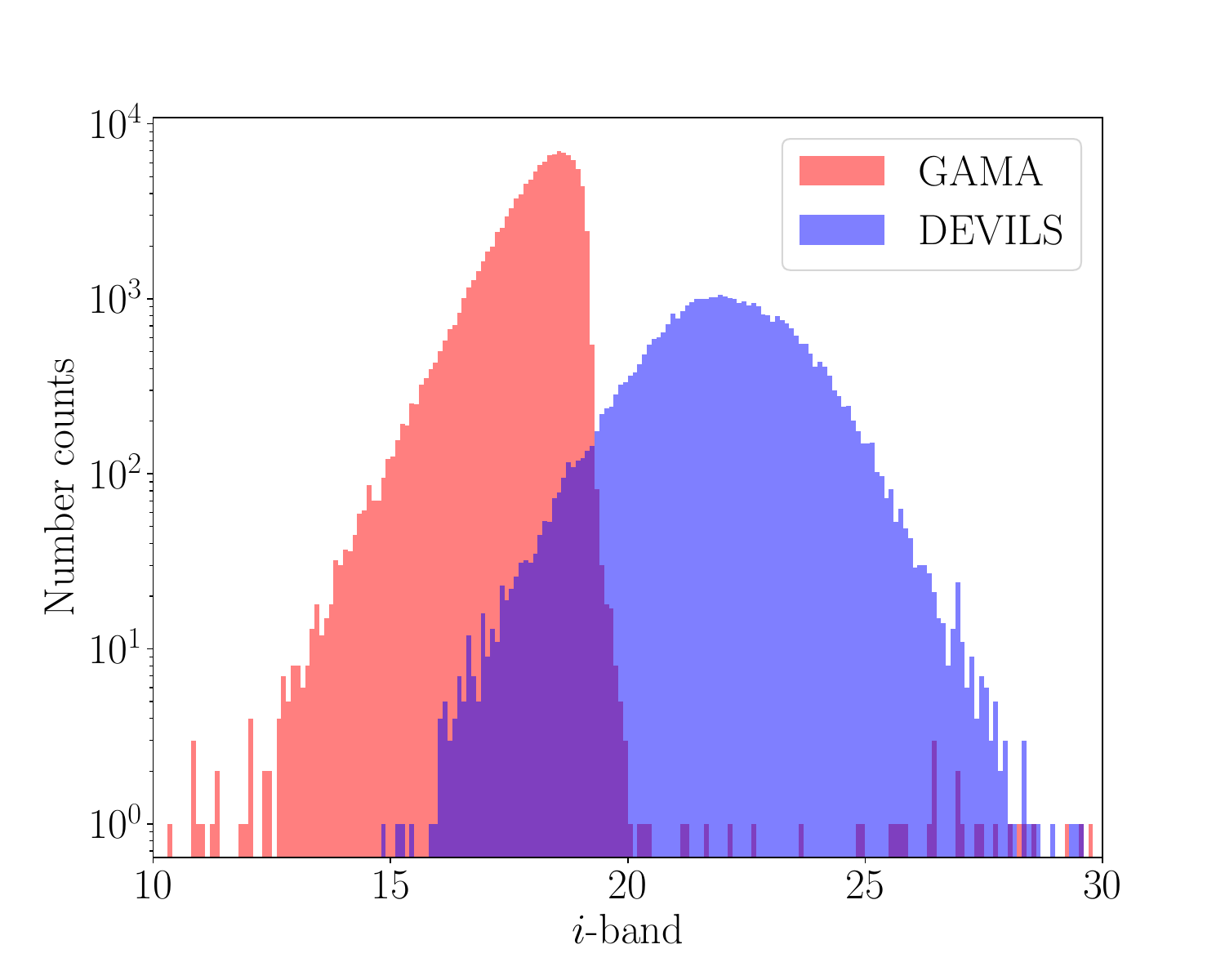}
      \caption{Histogram of i-band magnitudes for the mass complete samples of GAMA and DEVILS galaxies. GAMA $i$-band magnitude refers to GAMA-reprocessed version of the KiDS-VIKING images, while DEVILS $i$-band magnitude refers to the HSC version.
              }
    \label{fig:GAMA_DEVILS_iband_hist}
\end{figure}

The DEVILS data in the COSMOS field contains stellar masses, SFRs, gas metallicities, dust parameters and AGN luminosities for $494,084$ sources, of which $\sim 24$k  have spectroscopic redshifts from DEVILS, while for the remaining ones have photometric-estimated redshifts. In order to select an equally unbiased sample, we apply the stellar mass completeness limit reported in Fig.  4 of \cite{Thorne2022}. The sample of galaxies above the completeness limit (black line) is shown in Fig. \ref{fig:DEVILS_completeness_limit}. This mass completed sample contains $38,066$ galaxies. We report in Figs. \ref{fig:GAMA_DEVILS_z_hist} and \ref{fig:GAMA_DEVILS_iband_hist} the redshift and $i$-band magnitude distributions of this sample. GAMA galaxies are $13,6,4,3,3$ times more than DEVILS galaxies at $i \le 21,22,23,24,25$, respectively.

\section{Galaxy surviving stellar mass emulator}
\label{appendix:surviving_mass_emulator}

\begin{figure}[h]
   \centering
   \includegraphics[width=8cm]{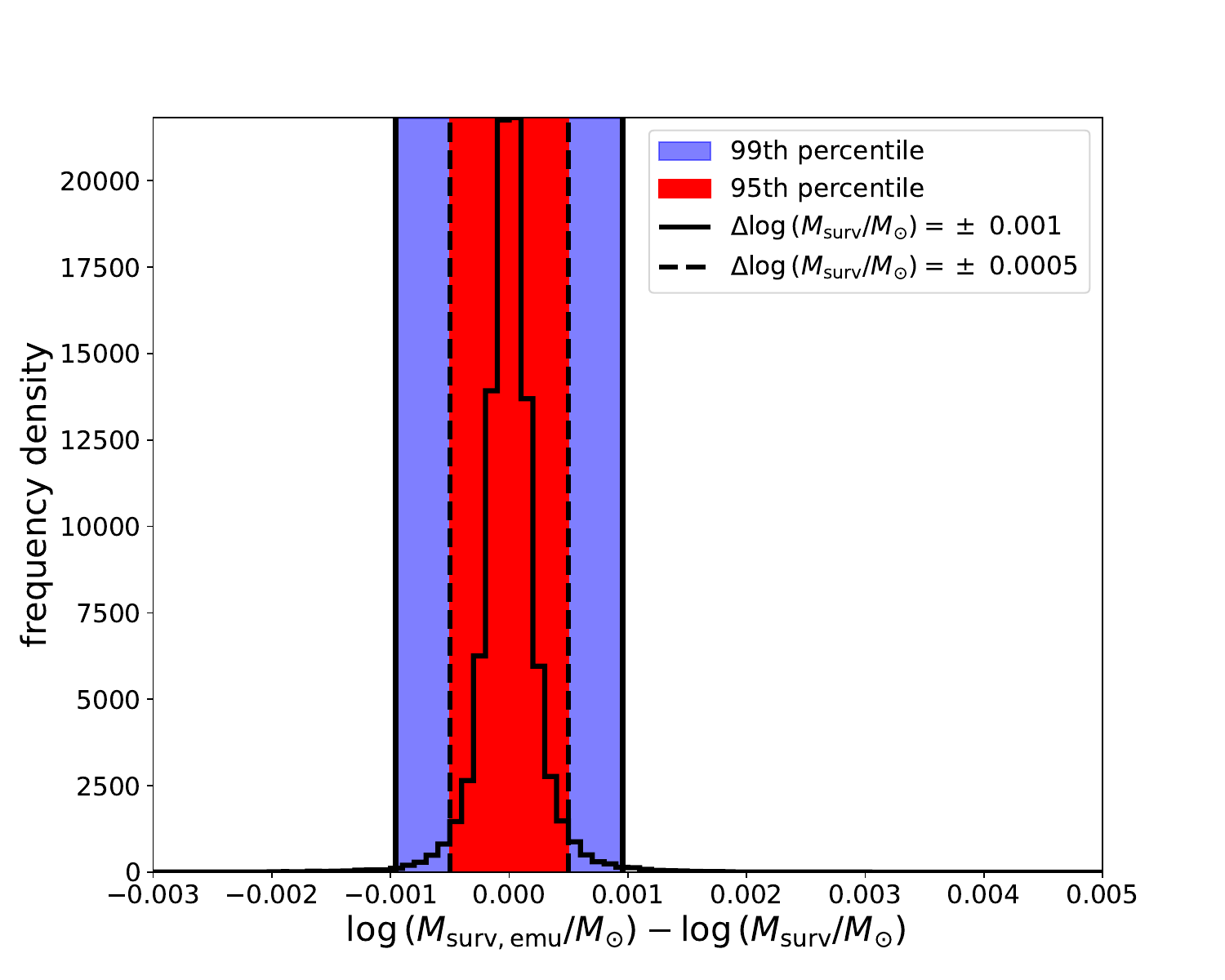}
      \caption{Histogram showing the prediction accuracy of the galaxy surviving stellar mass emulator.  Red and blue bands represent the ranges containing $95\%$ and $99\%$ of the samples from the test set,  respectively,  while dashed and solid vertical lines represent the $95$th and $99$th percentile values.  The trained emulator produces per-mille level accurate predictions for $99\%$ of the test set.}
    \label{fig:surviving_mass_emulator_hist}
\end{figure}

The galaxy surviving stellar mass is a quantity that is more closely tied to what we estimate from the observed galaxy light, since it is only the existing stars that contribute to it. \textsc{ProSpect} is able to compute the galaxy surviving stellar mass, however this increases the computational time of individual galaxy SEDs. We therefore train a machine learning-based emulator to accelerate the computation of galaxy surviving stellar masses.  The emulator predicts galaxy surviving masses from their formed masses,  redshifts,  SFH parameters and gas-phase metallicities at the time of observation.  We train the emulator on a sample of $10^6$ objects,  split in $80\%$ for training,  $10\%$ for validation and $10\%$ for testing. The results in Fig. \ref{fig:surviving_mass_emulator_hist} shows the prediction accuracy of the emulator.  We are able to recover the input galaxy surviving stellar masses with a per-mille level accuracy on $99\%$ of the test set.

\section{Faint sample comparison}
\label{appendix:faint_sample}

\begin{figure}[h]
   \centering
   \includegraphics[width=9cm]{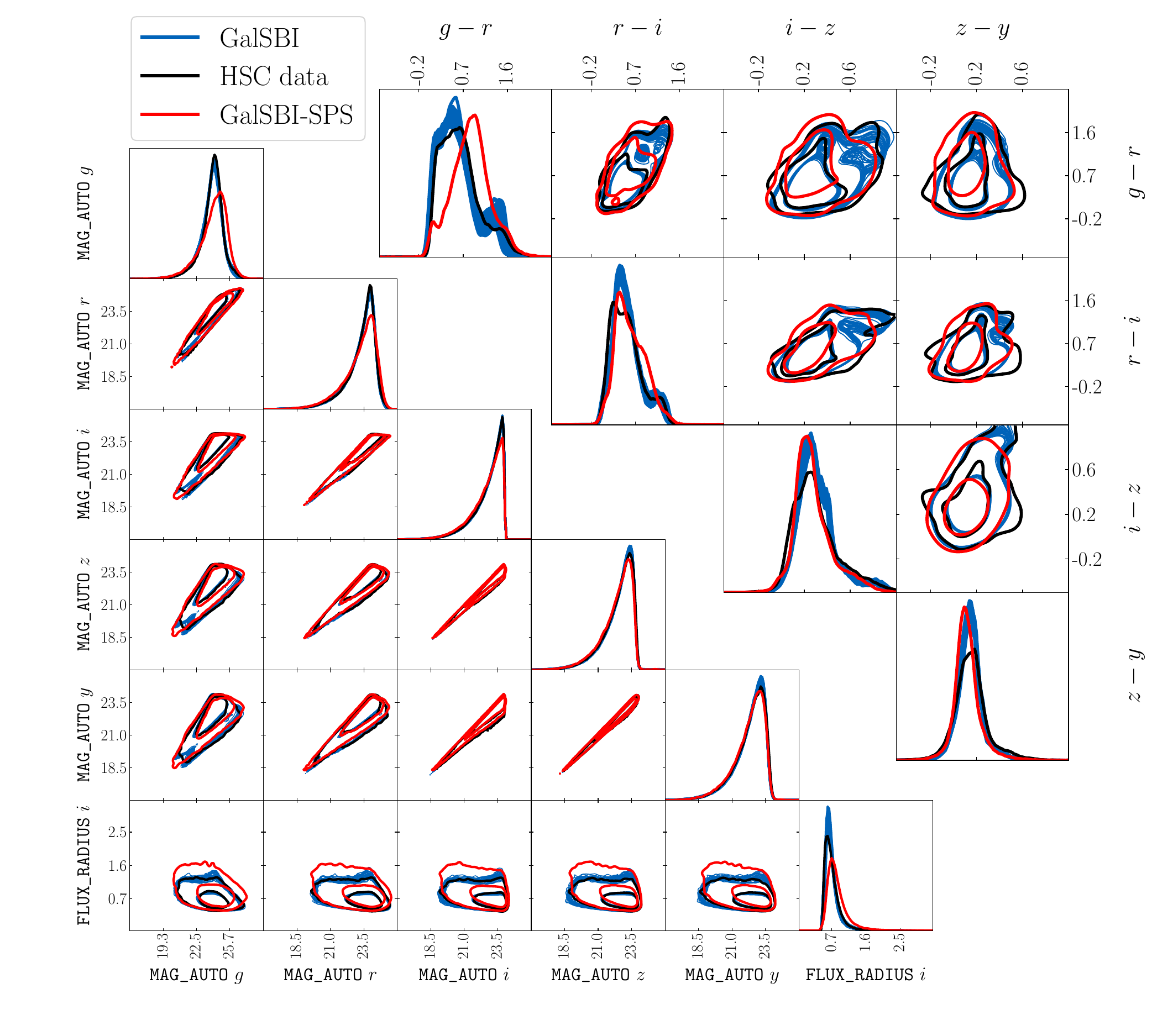}
   \includegraphics[width=9cm]{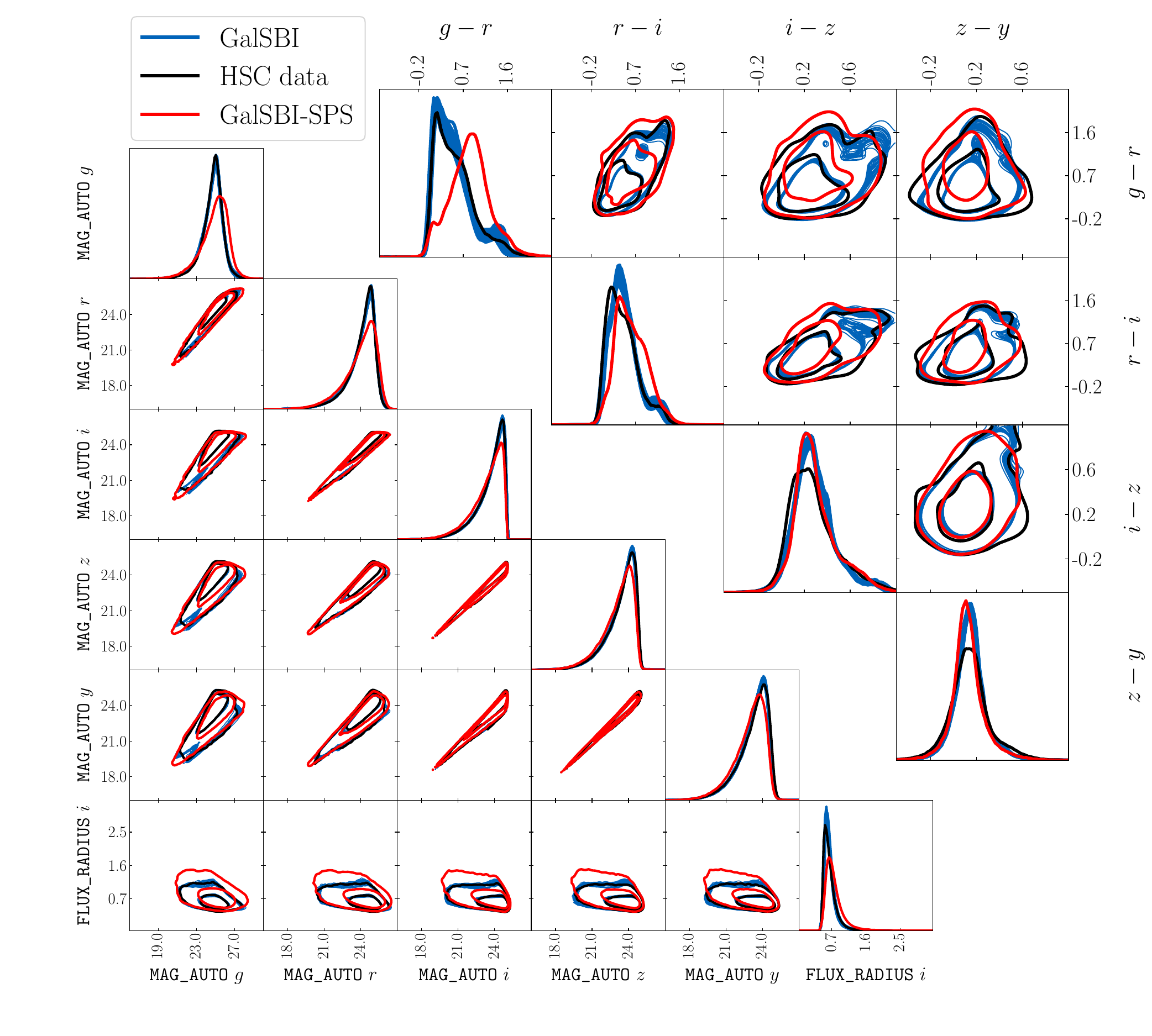}
      \caption{Comparison of magnitude, size, and colour distributions between the observed galaxies (black contours) and those from \textsc{GalSBI} (blue contours) and \textsc{GalSBI-SPS} (red contours) models for the $i \le 24$ (left panel) and $i \le 25$ (right panel) cuts. The lower left contours refer to the magnitude and size distributions, while the upper right to colour distributions. $\mathrm{MAG\_AUTO}$ refers to the galaxy magnitudes as measured by \textsc{Source Extractor}, $\mathrm{FLUX\_RADIUS}$ to the galaxy effective radius in $\mathrm{arcsec}$, while the colours are obtained as difference between the \textsc{Source Extractor} magnitudes. We report in the figure the \textsc{GalSBI} contours obtained from each posterior distribution sample in \cite{Fischbacher2025a}.}
    \label{fig:magnitude_size_colour_contours_24_25}
\end{figure}

We report in Fig. \ref{fig:magnitude_size_colour_contours_24_25} the magnitude, size, and colour distributions for the fainter samples at $i \le 24, 25$. By going at fainter magnitudes, we are including in the sample galaxies that are probing a regime where \textsc{GalSBI-SPS} is uninformed by the data we use to create the model. At both $i \le 24$ and $i \le 25$, the $i,z,y$ magnitude distributions are in very good agreement, while for the $g$ and $r$-band, \textsc{GalSBI-SPS} magnitude distributions are marginally fainter than their observed counterparts. The samples span a similar colour space, however the \textsc{GalSBI-SPS} $g-r$ colour distributions produces a population of redder $g-r$ colours than the observations. As we have seen in Fig. \ref{fig:stellarmass_sfr_size_plane}, the SFR distribution does not scale with fainter magnitude cuts as simulations do. The lack of strongly star-forming objects means redder colours than what the observations produce. Since in the $i \le 25$ galaxy sample we include more higher redshift objects whose bluer emission moves in redder wavebands, this lack of strongly star-forming objects is also marginally visibile in the $r-i$ colour. \textsc{GalSBI} magnitudes, sizes, and colours are in a better agreement with the observations since the distance metrics used to match simulations to observations have been computed using samples cut at these $i$-band thresholds.

\end{appendix}

\end{document}